\pdfoutput=1
 
\newcommand*{\ATLASLATEXPATH}{}
\documentclass[cernpreprint, UKenglish, texlive=2016, block=none]{\ATLASLATEXPATH atlasdoc}
\usepackage{\ATLASLATEXPATH atlaspackage}
\usepackage{\ATLASLATEXPATH atlasbiblatex}
\usepackage{\ATLASLATEXPATH atlasphysics}
 
\graphicspath{{logos/}{figures/}}
 
\usepackage{atlas-paper-defs}
\usepackage[utf8]{inputenc}
\usepackage{comment}
\usepackage{arydshln}
 
\AtlasTitle{Search for pair production of up-type vector-like quarks and for four-top-quark events in final states with multiple $b$-jets with the ATLAS detector}
 
\AtlasAbstract{
A search for pair production of up-type vector-like quarks ($T$) with a significant branching ratio into a top quark and either a
Standard Model Higgs boson or a $Z$ boson is presented.
The same analysis is also used to search for four-top-quark production in several new physics scenarios.
The search is based on a dataset of $pp$ collisions at $\sqrt{s}=13~\tev$ recorded in 2015 and 2016 with the ATLAS detector at the
CERN Large Hadron Collider and corresponds to an integrated luminosity of 36.1 fb$^{-1}$.
Data are analysed in the lepton+jets final state, characterised by an isolated
electron or muon with high transverse momentum, large missing transverse momentum and multiple jets,
as well as the jets+$E_{\text{T}}^{\text{miss}}$ final state, characterised by multiple jets and large missing transverse momentum.
The search exploits the high multiplicity of jets identified as originating from $b$-quarks, and
the presence of boosted, hadronically decaying top quarks and Higgs bosons reconstructed as large-radius jets, characteristic of signal events.
No significant excess above the Standard Model expectation is observed, and 95\% CL upper limits are set
on the production cross sections for the different signal processes considered. These cross-section limits
are used to derive lower limits on the mass of a vector-like $T$ quark under several branching ratio hypotheses assuming contributions
from $T \to Wb$, $Zt$, $Ht$ decays. The 95\% CL observed lower limits on the $T$ quark mass range between $0.99~\tev$ and $1.43~\tev$
for all possible values of the branching ratios into the three decay modes considered, significantly extending the reach beyond that of previous searches.
Additionally, upper limits on anomalous four-top-quark production are set in the context of an effective field
theory model, as well as in an universal extra dimensions model.}
 
\author{The ATLAS Collaboration}
 
\PreprintIdNumber{CERN-EP-2018-031}
\AtlasDate{\today}
\AtlasJournalRef{JHEP 07 (2018) 089}
\AtlasDOI{10.1007/JHEP07(2018)089}

\hypersetup{pdftitle={ATLAS document},pdfauthor={The ATLAS Collaboration}}
 
\begin{document}
 
\maketitle
 
% The next lines are included from the .//sections/introduction.tex input file
\section{Introduction}
\label{sec:intro}
 
The discovery of a new particle consistent with the Standard Model (SM) Higgs boson by the ATLAS~\cite{Aad:2012tfa} and CMS~\cite{Chatrchyan:2012ufa}
experiments at the Large Hadron Collider (LHC) represents a milestone in high-energy physics. A comprehensive programme of measurements of the Higgs boson
properties to unravel its nature is underway at the LHC,  so far yielding results compatible with the SM predictions.
This makes it more urgent than ever before to provide an explanation for why the electroweak mass scale (and the Higgs boson mass along with it) is so small
compared to the Planck scale, a situation known as the hierarchy problem. Naturalness arguments~\cite{Susskind:1978ms} require that quadratic divergences that
arise from radiative corrections to the Higgs boson mass are cancelled out by some new mechanism in order to avoid fine-tuning. To that effect, several explanations
have been proposed in theories beyond the SM (BSM).
 
One such solution involves the existence of a new strongly interacting sector, in which the Higgs boson would be a pseudo--Nambu--Goldstone boson~\cite{Hill:2002ap}
of a spontaneously broken global symmetry. One particular realisation of this scenario, referred to as Composite Higgs~\cite{Kaplan:1983sm,Agashe:2004rs},
addresses many open questions in the SM, such as the stability of the Higgs boson mass against quantum corrections,
and the hierarchy in the mass
spectrum of the SM particles, which would be explained by partial compositeness. In this scenario, the top quark would be a mostly composite particle, while all other SM fermions would be mostly elementary. A key prediction is the existence of new fermionic resonances referred to as vector-like quarks, which are also common in many other BSM
scenarios. Vector-like quarks are defined as colour-triplet spin-1/2 fermions whose left- and right-handed chiral components have the same transformation properties under
the weak-isospin SU(2) gauge group~\cite{delAguila:1982fs,AguilarSaavedra:2009es}.
Depending on the model, vector-like quarks are classified as SU(2) singlets, doublets or triplets of flavours $T$, $B$, $X$ or $Y$, in which the first two have the same
charge as the SM top and bottom quarks while the vector-like $Y$ and $X$ quarks have charge $-4/3e$ and $5/3e$. In addition, in these models, vector-like quarks
are expected to couple preferentially to third-generation quarks~\cite{delAguila:1982fs,Aguilar-Saavedra:2013wba} and can have flavour-changing
neutral-current decays in addition to the charged-current decays characteristic of chiral quarks. As a result, an up-type $T$ quark can decay not only into a $W$ boson and a
$b$-quark, but also into a $Z$ or Higgs boson and a top quark ($T \to Wb$, $Zt$, and $Ht$).
Similarly, a down-type $B$ quark can decay into a $Z$ or Higgs boson and a $b$ quark, in addition to decaying into a $W$ boson and a top quark
($B \to Wt$, $Zb$ and $Hb$).
Vector-like $Y$ quarks decay exclusively into $Wb$ and vector-like $X$ quarks decay exclusively into $Wt$.
To be consistent with the results from precision electroweak measurements a small mass-splitting between vector-like quarks belonging to
the same SU(2) multiplet is required, but no requirement is placed on which member of the multiplet is heavier~\cite{Aguilar-Saavedra:2013qpa}.
At the LHC, vector-like quarks with masses below $\sim$1$~\tev$ would be predominantly produced in pairs via the strong interaction. For higher masses, single production,
mediated by the electroweak interaction, may dominate depending on the coupling strength of the interaction between the vector-like quark and the SM quarks.
 
Another prediction of the Composite Higgs paradigm, as well as other BSM scenarios, such as Randall--Sundrum extra dimensions,
is the existence of new heavy vector resonances, which would predominantly couple to the third-generation quarks and thus lead to enhanced
four-top-quark production at high energies~\cite{Pomarol:2008bh,Lillie:2007hd,Kumar:2009vs,Cacciapaglia:2015eqa,Guchait:2007jd}.
In particular, the class of models where such vector particles are strongly coupled to the right-handed top quark are much less constrained by precision electroweak measurements than in the case of couplings to the left-handed top quark~\cite{Georgi:1994ha}. In the limit of sufficiently heavy particles, these models can be described via an effective field theory (EFT) involving a four-fermion contact interaction~\cite{Degrande:2010kt}.
The  corresponding Lagrangian is
 
\begin{equation*}
{\cal L}_{4t} = \frac{|C_{4t}|}{\Lambda^2} (\tbar_{\textrm{R}} \gamma^\mu t_{\textrm{R}}) (\tbar_{\textrm{R}} \gamma_\mu t_{\textrm{R}}),
\end{equation*}
 
\noindent where $t_{\textrm{R}}$ is the right-handed top quark spinor,  $\gamma_\mu$ are the Dirac matrices, $C_{4t}$ is the coupling constant,
and $\Lambda$ is the energy scale above which the effects of direct production of new vector particles must be considered.
Anomalous four-top-quark production also arises in Universal Extra Dimensions (UED) models, which involve new heavy particles.
For instance, in an UED model with two extra dimensions that are compactified using the geometry of the real projective plane
(2UED/RPP)~\cite{Cacciapaglia:2009pa}, the momenta of particles are discretised along the directions of the extra dimensions.
A tier of Kaluza--Klein (KK) towers is labelled by two integers, $k$ and $\ell$, referred to as ``tier $(k,\ell)$''. Within a given tier, the squared
masses of the particles are given at leading order by  $m^2 = k^2/R_4^2+\ell^2/R_5^2$, where $\pi R_4$ and $\pi R_5$
are the sizes of the two  extra dimensions. The model is parameterised by $R_4$ and $R_5$ or, alternatively, by
$m_{\KK}=1/R_4$ and $\xi=R_4/R_5$. Four-top-quark production can arise from tier (1,1), where particles
from this tier have to be pair produced because of symmetries of the model. Then they chain-decay into the
lightest particle of this tier, the heavy photon $A^{(1,1)}$, by emitting SM particles.
The branching ratios of $A^{(1,1)}$ into SM particles are not predicted by the model, although the decay
into $\ttbar$ is expected to be dominant~\cite{Cacciapaglia:2011kz}.
 
\begin{figure*}[t!]
\centering
\subfloat[]{\includegraphics[width=0.35\textwidth]{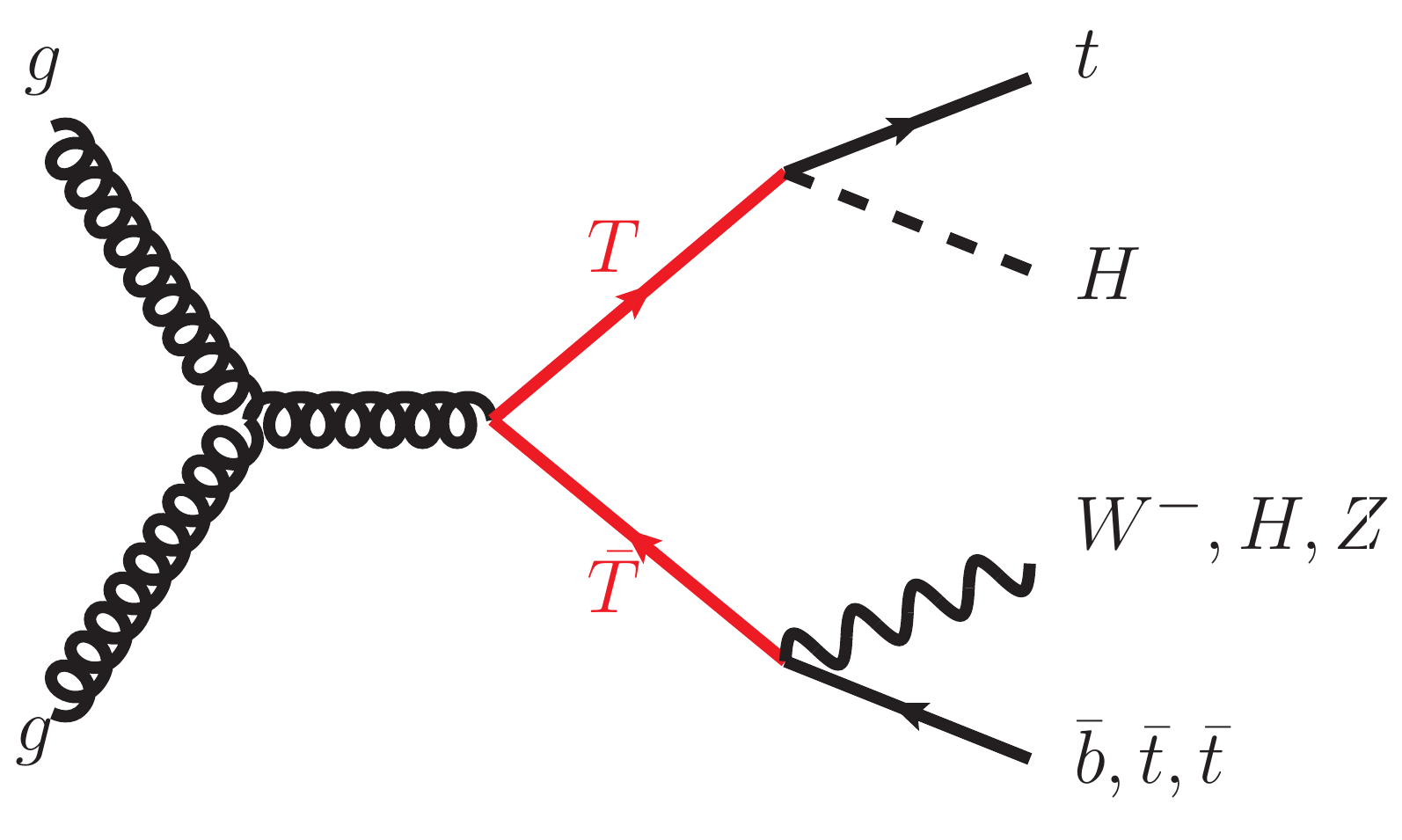}}
\subfloat[]{\includegraphics[width=0.27\textwidth]{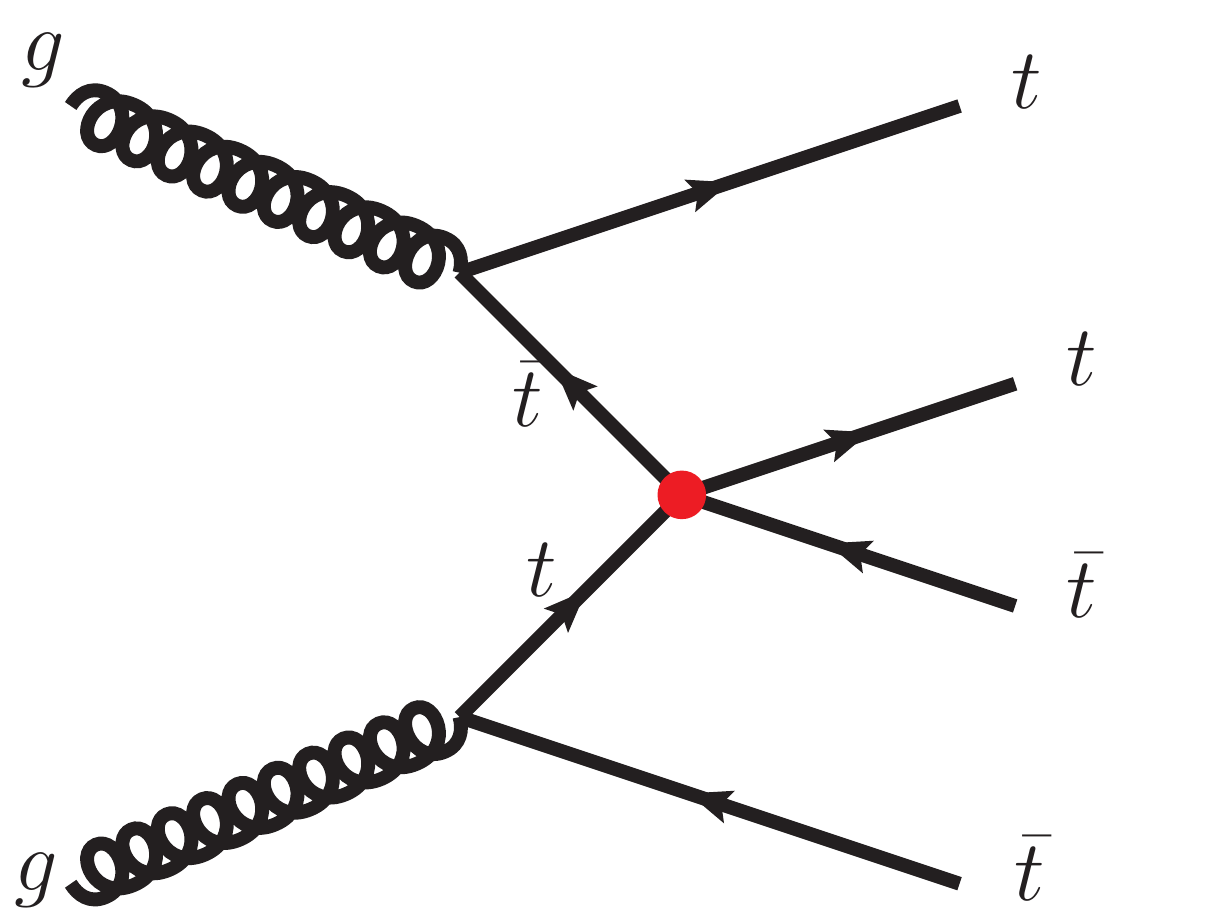}}
\subfloat[]{\includegraphics[width=0.37\textwidth]{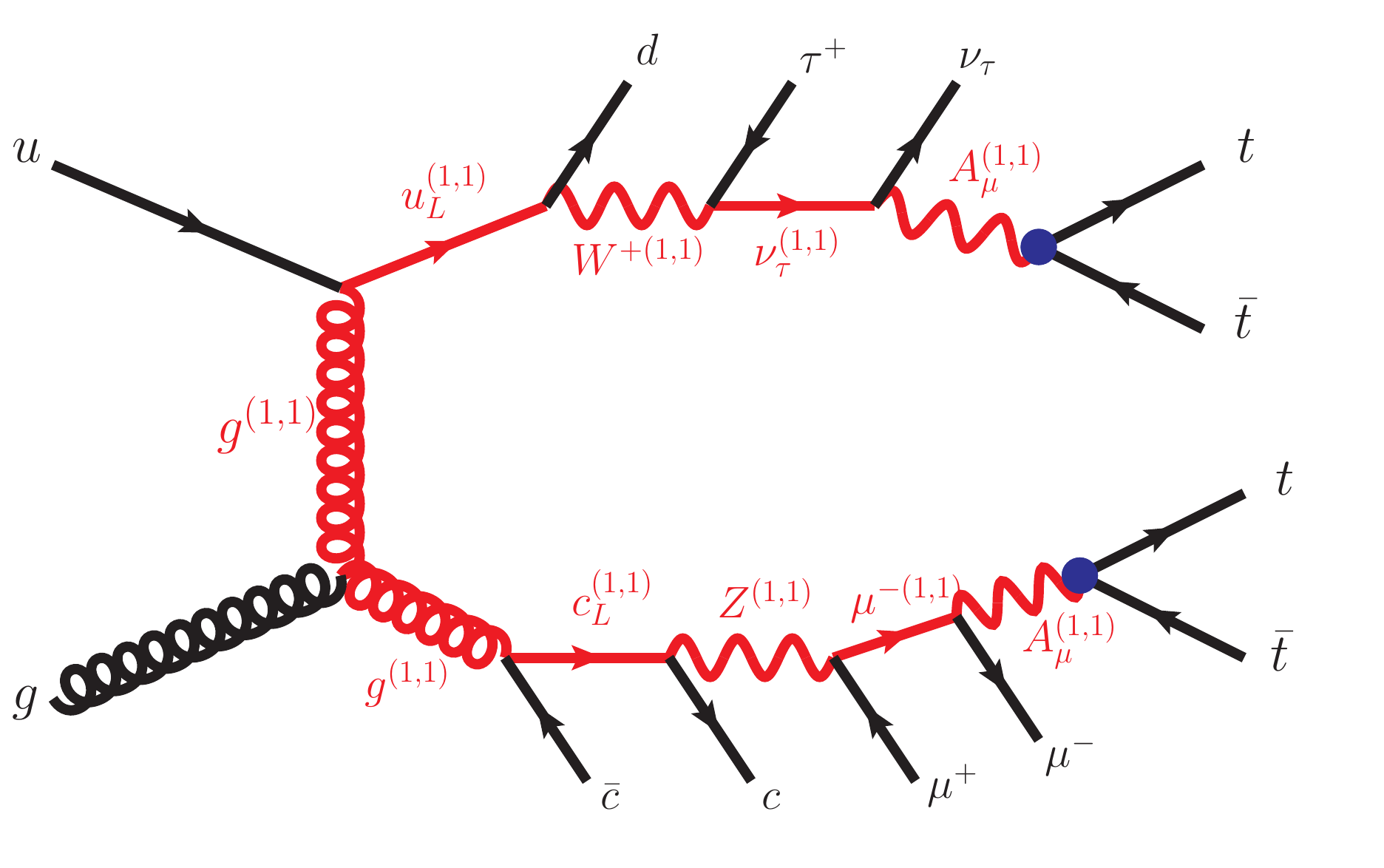}}
\caption{Representative leading-order Feynman diagrams for the signals probed by this search:
(a) $T\bar{T}$ production, and (b) four-top-quark production via an effective four-top-quark interaction in an effective field theory model,
and (c) four-top-quark production via cascade decays from Kaluza--Klein excitations in a universal extra dimensions
model with two extra dimensions compactified using the geometry of the real projective plane. }
\label{fig:signals_FD}
\end{figure*}

This paper presents a search for $T\bar{T}$ production with at least one $T$ quark decaying into $Ht$ with $H \to \bbbar$, or into $Zt$ with $Z \to \nu\bar{\nu}$,
as well as for anomalous four-top-quark production within an EFT model and within the 2UED/RPP model (see Figure~\ref{fig:signals_FD}).
Recent searches for $T\bar{T}$ production have been performed by the ATLAS~\cite{Aaboud:2017qpr,Aaboud:2017zfn} and
CMS~\cite{Sirunyan:2017usq,Sirunyan:2017pks} collaborations using up to 36.1 fb$^{-1}$ of $pp$ collisions at $\sqrt{s}=13~\tev$.
The most restrictive 95\% CL lower limits on the $T$ quark mass obtained are $1.35~\tev$ and $1.16~\tev$, corresponding to branching ratio
assumptions of ${\mathcal{B}}(T \to Wb)=1$ and ${\mathcal{B}}(T \to Zt)=1$, respectively.
Previous searches for anomalous $\fourtop$ production have been performed by the ATLAS Collaboration using the
full Run-1 dataset~\cite{Aad:2015gdg,Aad:2015kqa}, where
95\% CL limits of $|C_{4t}|/\Lambda^2<6.6~\tev^{-2}$ and
$m_{KK}>1.1~\tev$
were obtained
in the case of the EFT and the 2UED/RPP models, respectively. A recent search by the CMS Collaboration~\cite{Sirunyan:2017roi}
using 35.9 fb$^{-1}$ of $pp$ collisions at $\sqrt{s}=13~\tev$ has set an upper limit of 41.7 fb on the SM $\fourtop$ production cross section, about
4.5 times the SM prediction, thus placing some constraints on anomalous production with kinematics like in the SM.
 
This search uses 36.1 fb$^{-1}$ of data at $\sqrt{s}=13~\tev$ recorded in 2015 and 2016 by the ATLAS Collaboration,
and it closely follows the strategy developed in Run 1~\cite{Aad:2015kqa}, although it incorporates new ingredients, such as the identification of boosted objects, to substantially enhance sensitivity for heavy resonances.
Data are analysed in the lepton+jets final state, characterised by an isolated electron or muon with high transverse momentum,
large missing transverse momentum and multiple jets and, for the first time in searches for vector-like quarks, also in the jets+$\met$ final state,
characterised by multiple jets and large missing transverse momentum.
 
% End of text imported from the .//sections/introduction.tex input file
 
% The next lines are included from the .//sections/detector.tex input file
\section{ATLAS detector}
\label{sec:detector}
 
The ATLAS detector~\cite{PERF-2007-01} at the LHC covers almost the entire solid angle around the collision point,\footnote{ATLAS
uses a right-handed coordinate system with its origin at the nominal interaction point (IP) in the
centre of the detector.
The $x$-axis points from
the IP to the centre of the LHC ring,
the $y$-axis points upward,
and the $z$-axis coincides with the axis of the beam pipe.
Cylindrical coordinates ($r$,$\phi$) are used
in the transverse plane, $\phi$ being the azimuthal angle around the beam pipe. The pseudorapidity is defined in
terms of the polar angle $\theta$ as $\eta = - \ln \tan(\theta/2)$.
Angular distance is measured in units of $\Delta R\equiv \sqrt{(\Delta\eta)^2+(\Delta\phi)^2}$.} and
consists of an inner tracking detector surrounded by a thin superconducting solenoid producing a
2~T axial magnetic field, electromagnetic and hadronic calorimeters, and a muon spectrometer
incorporating three large toroid magnet assemblies. The inner detector consists of a high-granularity silicon pixel detector,
including the
insertable B-layer~\cite{IBL}, installed in 2014,
and a silicon microstrip tracker,
together providing a precise reconstruction of tracks of charged particles in the pseudorapidity range $|\eta|<2.5$, complemented by a
transition radiation tracker providing tracking and electron identification information for $|\eta|<2.0$.
The calorimeter system covers the pseudorapidity range $|\eta| < 4.9$.
Within the region $|\eta|< 3.2$, electromagnetic (EM) calorimetry is provided by barrel and
endcap high-granularity lead/liquid-argon (LAr) electromagnetic calorimeters,
with an additional thin LAr presampler covering $|\eta| < 1.8$,
to correct for energy loss in material upstream of the calorimeters.
Hadronic calorimetry is provided by
a
steel/scintillator-tile calorimeter,
segmented into three barrel structures within $|\eta| < 1.7$, and two copper/LAr hadronic endcap calorimeters.
The solid angle coverage is completed with forward copper/LAr and tungsten/LAr calorimeter modules
optimised for electromagnetic and hadronic measurements, respectively.
The muon spectrometer measures the
trajectories
of muons with $|\eta|<2.7$ using multiple layers of high-precision tracking
chambers located in a toroidal field of approximately 0.5~T and 1~T in the central and endcap regions of ATLAS, respectively.
The muon spectrometer is also instrumented with separate trigger chambers covering $|\eta|<2.4$.
A two-level trigger system~\cite{Aaboud:2016leb}, consisting of a hardware-based Level-1 trigger followed by
a software-based High-Level Trigger (HLT), is used to reduce the event rate to a maximum of around 1 kHz for offline storage.
% End of text imported from the .//sections/detector.tex input file
 
% The next lines are included from the .//sections/objects.tex input file
\section{Object reconstruction}
\label{sec:objects}
 
Interaction vertices from the proton--proton collisions are reconstructed from at least two tracks
with transverse momentum ($\pt$) larger than $400~\mev$ that are consistent with originating from the beam collision region in the $x$--$y$ plane.
If more than one primary vertex candidate is found, the
candidate whose associated tracks form the largest sum of squared $\pt$~\cite{ATL-PHYS-PUB-2015-026}
is selected as the hard-scatter primary vertex.
 
Electron candidates~\cite{ATLAS-CONF-2016-024,ATL-PHYS-PUB-2016-015} are reconstructed from energy
clusters in the EM calorimeter that are matched to reconstructed tracks in the inner detector and have $\pt>30~\gev$ and
$|\eta_{\textrm{cluster}}| < 2.47$;
candidates in the transition region between the EM barrel and endcap calorimeter ($1.37 < |\eta_{\textrm{cluster}}| < 1.52$) are excluded.
They are also required to satisfy the ``tight'' likelihood-based identification criteria~\cite{ATLAS-CONF-2016-024} based
on calorimeter, tracking and combined variables that provide separation between electrons and jets.
Muon candidates~\cite{Aad:2016jkr} are reconstructed by matching track segments in 
different
layers of the muon spectrometer
to tracks found in the inner detector.  The resulting muon candidates are refitted using the complete
track information from both detector systems and are required to have $\pt>30~\gev$ and $|\eta|<2.5$.
Electron (muon) candidates are matched to the primary vertex by requiring that the significance of their transverse impact parameter, $d_0$,
satisfies $|d_0/\sigma(d_0)|<5(3)$, where $\sigma(d_0)$ is the measured uncertainty in $d_0$,
and by requiring that their longitudinal impact parameter, $z_0$, satisfies $|z_0 \sin\theta|<0.5$~mm.
To further reduce the background from non-prompt leptons, photon conversions and hadrons, lepton candidates are also required to be isolated.
A lepton isolation criterion is defined by calculating the quantity $I_R = \sum \pt^{\textrm{trk}}$, where
the sum includes all tracks (excluding the lepton candidate itself) within the cone defined by
$\Delta R<R_{\textrm{cut}}$ about the 
direction
of the lepton.
The value of $R_{\textrm{cut}}$ is the smaller of $r_{\textrm{min}}$ and $10~\gev/\pt^\ell$, where
$r_{\textrm{min}}$ is set to 0.2 (0.3) for electron (muon) candidates,
and $\pt^\ell$ is the lepton $\pt$.
All lepton candidates must satisfy $I_R/\pt^\ell < 0.06$.
 
Candidate jets are reconstructed with the anti-$k_t$ algorithm~\cite{Cacciari:2008gp,Cacciari:2005hq,Cacciari:2011ma} with a
radius parameter $R=0.4$ (referred to as ``small-$R$ jets''), using topological clusters~\cite{Aad:2016upy}
built from energy deposits in the calorimeters calibrated to the electromagnetic scale.
The reconstructed jets are then calibrated to the particle level by the application of a jet energy scale
derived from simulation and {\textit{in situ}} corrections based on $\sqrt{s}=13~\tev$ data~\cite{Aaboud:2017jcu}.
Calibrated
jets are required to have $\pt > 25~\gev$ and $|\eta| < 2.5$.
Quality criteria are imposed to reject events that contain any jets arising from non-collision sources
or detector noise~\cite{ATLAS-CONF-2015-029}.  To reduce the contamination due to jets originating from pile-up interactions,
an additional requirement on the Jet Vertex Tagger (JVT)~\cite{Aad:2015ina} output
is made for jets with $\pt<60~\gev$ and $|\eta| < 2.4$.
 
Jets containing $b$-hadrons are identified ($b$-tagged) via an algorithm~\cite{Aad:2015ydr,ATL-PHYS-PUB-2016-012}
that uses multivariate techniques to combine information about the impact parameters of displaced tracks and the  topological properties
of secondary and tertiary decay vertices reconstructed within the jet. For each jet, a value for the multivariate $b$-tagging discriminant is
calculated. In this analysis, a jet is considered $b$-tagged if this value is above the threshold corresponding to
an average 77\% efficiency to tag a $b$-quark jet, with a light-jet\footnote{Light-jet refers to a jet originating from the hadronisation of a light quark
($u$, $d$, $s$) or a gluon.} rejection factor of $\sim$134 and a charm-jet rejection factor of $\sim$6.2, as determined for jets with
$\pt >20~\gev$ and $|\eta|<2.5$ in simulated $\ttbar$ events.
 
Overlaps between candidate objects are removed sequentially.  Firstly, electron candidates that lie within $\Delta R = 0.01$ of a muon
candidate are removed to suppress contributions from muon bremsstrahlung.  Overlaps between electron and jet candidates are resolved
next, and finally, overlaps between remaining jet candidates and muon candidates are removed.
Clusters from identified electrons are not excluded during jet reconstruction.
In order to avoid double-counting of electrons as jets, the closest jet whose axis is within ${\Delta}R = 0.2$ of an electron is discarded. If
the electron is within ${\Delta}R = 0.4$ of the axis of any jet after this initial removal, the jet is retained and  the electron is removed.
The overlap removal procedure between the remaining jet candidates and muon candidates is designed to remove those muons
that are likely to have arisen in the decay chain of hadrons and to retain the overlapping jet instead.
Jets and muons may also appear in close proximity when the jet results from high-$\pt$ muon bremsstrahlung,
and in such cases the jet should be removed and the muon retained. Such jets are characterised by having very
few matching inner-detector tracks. Selected muons that satisfy $\Delta R(\mu,{\textrm{jet}}) < 0.04+10~\gev/\pt^\mu$ are rejected
if the jet has at least three tracks originating from the primary vertex; otherwise the jet is removed and the muon is kept.
 
The candidate small-$R$ jets surviving the overlap removal procedure discussed above are used as inputs for further
jet reclustering~\cite{Nachman:2014kla} using the anti-$k_t$ algorithm with a radius parameter $R=1.0$.
In this way it is possible to evaluate the uncertainty in the mass
of the large-$R$ jets
that arises from the uncertainties in the energy scale and resolution of its constituent small-$R$ jets.
In order to suppress contributions from pile-up and soft radiation, the reclustered large-$R$ (RCLR)
jets are trimmed~\cite{Krohn:2009th} by removing all small-$R$ (sub)jets within a RCLR jet that have $\pt$ below 5\% of the $\pt$ of the
reclustered jet.
Due to the pile-up suppression and $\pt>25~\gev$ requirements made on the small-$R$ jets,  the average fraction of small-$R$ jets removed
by the trimming requirement is less than 1\%.
The resulting RCLR jets are required to have $|\eta|<$~2.0 and are used to identify high-$\pt$ hadronically decaying
top quark or Higgs boson candidates by making requirements on their transverse momentum, mass, and number of constituents.
Hadronically decaying top quark candidates are reconstructed as RCLR jets with $\pt > 300~\gev$, mass larger than 140~\gev, and
at least two subjets.
Higgs boson candidates are reconstructed as RCLR jets with $\pt > 200~\gev$, a mass between 105 and
140~\gev, and a $\pt$-dependent requirement on the number of subjets (exactly two for $\pt < 500~\gev$,
and one or two for $\pt >500~\gev$).
In the following, these are referred to as ``top-tagged'' and ``Higgs-tagged'' jets, respectively, while the term ``jet'' without further qualifiers
is used to refer to small-$R$ jets.
 
The missing transverse momentum $\mpt$ (with magnitude $\met$) is defined as the negative vector sum of the
$\pt$ of all selected and calibrated objects in the event, including a term to account for energy from soft particles
in the event which are not associated with any of the selected objects.
This soft term is calculated from inner-detector tracks matched to the selected primary vertex to make it more resilient to
contamination from pile-up interactions~\cite{Aad:2016nrq,ATL-PHYS-PUB-2015-027}.

% End of text imported from the .//sections/objects.tex input file
 
% The next lines are included from the .//sections/data_presel.tex input file
\section{Data sample and event preselection}
\label{sec:data_presel}
 
This search is based on a dataset of $pp$ collisions at $\sqrt{s}=13~\tev$ with 25 ns bunch spacing collected by the ATLAS experiment in 2015 and 2016,
corresponding to an integrated luminosity of $36.1~\ifb$.
Only events recorded with a single-electron trigger, a single-muon trigger, or a $\met$ trigger under stable beam conditions and for which
all detector subsystems were operational are considered.
 
Single-lepton triggers with low $\pt$ threshold and lepton isolation requirements are combined in a logical OR
with higher-threshold triggers without isolation requirements to give maximum efficiency. For muon triggers, the lowest $\pt$ threshold is
20 (26)~\gev\ in 2015 (2016), while the higher $\pt$ threshold is 50~\gev\ in both years. For electrons,
triggers
with a $\pt$ threshold of 24 (26)~\gev\ in 2015 (2016)
and isolation requirements
are used
along with triggers with a 60~\gev\ threshold and no isolation requirement, and with a
120 (140)~\gev\
threshold with
looser identification criteria.
The {\met} trigger~\cite{Aaboud:2016leb} considered uses an $\met$ threshold of 70~\gev\ in the HLT in 2015
and a run-period-dependent {\met} threshold varying between 90~\gev\ and 110~\gev\ in 2016.
 
Events satisfying the trigger selection are required to have at least one primary vertex candidate.
They are then classified into the ``1-lepton'' or ``0-lepton'' channels depending on the multiplicity of selected leptons.
Events in the 1-lepton channel are required to satisfy a single-lepton trigger and to have exactly one selected electron or muon
that matches, with $\Delta R < 0.15$, the lepton reconstructed by the trigger. In the following, 1-lepton events satisfying either the electron
or muon selections are combined and treated as a single analysis channel.
Events in the 0-lepton channel are required to satisfy the $\met$ trigger and to have no selected leptons.
In addition, events in the 1-lepton (0-lepton) channel are required to have $\geq$5 ($\geq$6) small-$R$ jets.
In the following, all selected small-$R$ jets are considered, including those used to build large-$R$ jets.
For both channels, backgrounds
that do not include
$b$-quark jets are suppressed by requiring at least two $b$-tagged jets.
 
Additional requirements are made to suppress the background from multijet production.
In the case of the 1-lepton channel, requirements are made on $\met$ as well as on the transverse mass of the lepton and $\met$ system ($\mtw$):\footnote{$\mtw = \sqrt{2 p^\ell_{\mathrm T} \met (1-\cos\Delta\phi)}$, where $p^\ell_{\mathrm T}$  is the transverse momentum (energy) of the muon (electron) and $\Delta\phi$ is the
azimuthal angle separation between the lepton and the direction of the missing transverse momentum.} $\met >20~\gev$ and $\met +\mtw>60~\gev$.
In the case of the 0-lepton channel, the requirements are $\met >200~\gev$ (for which the $\met$ trigger is fully efficient) and
$\Delta\phi_{\textrm{min}}^\textrm{4j}>0.4$, where $\Delta\phi_{\textrm{min}}^\textrm{4j}$ is the minimum azimuthal separation between $\mpt$ and each of the four highest-$\pt$ jets. The latter requirement in the 0-lepton channel is very effective in suppressing multijet events, where the large $\met$
results from the mismeasurement of a high-$\pt$ jet or the presence of neutrinos emitted close to a jet axis.
 
The above requirements are referred to as the ``preselection'' and are summarised in Table~\ref{tab:preselection}.
 
\begin{table*}[t!]
\begin{center}
\begin{tabular}{l|c|c}
\toprule\toprule
\multicolumn{3}{c}{Preselection requirements } \\
\midrule
Requirement & 1-lepton channel & 0-lepton channel \\
\midrule
Trigger & Single-lepton trigger & $\met$ trigger \\
Leptons  & =1 isolated $e$ or $\mu$ & =0 isolated $e$ or $\mu$ \\
Jets  & $\geq$5 jets & $\geq$6 jets \\
$b$-tagging & $\geq$2 $b$-tagged jets & $\geq$2 $b$-tagged jets \\
$\met$ & $\met >20~\gev$ & $\met >200~\gev$ \\
Other $\met$-related & $\met +\mtw>60~\gev$ & $\Delta\phi_{\textrm{min}}^\textrm{4j}>0.4$ \\
\bottomrule\bottomrule
\end{tabular}
\caption{\small{Summary of preselection requirements for the 1-lepton and 0-lepton channels.
Here $\mtw$ is the transverse mass of the lepton and the $\met$ vector, and $\Delta\phi_{\textrm{min}}^\textrm{4j}$ is the
minimum azimuthal separation between the $\met$ vector and each of the four highest-$\pt$ jets.}}
\label{tab:preselection}
\end{center}
\end{table*}
% End of text imported from the .//sections/data_presel.tex input file
 
% The next lines are included from the .//sections/signal_background_modeling.tex input file
\section{Signal and background modelling}
\label{sec:signal_background_model}
 
Signal and most background processes were modelled using Monte Carlo (MC) simulations.
In the simulation, the top quark and SM Higgs boson masses were set to $172.5~\gev$ and $125~\gev$, respectively.
All simulated samples, except those produced with the {\sherpa}~\cite{Gleisberg:2008ta} event generator,
utilised \textsc{EvtGen} v1.2.0~\cite{Lange:2001uf} to model the decays of heavy-flavour hadrons.
To model the effects of pile-up, events from minimum-bias interactions were generated using the \textsc{Pythia} 8.186~\cite{Sjostrand:2007gs}
event generator and overlaid onto the simulated hard-scatter events according to the luminosity profile of the recorded data.
The generated events were processed through a simulation~\cite{Aad:2010ah} of the ATLAS detector geometry and response
using \textsc{Geant4}~\cite{Agostinelli:2002hh}. A faster simulation, where the full \textsc{Geant4} simulation of
the calorimeter response is replaced by a detailed parameterisation of the shower shapes~\cite{FastCaloSim},
was adopted for some of the samples used to estimate
systematic uncertainties.
Simulated events are processed through the same reconstruction software as the data, and
corrections are applied so that the object identification efficiencies, energy
scales and energy resolutions match those determined from data control samples.
 
\subsection{Signal modelling}
\label{sec:signal_model}
 
Samples of simulated $T\bar{T}$ events were generated with the leading-order (LO) generator\footnote{In the following, the order of a generator should be understood
as referring to the order in the strong coupling constant at which the matrix element calculation is performed.} \textsc{Protos}~2.2~\cite{protos,AguilarSaavedra:2009es}
using the NNPDF2.3 LO~\cite{Ball:2012cx} parton distribution function (PDF) set and passed to \textsc{Pythia} 8.186~\cite{Sjostrand:2007gs} for parton showering and fragmentation.
The A14~\cite{ATLASUETune4} set of optimised parameters for the underlying event (UE) description using the NNPDF2.3 LO PDF set,
referred to as the ``A14 UE tune'', was used. The samples were generated assuming singlet couplings and for heavy-quark masses between $350~\gev$ and $1.5~\tev$ in steps of $50~\gev$.
Additional samples were produced at three mass points ($700~\gev$, $950~\gev$ and $1.2~\tev$) assuming doublet couplings 
in order to confirm that,
at fixed branching fraction,
kinematic differences
arising from the different chirality of singlet and doublet couplings
have negligible impact on this search.
The vector-like quarks were forced to decay with a branching ratio of $1/3$ into each of the three modes ($W,Z,H$). These samples were reweighted using generator-level information
to allow results to be interpreted for arbitrary sets of branching ratios that are consistent with the three decay modes summing to unity.
The generated samples were normalised to the theoretical cross sections computed using \textsc{Top++} v2.0~\cite{Czakon:2011xx} at
next-to-next-to-leading order (NNLO) in quantum chromodynamics (QCD), including resummation of next-to-next-to-leading logarithmic (NNLL) soft gluon
terms~\cite{Cacciari:2011hy,Baernreuther:2012ws,Czakon:2012zr,Czakon:2012pz,Czakon:2013goa},
and using the MSTW 2008 NNLO~\cite{Martin:2009iq,Martin:2009bu} set of PDFs.
The predicted pair-production cross section at $\sqrt{s}=13~\tev$ ranges from 24~pb for a vector-like quark mass of $350~\gev$ to 2.0~fb for a mass of $1.5~\tev$,
with an uncertainty that increases from 8\% to 18\% over this mass range.
The theoretical uncertainties result from variations of the factorisation and renormalisation scales, as well as from uncertainties in the
PDF and $\alpha_{\textrm{S}}$. The latter two represent the largest contribution to the overall theoretical uncertainty in the cross section
and were calculated using the PDF4LHC prescription~\cite{Botje:2011sn}
with the MSTW 2008 68\% CL NNLO, CT10 NNLO~\cite{Lai:2010vv,Gao:2013xoa} and NNPDF2.3 5f FFN~\cite{Ball:2012cx} PDF sets.
 
Samples of simulated four-top-quark events produced via an EFT and within the 2UED/RPP model were generated at LO with the {\amcatnlolong}~\cite{Alwall:2014hca}
generator (referred to in the following as {\amcatnlo}; the versions used are 2.2.3 and 1.5.14 for EFT and 2UED/RPP, respectively) and the NNPDF2.3 LO PDF set,
interfaced to \textsc{Pythia} 8 (the versions used are 8.205 and 8.186 for EFT and 2UED/RPP, respectively) and the A14 UE tune. The EFT $\fourtop$ sample was normalised
assuming $|C_{4t}|/\Lambda^2 = 4\pi~\tev^{-2}$, where $C_{4t}$ denotes the coupling constant and $\Lambda$ the energy scale of new physics, which yields a cross section
of 928~fb computed using \amcatnlo. In the case of the 2UED/RPP model, samples were generated for four different values of $m_{\KK}$ (from $1~\tev$ to $1.8~\tev$ in steps of $200~\gev$) and the
\textsc{Bridge}~\cite{Meade:2007js} generator was used to decay the pair-produced excitations from tier (1,1) generated by \textsc{Madgraph5}.
The corresponding predicted cross section ranges from 343~fb for $m_{\KK}=1~\tev$ to 1.1~fb for $m_{\KK}=1.8~\tev$.
 
\subsection{Background modelling}
\label{sec:bkg_model}
 
After the event preselection, the main background is $\ttbar$ production, often in association with jets, denoted by $\ttbar$+jets in the following.
Small contributions arise from single-top-quark, $W/Z$+jets, multijet and diboson ($WW,WZ,ZZ$) production, as well as from the associated
production of a vector boson $V$ ($V=W,Z$) or a Higgs boson and a $\ttbar$ pair ($\ttbar V$ and $\ttbar H$). All backgrounds are estimated using
samples of simulated events and initially normalised to their theoretical cross sections, with the exception of the multijet background,
which is estimated using data-driven methods. The background prediction is further improved during the statistical analysis by performing a likelihood
fit to data using multiple signal-depleted search regions, as discussed in Section~\ref{sec:strategy}.
 
The nominal sample used to model the $\ttbar$ background was generated with the NLO generator {\powheg}~v2 \cite{Frixione:2007nw,Nason:2004rx,Frixione:2007vw,Alioli:2010xd}
using the CT10 PDF set~\cite{Lai:2010vv}. The {\powheg} model parameter $h_{\textrm{damp}}$, which controls matrix element to parton shower matching
and effectively regulates the high-$\pt$ radiation, was set to the top quark mass, a setting that was found to describe the $\ttbar$ system's $\pt$ at
$\sqrt{s} = 7~\tev$~\cite{ATL-PHYS-PUB-2015-002}.
The nominal $\ttbar$ sample was interfaced to {\pythia}~6.428~\cite{Sjostrand:2006za} with the CTEQ6L PDF set and the
Perugia 2012 (P2012) UE tune~\cite{Skands:2010ak}.  Alternative $\ttbar$ simulation samples used to derive systematic uncertainties are
described in Section~\ref{sec:syst_bkgmodeling}.
 
All $\ttbar$ samples were generated inclusively, but events are categorised depending
on the flavour content of additional particle jets not originating from the decay of the $\ttbar$ system (see Ref.~\cite{Aad:2015gra} for details).
Events labelled as either \ttbin\ or \ttcin\ are generically referred in the following as $\ttbar$+HF events, where HF stands for ``heavy flavour''.
A finer categorisation of \ttbin\ events is considered for the purpose of applying further corrections and
assigning systematic uncertainties associated with the modelling of heavy-flavour production in different topologies~\cite{Aad:2015gra}.
The remaining events are labelled as $\ttbar$+light-jets events, including those with no additional jets.
In previous studies, better agreement between data and prediction was observed, particularly for the top quark $\pt$ distribution,
when comparing to NNLO calculations~\cite{Aad:2015mbv}.
These small
improvements to the modelling are incorporated by reweighting all $\ttbar$ samples to match their top quark $\pt$ distribution
to that predicted at NNLO accuracy in QCD~\cite{Czakon:2015owf,Czakon:2016dgf}.
This correction is not applied to \ttbin\ events, which instead
are reweighted to an NLO prediction in the four-flavour (4F) scheme of \ttbin\ including parton showering~\cite{Cascioli:2013era}, based on
{\ShOLlong}~\cite{Gleisberg:2008ta, Cascioli:2011va} (referred to as {\ShOL} in the following) using the CT10 PDF set.  This reweighting is performed
separately for each of the \ttbin\ categories in such a way that their inter-normalisation and the shape of the relevant
kinematic distributions are at NLO accuracy, while preserving the nominal \ttbin\ cross section in {\powheg}+{\pythia}.
The corrections described in this paragraph
are applied to the nominal as well as the alternative $\ttbar$ samples.
 
Samples of single-top-quark events corresponding to the $t$-channel production mechanism were generated with the
{\powheg}~v1~\cite{Frederix:2012dh} generator that uses the 4F scheme  for the NLO matrix element calculations
and the fixed 4F \textsc{CT10}f\textsc{4}~\cite{Lai:2010vv} PDF set.
Samples corresponding to the $Wt$- and $s$-channel production mechanisms were generated
with {\powheg}~v2 using the CT10 PDF set. Overlaps between the $\ttbar$ and $Wt$ final states are avoided by using
the ``diagram removal'' scheme~\cite{Frixione:2005vw}.
The parton shower, hadronisation and the underlying event are modelled using {\pythia} 6.428 with the CTEQ6L1 PDF set
in combination with the P2012 UE tune.
The single-top-quark samples were normalised to the approximate NNLO theoretical cross
sections~\cite{Kidonakis:2011wy,Kidonakis:2010ux,Kidonakis:2010tc}.
 
Samples of $W/Z$+jets events were generated with the {\sherpa}~2.2~\cite{Gleisberg:2008ta} generator.
The matrix element was calculated for up to two partons at NLO and up to four partons at LO using
\textsc{Comix}~\cite{Gleisberg:2008fv} and \textsc{OpenLoops}~\cite{Cascioli:2011va}. The matrix element calculation
was merged with the {\sherpa} parton shower~\cite{Schumann:2007mg} using the ME+PS@NLO prescription~\cite{Hoeche:2012yf}. The PDF set used for the matrix-element calculation is NNPDF3.0NNLO~\cite{Ball:2014uwa} with a dedicated parton shower tuning developed for {\sherpa}.
Separate samples were generated for different $W/Z$+jets categories using filters for a $b$-jet
($W/Z+\geq$1$b$+jets), a $c$-jet and no $b$-jet ($W/Z+\geq$1$c$+jets), and with a veto on $b$- and $c$-jets
($W/Z$+light-jets), which were combined into the inclusive $W/Z$+jets samples.
Both the $W$+jets and $Z$+jets samples were normalised to their respective inclusive NNLO theoretical
cross sections in QCD calculated with \textsc{FEWZ}~\cite{Anastasiou:2003ds}.
 
Samples of $WW/WZ/ZZ$+jets events were generated with {\sherpa}~2.1.1 using the CT10 PDF set
and include processes containing up to four electroweak vertices. The matrix element includes zero additional partons
at NLO and up to three partons at LO using the same procedure as for the $W/Z$+jets samples.
The final states simulated require one of the bosons to decay leptonically and the other hadronically.
All diboson samples were normalised to their NLO theoretical cross sections provided by {\sherpa}.
 
Samples of $\ttbar V$ and $\ttbar H$ events were generated with {\amcatnlo}~2.3.2, using
NLO matrix elements and the NNPDF3.0NLO~\cite{Ball:2014uwa} PDF set.
Showering was performed using {\pythia} 8.210 and the A14 UE tune.
The $\ttbar V$ samples were normalised to the NLO cross section computed with {\amcatnlo}.
The $\ttbar H$ sample was normalised using the NLO cross section~\cite{Raitio:1978pt,Beenakker:2002nc,Dawson:2003zu,Yu:2014cka,Frixione:2015zaa}
and the Higgs boson decay branching ratios calculated using \textsc{Hdecay}~\cite{Djouadi:1997yw}.
 
The production of four-top-quark events in the SM was simulated by samples generated at LO using {\amcatnlo}~2.2.2
and the NNPDF2.3 LO PDF set, interfaced to {\pythia} 8.186 in combination with the A14 UE tune.
The sample was normalised to a cross section of 9.2 fb, computed at NLO~\cite{Alwall:2014hca}.
 
The background from multijet production (``multijet background'' in the following) in the 1-lepton channel contributes to the selected
data sample via  several production and misreconstruction mechanisms.
In the electron channel, it consists of non-prompt electrons (from semileptonic $b$- or $c$-hadron decays) as well as
misidentified photons (e.g.~from a conversion of a photon into an $e^+e^-$ pair) or jets with a high fraction of
their energy deposited in the EM calorimeter.  In the muon channel, the multijet background is predominantly from
non-prompt muons.  The multijet background normalisation and shape are estimated directly from data by using the ``matrix method''
technique~\cite{Aad:2010ey}, which exploits differences in lepton identification and isolation properties between prompt leptons
and leptons that are either non-prompt or result from the misidentification of photons or jets.
Further details can be found in Ref.~\cite{Aad:2015kqa}. The main type of multijet background that contributes to the 0-lepton
channel are events in which the energy of a high-$p_\mathrm{T}$ jet is mismeasured, consequently leading to a large missing transverse momentum in
the final state. Most of this background is suppressed by selecting events satisfying $\Delta\phi_{\textrm{min}}^\textrm{4j}>0.4$. The
remaining multijet background in each search region is estimated from a control region defined with the same selection as the
search region, but with the selection on $\Delta\phi_{\textrm{min}}^\textrm{4j}$ changed to $\Delta\phi_{\textrm{min}}^\textrm{4j}<0.1$.
The normalisation of the multijet background is extrapolated from the control region to its corresponding search region by performing an exponential fit to the $\Delta\phi_{\textrm{min}}^\textrm{4j}$ distribution in the range $0< \Delta\phi_{\textrm{min}}^\textrm{4j}<0.4$. The background prediction
is validated by comparing the data and total prediction in multijet-rich samples selected by choosing ranges of $\Delta\phi_{\textrm{min}}^\textrm{4j}$
(e.g.  $0.3<\Delta\phi_{\textrm{min}}^\textrm{4j}< 0.4$).
 
% End of text imported from the .//sections/signal_background_modeling.tex input file
 
% The next lines are included from the .//sections/strategy.tex input file
\section{Search strategy}
\label{sec:strategy}
 
\begin{figure*}[t!]
\centering
\subfloat[]{\includegraphics[width=0.45\textwidth]{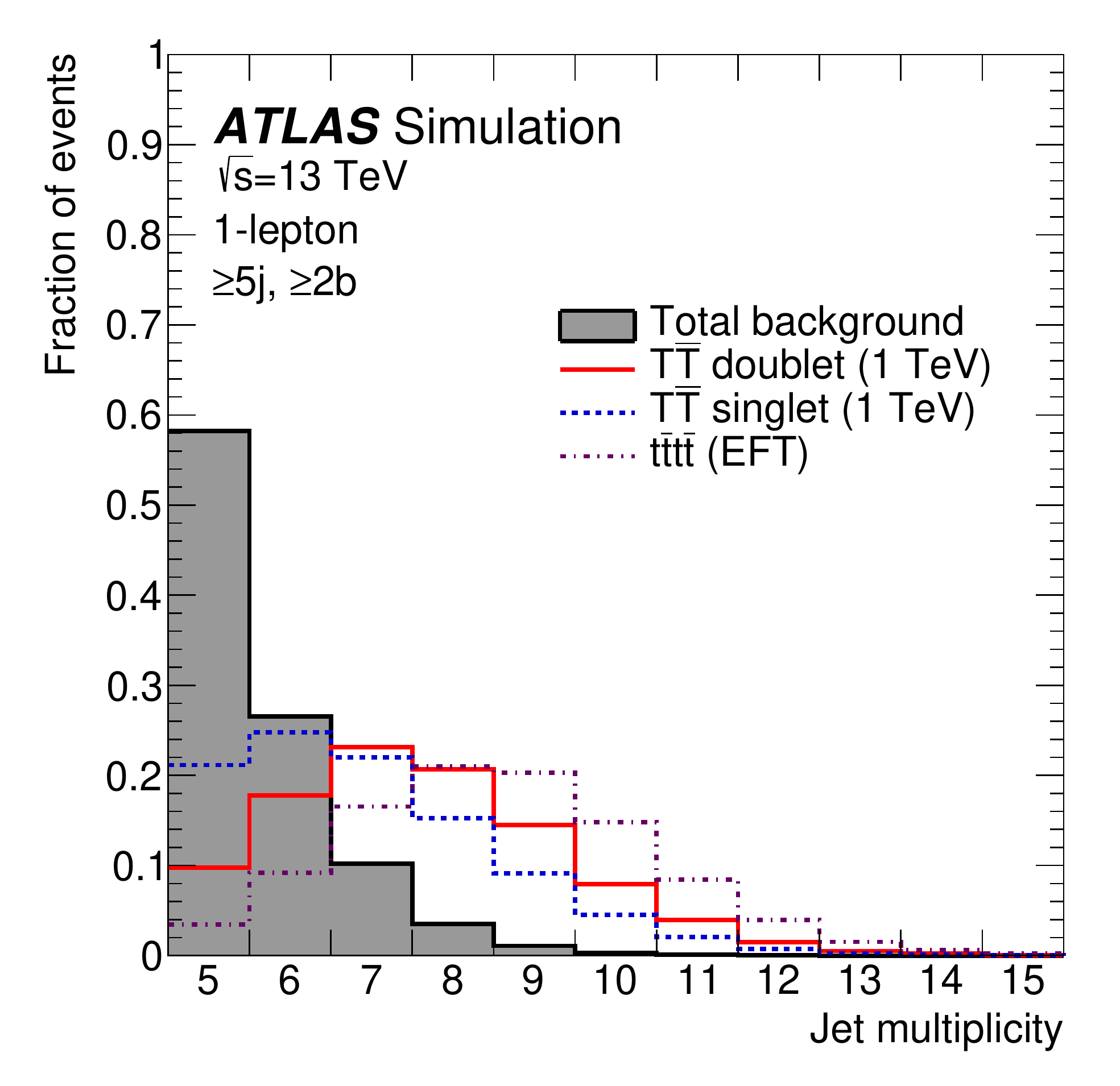}}
\subfloat[]{\includegraphics[width=0.45\textwidth]{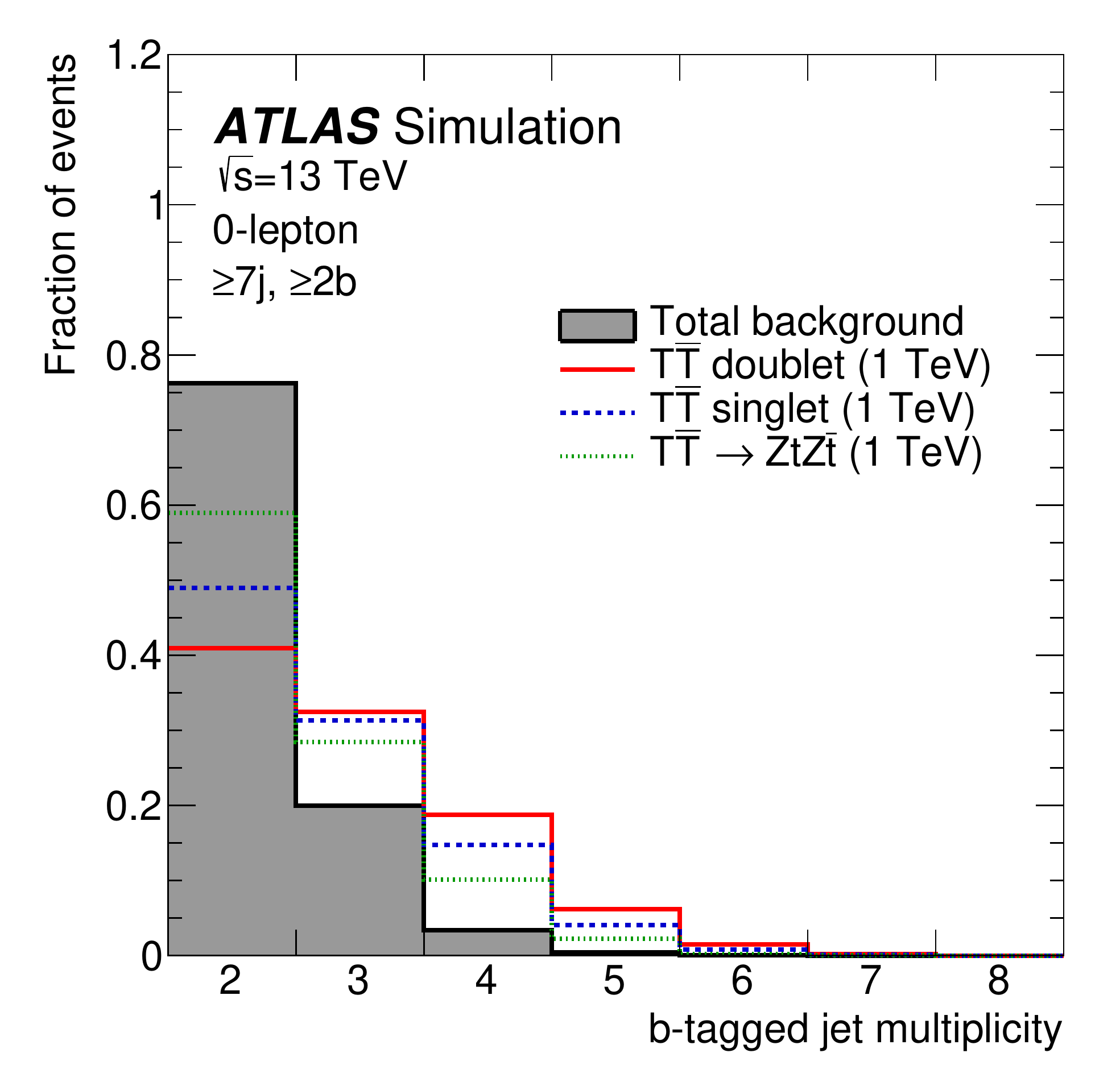}}
\caption{Comparison of the distribution of (a) the jet multiplicity, and (b) the $b$-tagged jet multiplicity,
between the total background (shaded histogram) and several signal scenarios considered in this search.
The selection used in (a) corresponds to events in the 1-lepton channel satisfying the preselection requirements,
whereas the selection used in (b) corresponds to events in the 0-lepton channel satisfying the preselection requirements and $\geq$7 jets.
The signals shown correspond to:
$T\bar{T}$ production in the weak-isospin doublet and singlet scenarios, and
in the ${\mathcal{B}}(T \to Zt)=1$ case, assuming  $m_T=1~\tev$;
and $\fourtop$ production within an EFT model.
}
\label{fig:shape_njet_nbtag}
\end{figure*}
 
The searches discussed in this paper primarily target $T\bar{T}$ production where at least one of the
$T$ quarks decays into a Higgs boson and a top quark resulting in the following processes:  $T\bar{T} \to HtHt$, $HtZt$ and $HtWb$.\footnote{In the following,
$HtZt$ is used to denote both $HtZ\bar{t}$ and its charge conjugate, $H\bar{t}Zt$. Similar notation is used for other processes, as appropriate.}
For the dominant $H\to b\bar{b}$ decay mode, the final-state signatures in both the 1-lepton and 0-lepton searches are characterised by high jet
and $b$-tagged jet multiplicities, which provide a powerful experimental handle to suppress the background.
The presence of high-momentum $Z$ bosons decaying into $\nu\bar{\nu}$ or $W$ bosons decaying leptonically,
either to an electron or muon that is not reconstructed or to a hadronically decaying $\tau$-lepton that is identified as a jet,
yields
high $\met$, which is exploited by the 0-lepton search.
Both searches have
some sensitivity to $T\bar{T} \to ZtZt$ and $ZtWb$, with $Z\to b\bar{b}$.
Possible contributions from pair production of the $B$ or $X$ quarks that would be included, along with the $T$ quark, in a weak-isospin doublet
are ignored. Such particles are expected to decay primarily through $X, B \to Wt$~\cite{AguilarSaavedra:2009es},
and thus not lead to high $b$-tagged jet multiplicity, which is the primary focus of these searches.
High jet and $b$-tagged jet multiplicities are also characteristic of $\fourtop$ events (both within the SM and in BSM scenarios); this search is sensitive to these events.
The four-top-quark production scenarios considered here
do not feature large $\met$,
so only the 1-lepton search is used to probe them.
No dedicated re-optimisation for $\fourtop$ events was performed.
 
In Figure~\ref{fig:shape_njet_nbtag}(a) the jet multiplicity distribution in the 1-lepton channel
after preselection (described in Section~\ref{sec:data_presel}) is compared between the total background and several signal scenarios,
chosen to illustrate differences among various types of signals the search is sensitive to.
A similar comparison for the $b$-tagged jet multiplicity distribution is shown in Figure~\ref{fig:shape_njet_nbtag}(b) for events in the
0-lepton channel after preselection plus the requirement of $\geq$7 jets.

\begin{figure*}[t!]
\centering
\subfloat[]{\includegraphics[width=0.45\textwidth]{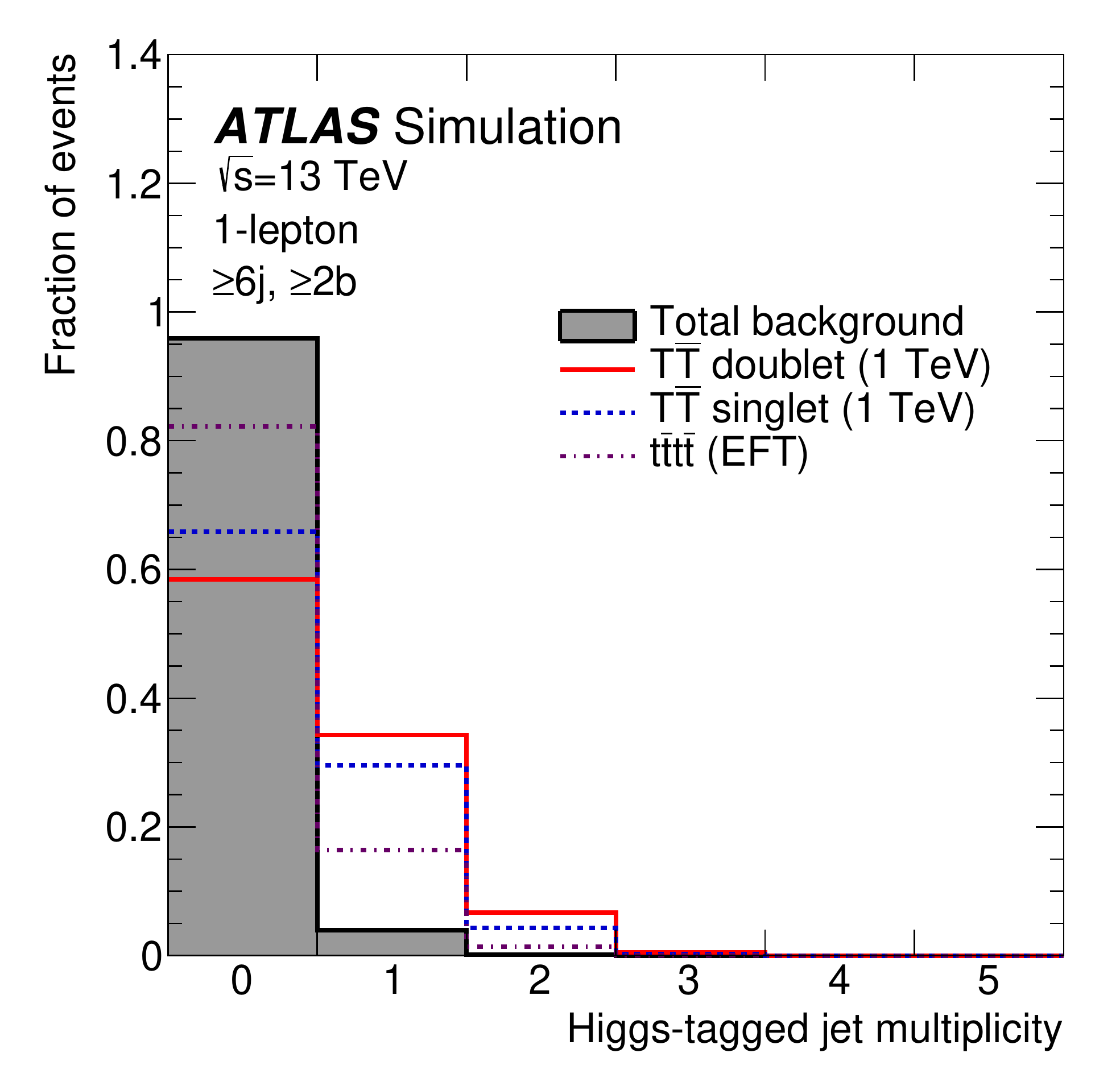}}
\subfloat[]{\includegraphics[width=0.45\textwidth]{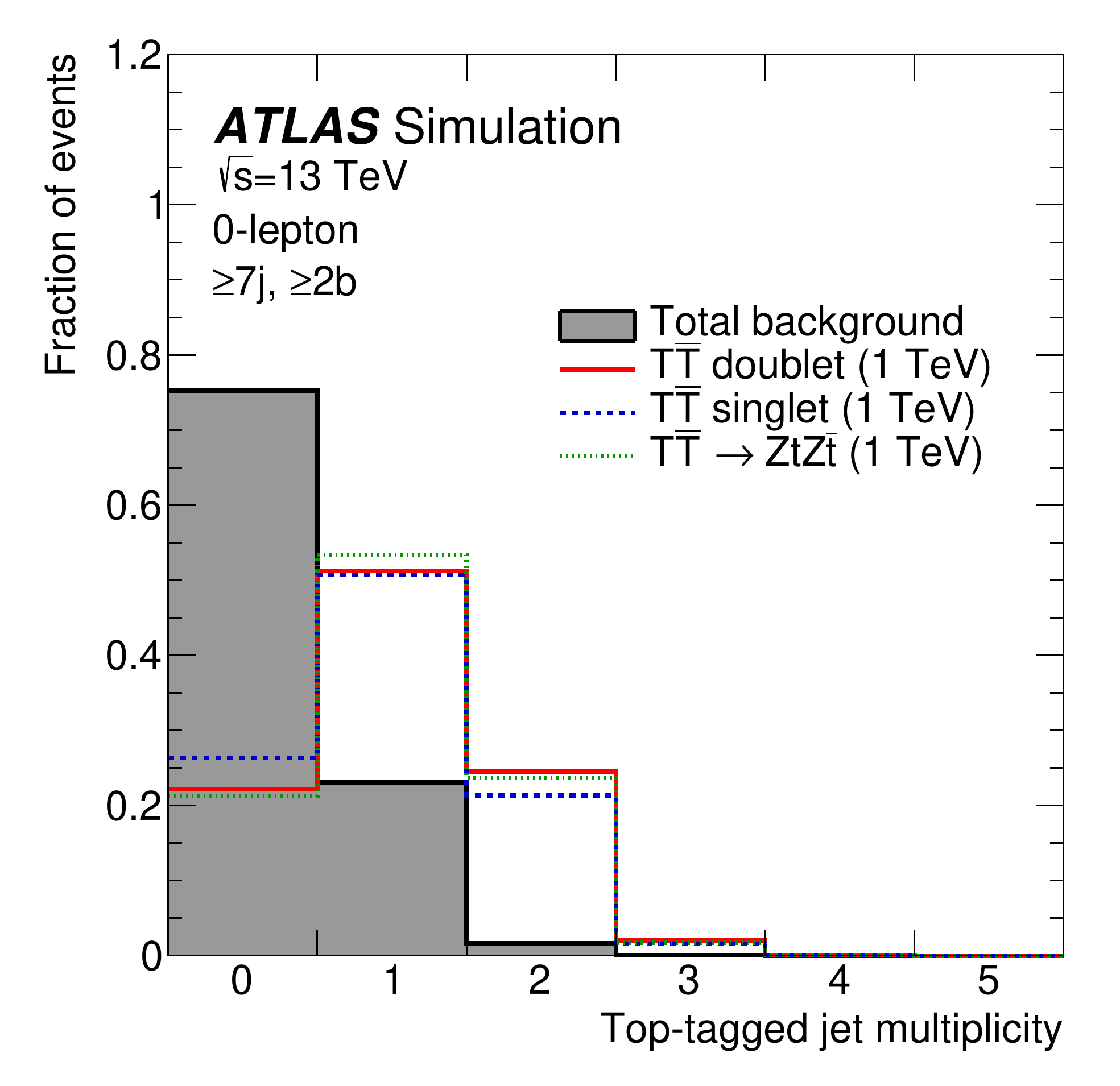}}
\caption{Comparison of the distribution of (a) the Higgs-tagged jet multiplicity and (b) the top-tagged jet multiplicity,
between the total background (shaded histogram) and several signal scenarios considered in this search.
The selection used in (a) corresponds to events in the 1-lepton channel satisfying the preselection requirements and $\geq$6 jets,
whereas the selection used in (b) corresponds to events in the 0-lepton channel satisfying the preselection requirements and $\geq$7 jets.
The signals shown correspond to: $T\bar{T}$ production in the weak-isospin doublet and singlet scenarios, and
in the ${\mathcal{B}}(T \to Zt)=1$ case, assuming $m_T=1~\tev$;
and $\fourtop$ production within an EFT model.
}
\label{fig:shape_topHiggsjets}
\end{figure*}
 
Compared to Run 1, the larger centre-of-mass energy in Run 2
provides sensitivity to
higher-mass signals, which decay into
boosted heavy SM particles (particularly Higgs bosons and top quarks).
These potentially give rise to a high multiplicity of
large-$R$ jets
that capture
their decay products (see Section~\ref{sec:objects}).
While $\ttbar$+jets events in the 1-lepton and 0-lepton channels are expected to typically contain one top-tagged jet,
the signal events of interest are characterised by higher Higgs-tagged jet and top-tagged jet multiplicities, as illustrated in Figures~\ref{fig:shape_topHiggsjets}(a)
and~\ref{fig:shape_topHiggsjets}(b). The small fraction (about 5\%) of background events with $\geq$2 top-tagged jets or $\geq$1 Higgs-tagged jets
results from the misidentification of at least one large-$R$ jet where
initial- or final-state radiation was responsible for a large fraction of the constituents.
 
In order to optimise the sensitivity of the searches, the selected events are categorised into different regions
depending on the jet multiplicity (5 and $\geq$6 jets in the 1-lepton channel;  6 and $\geq$7 jets in the 0-lepton channel),
$b$-tagged jet multiplicity  (3 and $\geq$4 in the 1-lepton channel; 2, 3 and $\geq$4 in the 0-lepton channel) and Higgs- and top-tagged jet multiplicity (0, 1 and $\geq$2).
In the following, channels with $N_\textrm{t}$ top-tagged jets, $N_\textrm{H}$ Higgs-tagged jets, $n$ jets, and $m$ $b$-tagged jets
are denoted by ``$N_\textrm{t}$t, $N_\textrm{H}$H, $n$j, $m$b''. Whenever the top/Higgs-tagging requirement is made on the sum
$N_\textrm{t}+N_\textrm{H} \equiv N_\textrm{tH}$, the channel is denoted by ``$N_\textrm{tH}$tH, $n$j, $m$b''.
In addition, events in the 0-lepton channel are further categorised into two regions according to the value of $\mtbmin$, the minimum transverse mass
of $\met$ and any of the three (or two, in events with exactly two $b$-tagged jets) leading $b$-tagged jets in the event:
$\mtbmin<160~\gev$ (referred to as ``LM'', standing for ``low mass'') and $\mtbmin>160~\gev$ (referred to as ``HM'', standing for ``high mass'').
This kinematic variable is bounded from above by the top quark mass for semileptonic $t\bar{t}$ background events, while the signal can have higher
values of $\mtbmin$ due to the presence of high-$\pt$ neutrinos from $T \to Zt$, $Z\to \nu\nu$ or $T \to Wb$, $W\to \ell\nu$ decays.
Although the requirements of a minimum top/Higgs-tagged jet multiplicity reduces the value of $\mtbmin$ because of the resulting stronger collimation
of the top quark decay products, this variable still provides useful discrimination between signal and $t\bar{t}$ background, as shown in Figure~\ref{fig:shape_mtbmin}.
While the 1-lepton channel only considers regions with exactly 3 or $\geq$4 $b$-tagged jets, the 0-lepton channel also includes regions with
exactly two $b$-jets and $\mtbmin>160~\gev$, to gain sensitivity to $T\bar{T} \to ZtZt$ decays with at least one $Z \to \nu\bar{\nu}$ decay.
 
\begin{figure*}[t!]
\centering
\includegraphics[width=0.45\textwidth]{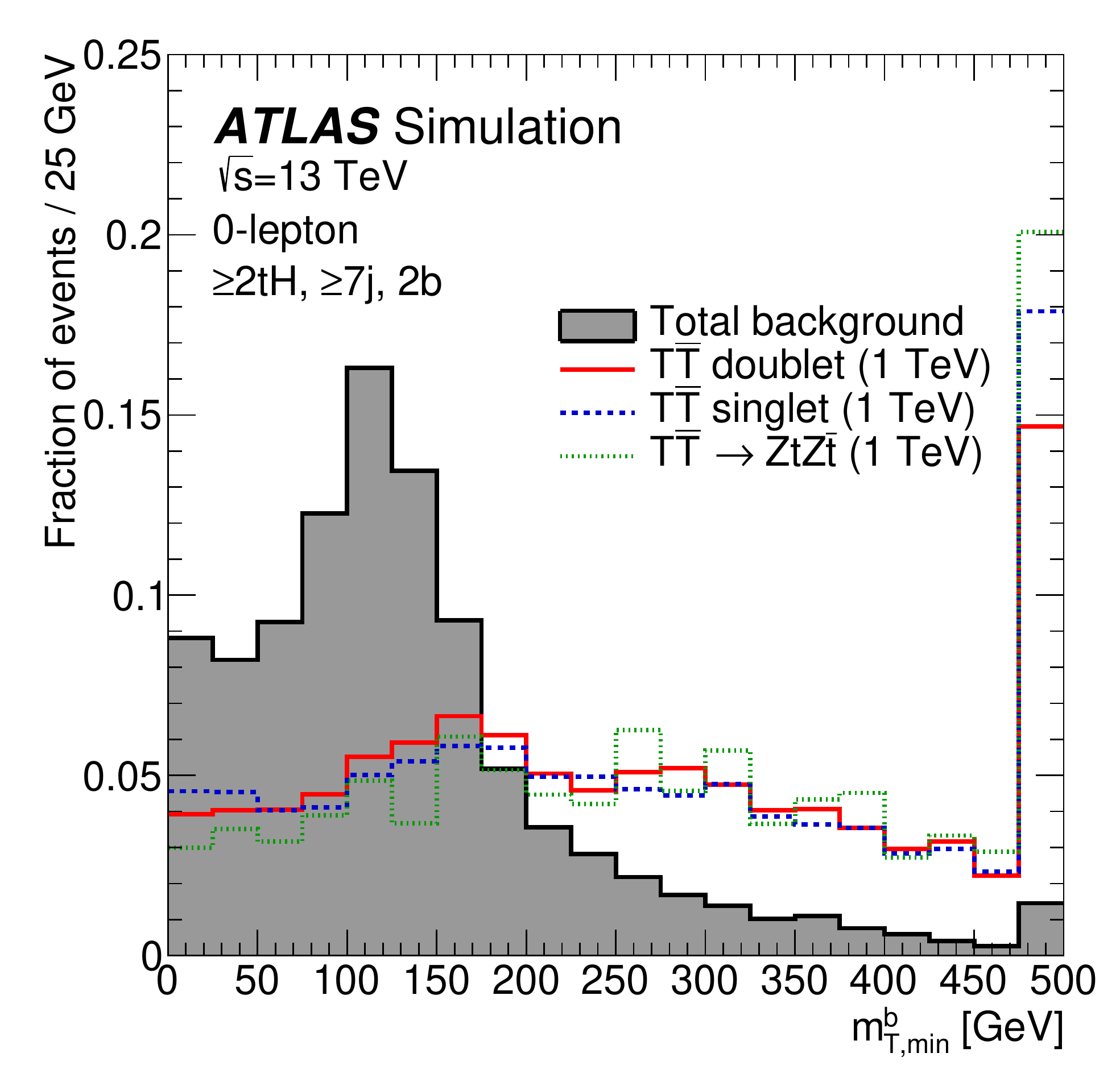}
\caption{Comparison of the distribution of the minimum transverse mass
of $\met$ and any of the three (or two, in events with exactly two $b$-tagged jets) leading $b$-tagged jets
in the event ($\mtbmin$), between the total background (shaded histogram) and several signal scenarios considered in this search.
The selection used corresponds to events in the ($\geq$2tH, $\geq$7j, 2b) region of the 0-lepton channel.
The signals shown correspond to $T\bar{T}$ production in the weak-isospin doublet and singlet scenarios, and
in the ${\mathcal{B}}(T \to Zt)=1$ case, assuming  $m_T=1~\tev$.
The last bin in the figure contains the overflow.}
\label{fig:shape_mtbmin}
\end{figure*}
 
To further improve the separation between the $T\bar{T}$ signal and background, the distinct kinematic features of the signal are exploited.
In particular, the large $T$ quark mass results in leptons and jets with large energy in the final state
and the effective mass ($\meff$), defined as the scalar sum of the transverse momenta of the lepton, the selected jets and the missing transverse
momentum,  provides a powerful discriminating variable between signal and background.  The $\meff$ distribution
peaks at approximately $2 m_{T}$ for signal events and at lower values for the $\ttbar$+jets background. For the same reasons, the various $\fourtop$ signals from BSM scenarios
also populate high values of $\meff$.
An additional selection requirement of $\meff>1~\tev$ is made in order to minimise the effect of possible mismodelling of the $\meff$ distribution at
low values originating from small backgrounds with large systematic uncertainties, such as multijet production.
Such a requirement is applied for regions with $N_\textrm{t}+N_\textrm{H}\leq 1$ in the 1-lepton channel, and for all regions in the 0-lepton channel.
Since the $T\bar{T}$ signal is characterised by having at least one top/Higgs-tagged jet and large values of $\meff$, this minimum
requirement on $\meff$
does not decrease the signal efficiency.
In Figure~\ref{fig:shape_meff}, the $\meff$ distribution is compared between signal and background for events in
signal-rich regions of the 1-lepton and 0-lepton channels. The kinematic requirements in these regions result in a significantly harder $\meff$ spectrum
for the background than in regions without top/Higgs-tagged jets, but this variable still shows good discrimination
between signal and background. Thus, the $\meff$ distribution is used as the final discriminating variable in all regions considered in this search.
 
The regions with $\geq$6 jets ($\geq$7 jets) are used to perform the search in the 1-lepton (0-lepton) channel (referred to as ``search regions''),
whereas the regions with exactly 5 jets (6 jets) are used to validate the background modelling in different regimes of
event kinematics and heavy-flavour content (referred to as ``validation regions'').
A total of 12 search regions and 10 validation regions are considered in the 1-lepton channel, whereas
22 search regions and 16 validation regions are considered in the 0-lepton channel,
defined in Tables~\ref{tab:channels_1L} and~\ref{tab:channels_0L} respectively.
In each channel, there are fewer validation regions than signal regions since
some validation regions are merged to ensure a minimum of about 10 expected events.
The level of possible signal contamination in the validation regions that have high event yields, and are therefore the regions that are most useful to validate the
background prediction, depends on the signal scenario considered but is typically well below 10\% for a 1~\tev\ $T$ quark.
 
\begin{figure*}[t!]
\centering
\subfloat[]{\includegraphics[width=0.45\textwidth]{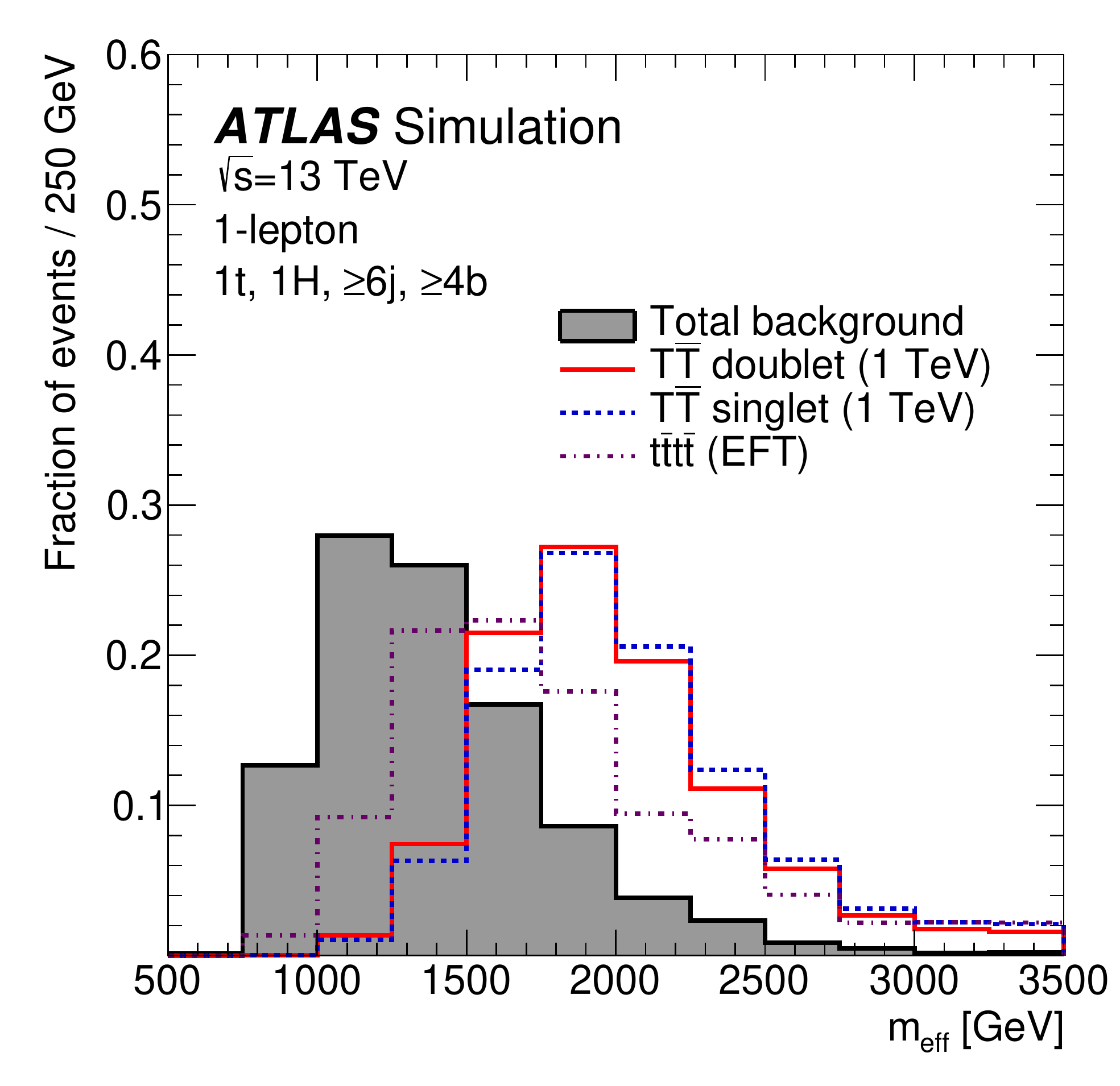}}
\subfloat[]{\includegraphics[width=0.45\textwidth]{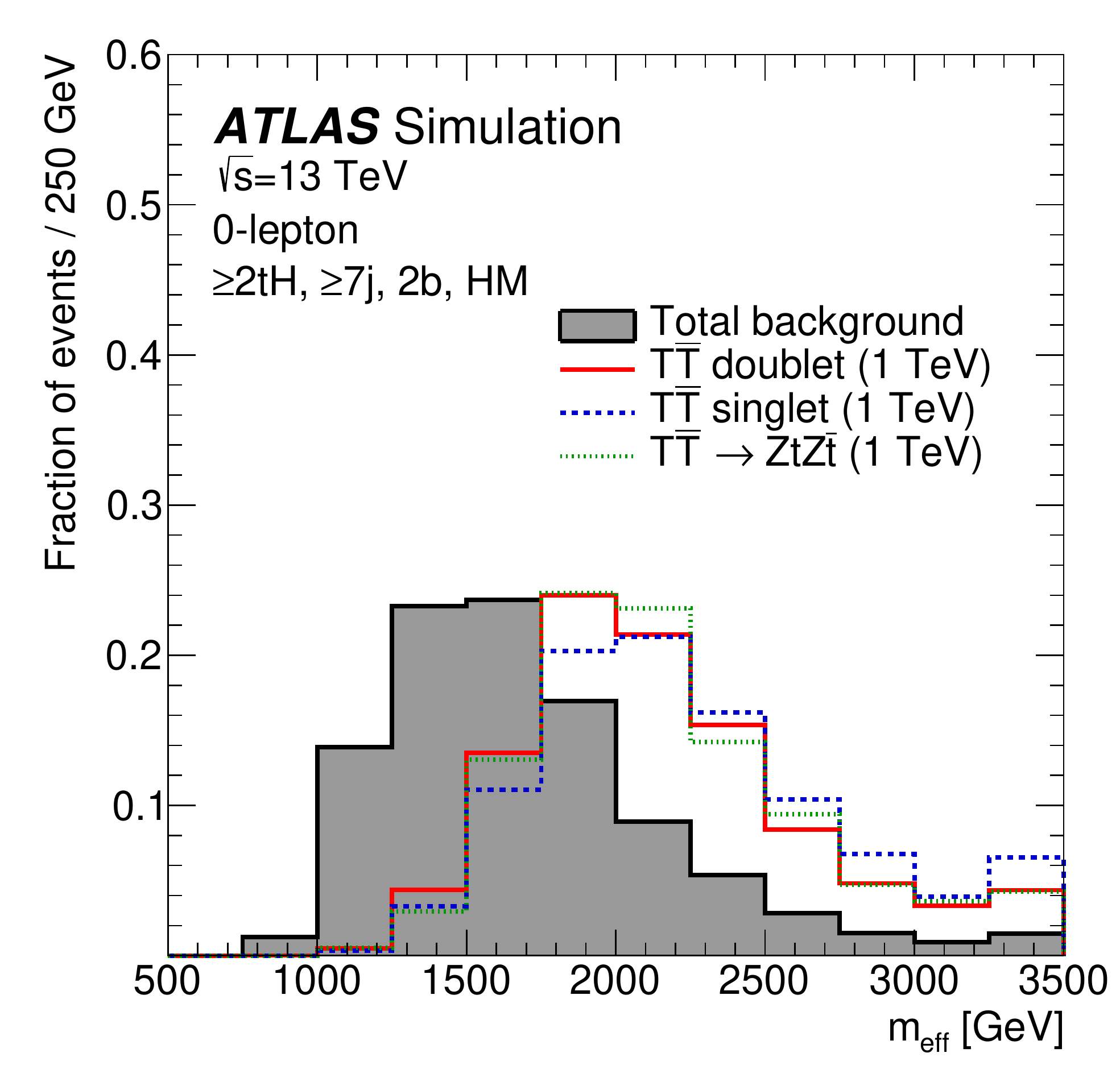}}
\caption{Comparison of the distribution of the effective mass ($\meff$), between the total background (shaded histogram) and several signal scenarios considered in this search.
The selection used in (a) corresponds to events in the (1t, 1H, $\geq$6j, $\geq$4b) region of the 1-lepton channel,
whereas the selection used in (b) corresponds to events in the ($\geq$2tH, $\geq$7j, 2b, HM) region of the 0-lepton channel.
The signals shown correspond to:
$T\bar{T}$ production in the weak-isospin doublet and singlet scenarios, and in the ${\mathcal{B}}(T \to Zt)=1$ case, assuming  $m_T=1~\tev$;
and $\fourtop$ production within an EFT model.
The last bin in 
each distribution contains the overflow.}
\label{fig:shape_meff}
\end{figure*}
 
The overall rate and composition of the $\ttbar$+jets background strongly depends on the jet and $b$-tagged jet multiplicities,
as illustrated in Figure~\ref{fig:Summary}.
The $\ttbar$+light-jets background is dominant in events with exactly two $b$-tagged jets, which typically correspond to the two $b$-quarks
from the top quark decays. It also contributes significantly to events with exactly three $b$-tagged jets, in which typically  a charm quark from
the hadronic $W$ boson decay is also $b$-tagged.
Contributions from \ttcin\ and \ttbin\ become significant as the $b$-tagged jet multiplicity
increases, with the \ttbin\ background being dominant for events with $\geq$4 $b$-tagged jets.
The regions with different top/Higgs-tagged jet multiplicities probe different kinematic regimes,
both soft (e.g.~low-mass $T$ quark) and hard (e.g.~high-mass $T$ quark or BSM $\fourtop$ production).
The search regions with the higher multiplicities of top-/Higgs-tagged jets and $b$-tagged jets in both the 1-lepton and 0-lepton channels,
as well as the HM regions in the 0-lepton channel, have the largest signal-to-background ratio, and therefore drive the sensitivity of the search.
The remaining
search regions have significantly lower signal-to-background ratios, but
are useful for checking and correcting
the $t\bar{t}$+jets background prediction and constraining the related systematic uncertainties (see Section~\ref{sec:systematics})
through a likelihood fit to data (see Section~\ref{sec:stat_analysis}). A summary of the signal-to-background ratio in the different search regions is displayed in Figure~\ref{fig:soverb} for the $T$ quark signal with various decay configurations.
A similar
fitting strategy was followed in the Run-1 search in the 1-lepton channel~\cite{Aad:2015kqa}.

A summary of the observed and expected yields before the fit to data in five of the most sensitive search regions
in the 1-lepton and 0-lepton channels can be found in Tables~\ref{tab:Prefit_Yields_1L_unblind} and~\ref{tab:Prefit_Yields_0L_unblind}, respectively.
The search regions shown in Table~\ref{tab:Prefit_Yields_1L_unblind} for the 1-lepton channel are a selection of some of the regions with the highest
$S/\sqrt{B}$ ratio (where $S$ and $B$ are the expected signal and background yields, respectively) across several
signal benchmark scenarios considered ($T\bar{T}$ in the ${\mathcal{B}}(T\to Ht)=1$,  $T$ doublet, and $T$ singlet scenarios, in all cases assuming $m_{T}=1~\tev$,
and $\fourtop$ within an EFT and the 2UED/RPP models).
Similarly, the search regions shown in Table~\ref{tab:Prefit_Yields_0L_unblind} for the 0-lepton channel are a superset of the regions
with the highest $S/\sqrt{B}$ ratio for different $T\bar{T}$ signal benchmark scenarios ($T$ doublet, $T$ singlet
and ${\mathcal{B}}(T\to Zt)=1$, also assuming $m_{T}=1~\tev$).

\begin{table*}[p]
\begin{center}
\begin{tabular}{ccccc}
\toprule\toprule
\multicolumn{5}{c}{1-lepton channel} \\
\midrule\midrule
\multicolumn{5}{c}{Search regions ($\geq$6 jets)} \\
\midrule
$N_\textrm{t}$ & $N_\textrm{H}$ & $b$-tag multiplicity & $\meff$ & Channel name \\
\midrule
0 & 0  & 3 & $>$1~\tev & 0t, 0H, $\geq$6j, 3b \\
0 & 0  & $\geq$4 & $>$1~\tev & 0t, 0H, $\geq$6j, $\geq$4b \\
\hdashline
1 & 0  & 3 & $>$1~\tev & 1t, 0H, $\geq$6j, 3b \\
1 & 0  & $\geq$4 & $>$1~\tev & 1t, 0H, $\geq$6j, $\geq$4b \\
0 & 1  & 3 & $>$1~\tev & 0t, 1H, $\geq$6j, 3b \\
0 & 1  & $\geq$4 & $>$1~\tev & 0t, 1H, $\geq$6j, $\geq$4b \\
\hdashline
1 & 1   & 3 & -- & 1t, 1H, $\geq$6j, 3b \\
1 & 1  & $\geq$4 & -- & 1t, 1H, $\geq$6j, $\geq$4b \\
$\geq$2 & 0 or 1  &  3 & -- & $\geq$2t, 0--1H, $\geq$6j, 3b \\
$\geq$2 & 0 or 1  &  $\geq$4 & -- & $\geq$2t, 0--1H, $\geq$6j, $\geq$4b \\
$\geq$0& $\geq$2 & 3 & -- & $\geq$0t, $\geq$2H, $\geq$6j, 3b \\
$\geq$0& $\geq$2 & $\geq$4 & -- & $\geq$0t, $\geq$2H, $\geq$6j, $\geq$4b \\
\midrule\midrule
\multicolumn{5}{c}{Validation regions (5 jets)} \\
\midrule
$N_\textrm{t}$ & $N_\textrm{H}$ & $b$-tag multiplicity & $\meff$ & Channel name \\
\midrule
0 & 0  & 3 & $>$1~\tev & 0t, 0H, 5j, 3b \\
0 & 0  & $\geq$4 & $>$1~\tev & 0t, 0H, 5j, $\geq$4b \\
\hdashline
1 & 0  & 3 & $>$1~\tev & 1t, 0H, 5j, 3b \\
1 & 0  & $\geq$4 & $>$1~\tev & 1t, 0H, 5j, $\geq$4b \\
0 & 1  & 3 & $>$1~\tev & 0t, 1H, 5j, 3b \\
0 & 1  & $\geq$4 & $>$1~\tev & 0t, 1H, 5j, $\geq$4b \\
\hdashline
1 & 1   & 3 & -- & 1t, 1H, 5j, 3b \\
$\geq$2 & 0 or 1 & 3 & -- & $\geq$2t, 0--1H, 5j, 3b \\
$\geq$0 & $\geq$2 & 3 & -- & $\geq$0t, $\geq$2H, 5j, 3b \\
\multicolumn{2}{c}{$N_\textrm{t} + N_\textrm{H} \geq 2$} & $\geq$4 & -- & $\geq$2tH, 5j, $\geq$4b \\
\bottomrule\bottomrule
\end{tabular}
\caption{\small{Definition of the search and validation regions (see text for details) in the 1-lepton channel.}}
\label{tab:channels_1L}
\end{center}
\end{table*}
\begin{table*}[p]
\begin{center}
\begin{tabular}{cccccc}
\toprule\toprule
\multicolumn{6}{c}{0-lepton channel} \\
\midrule\midrule
\multicolumn{6}{c}{Search regions ($\geq$7 jets)} \\
\midrule\midrule
$N_\textrm{t}$ & $N_\textrm{H}$ & $b$-tag multiplicity  & $\mtbmin$ & $\meff$ & Channel name \\
\midrule
0 & 0& 2 & $>$160~\gev& $>$1~\tev & 0t, 0H, $\geq$7j, 2b, HM \\
0 & 0& 3 & $<$160~\gev& $>$1~\tev & 0t, 0H, $\geq$7j, 3b, LM \\
0 & 0& 3 & $>$160~\gev& $>$1~\tev & 0t, 0H, $\geq$7j, 3b, HM \\
0 & 0& $\geq$4 & $<$160~\gev& $>$1~\tev & 0t, 0H, $\geq$7j, $\geq$4b, LM \\
0 & 0& $\geq$4 & $>$160~\gev& $>$1~\tev & 0t, 0H, $\geq$7j, $\geq$4b, HM \\
\hdashline
1 & 0& 2 & $>$160~\gev& $>$1~\tev & 1t, 0H, $\geq$7j, 2b, HM \\
1 & 0& 3 & $<$160~\gev& $>$1~\tev & 1t, 0H, $\geq$7j, 3b, LM \\
1 & 0& 3 & $>$160~\gev& $>$1~\tev & 1t, 0H, $\geq$7j, 3b, HM \\
1 & 0& $\geq$4 & $<$160~\gev& $>$1~\tev & 1t, 0H, $\geq$7j, $\geq$4b, LM \\
1 & 0& $\geq$4 & $>$160~\gev& $>$1~\tev & 1t, 0H, $\geq$7j, $\geq$4b, HM \\
\hdashline
0 & 1& 2 & $>$160~\gev& $>$1~\tev & 0t, 1H, $\geq$7j, 2b, HM \\
0 & 1& 3 & $<$160~\gev& $>$1~\tev & 0t, 1H, $\geq$7j, 3b, LM \\
0 & 1& 3 & $>$160~\gev& $>$1~\tev & 0t, 1H, $\geq$7j, 3b, HM \\
0 & 1& $\geq$4 & $<$160~\gev& $>$1~\tev & 0t, 1H, $\geq$7j, $\geq$4b, LM \\
0 & 1& $\geq$4 & $>$160~\gev& $>$1~\tev & 0t, 1H, $\geq$7j, $\geq$4b, HM \\
\hdashline
1 & 1& 3 & $<$160~\gev& $>$1~\tev & 1t, 1H, $\geq$7j, 3b, LM \\
1 & 1& 3 & $>$160~\gev& $>$1~\tev & 1t, 1H, $\geq$7j, 3b, HM \\
\hdashline
$\geq$2 & 0 or 1& 3 & $<$160~\gev& $>$1~\tev & $\geq$2t, 0--1H, $\geq$7j, 3b, LM \\
$\geq$2 & 0 or 1& 3 & $>$160~\gev& $>$1~\tev & $\geq$2t, 0--1H, $\geq$7j, 3b, HM \\
$\geq$0 & $\geq$2& 3 & --& $>$1~\tev & $\geq$0t, $\geq$2H, $\geq$7j, 3b \\
\hdashline
\multicolumn{2}{c}{$N_\textrm{t} + N_\textrm{H} \geq 2$} & 2 & $>$160~\gev & $>$1~\tev & $\geq$2tH, $\geq$7j, 2b, HM \\
\multicolumn{2}{c}{$N_\textrm{t} + N_\textrm{H} \geq 2$} & $\geq$4 & -- & $>$1~\tev & $\geq$2tH, $\geq$7j, $\geq$4b \\
\midrule\midrule
\multicolumn{6}{c}{Validation regions (6 jets)} \\
\midrule
$N_\textrm{t}$ & $N_\textrm{H}$ & $b$-tag multiplicity  & $\mtbmin$ & $\meff$ & Channel name \\
\midrule
0 & 0& 2 & $>$160~\gev& $>$1~\tev & 0t, 0H, 6j, 2b, HM \\
0 & 0& 3 & $<$160~\gev& $>$1~\tev & 0t, 0H, 6j, 3b, LM \\
0 & 0& 3 & $>$160~\gev& $>$1~\tev & 0t, 0H, 6j, 3b, HM \\
0 & 0& $\geq$4 & $<$160~\gev& $>$1~\tev & 0t, 0H, 6j, $\geq$4b, LM \\
0 & 0& $\geq$4 & $>$160~\gev& $>$1~\tev & 0t, 0H, 6j, $\geq$4b, HM \\
\hdashline
1 & 0& 2 & $>$160~\gev& $>$1~\tev & 1t, 0H, 6j, 2b, HM \\
1 & 0& 3 & $<$160~\gev& $>$1~\tev & 1t, 0H, 6j, 3b, LM \\
1 & 0& 3 & $>$160~\gev& $>$1~\tev & 1t, 0H, 6j, 3b, HM \\
1 & 0& $\geq$4 & --& $>$1~\tev & 1t, 0H, 6j, $\geq$4b \\
\hdashline
0 & 1& 2 & $>$160~\gev& $>$1~\tev & 0t, 1H, 6j, 2b, HM \\
0 & 1& 3 & $<$160~\gev& $>$1~\tev & 0t, 1H, 6j, 3b, LM \\
0 & 1& 3 & $>$160~\gev& $>$1~\tev & 0t, 1H, 6j, 3b, HM \\
0 & 1& $\geq$4 & --& $>$1~\tev & 0t, 1H, 6j, $\geq$4b \\
\hdashline
\multicolumn{2}{c}{$N_\textrm{t} + N_\textrm{H} \geq 2$} & 2 & $>$160~\gev & $>$1~\tev & $\geq$2tH, 6j, 2b, HM \\
\multicolumn{2}{c}{$N_\textrm{t} + N_\textrm{H} \geq 2$} & 3 & -- & $>$1~\tev & $\geq$2tH, 6j, 3b \\
\multicolumn{2}{c}{$N_\textrm{t} + N_\textrm{H} \geq 2$} & $\geq$4 & -- & $>$1~\tev & $\geq$2tH, 6j, $\geq$4b \\
\bottomrule\bottomrule
\end{tabular}
\caption{\small{Definition of the search and validation regions (see text for details) in the 0-lepton channel.}}
\label{tab:channels_0L}
\end{center}
\end{table*}
 
\begin{figure*}[t!]
\begin{center}
\includegraphics[width=0.8\textwidth]{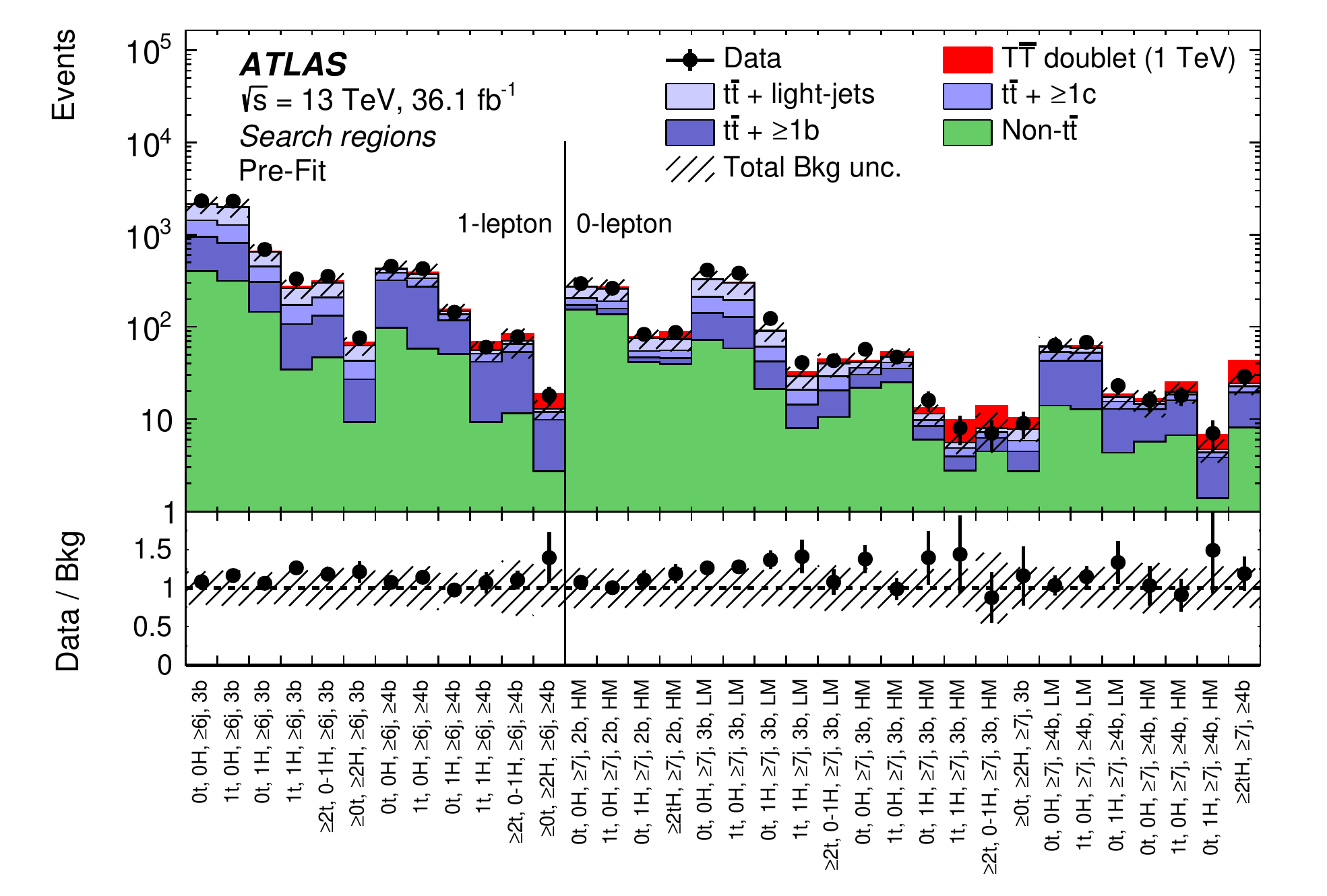}
\caption{\small {Comparison between the data and the background prediction for the yields in the search regions considered
in the 1-lepton and 0-lepton channels, before the fit to data (``Pre-fit'').
The small contributions from $\ttbar V$, $\ttbar H$, single-top, $W/Z$+jets, diboson, and multijet backgrounds are combined into a single background source
referred to as ``Non-$\ttbar\,$''.
The expected $T\bar{T}$ signal (solid red) corresponding to $m_{T}=1~\tev$ in the $T$ doublet scenario is also shown,
added on top of the background prediction.
The bottom panel displays the ratio of data to the SM background (``Bkg'') prediction.
The hashed area represents the total uncertainty of the background, excluding the normalisation uncertainty of the $\ttbin$ background,
which is determined via a likelihood fit to data.}}
\label{fig:Summary}
\end{center}
\end{figure*}

\begin{figure*}[t!]
\begin{center}
\includegraphics[width=0.8\textwidth]{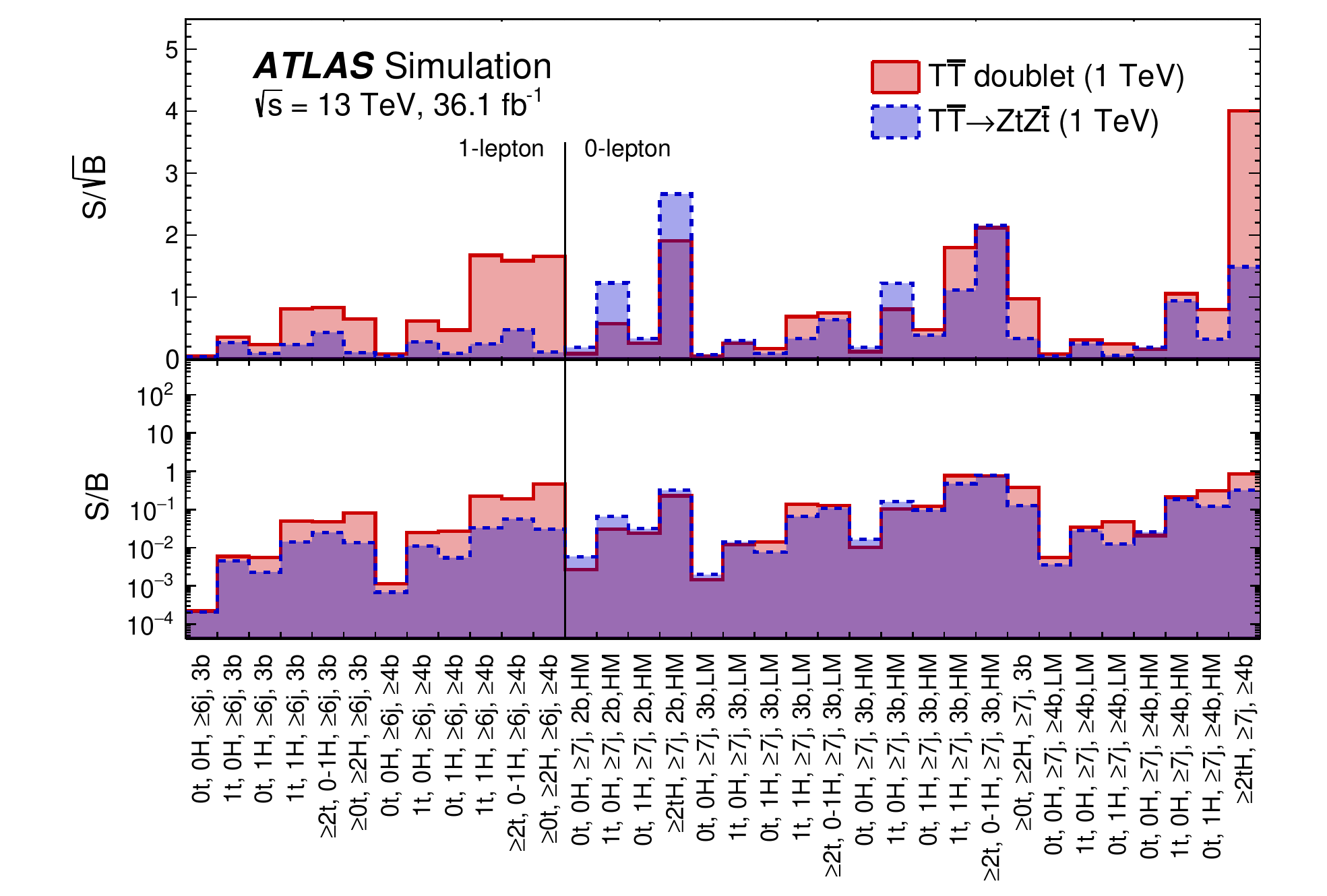}
\caption{\small {Signal-to-background ratio expressed as $S/\sqrt{B}$ (resp. $S/B$) in the top (resp. bottom) panel for each of the search regions. $B$ and $S$ stand for the total numbers of expected background and signal events in each region, respectively. For a 1~\tev\ $T$ quark mass hypothesis, two branching ratio configurations are displayed: the doublet model (red filled area) and ${\mathcal{B}}(T\to Zt)=1$ (blue filled area).}}
\label{fig:soverb}
\end{center}
\end{figure*}

\begin{table}[t!]
\begin{center}
\begin{tabular}{l*{5}{c}}
\toprule\toprule
1-lepton channel & $\geq$2t, 0--1H, & 1t, 0H,  & 1t, 1H,  & $\geq$2t, 0--1H, & $\geq$0t, $\geq$2H,  \\
& $\geq$6j, 3b & $\geq$6j, $\geq$4b & $\geq$6j, $\geq$4b &  $\geq$6j, $\geq$4b & $\geq$6j, $\geq$4b \\
\midrule\midrule
\multicolumn{5}{l}{$T\bar{T}$  ($m_{T}=1~\tev$)} \\
{\,} ${\mathcal{B}}(T\to Ht)=1$  & $ 19.6 \pm 1.5 $ & $ 21.5 \pm 2.6 $ &   $ 24.3 \pm 2.7 $ &   $ 23.9 \pm 2.8 $ &   $ 14.6 \pm 2.0 $ \\
{\,} $T$ doublet & $ 14.2 \pm 1.0 $ & $ 15.2 \pm 1.6 $ &   $ 12.5 \pm 1.4 $ &   $ 13.3 \pm 1.5 $ &   $ 5.96 \pm 0.62 $ \\
{\,} $T$ singlet  & $ 7.88 \pm 0.58 $ & $ 8.13 \pm 0.94 $ &   $ 5.47 \pm 0.62 $ &   $ 5.51 \pm 0.69 $ &   $ 2.18 \pm 0.23 $ \\  [0.2cm]
\multicolumn{5}{l}{$\fourtop$} \\
{\,} EFT ($|C_{4t}|/\Lambda^2 = 4\pi~\tev^{-2}$) & $ 535 \pm 30 $ & $ 706 \pm 80 $ &   $ 171 \pm 19 $ & $ 468 \pm 55 $ & $ 34.3 \pm 5.0 $  \\
{\,} 2UED/RPP ($m_{\KK}=1.6~\tev$) & $ 9.77 \pm 0.46 $ & $ 1.84 \pm 0.35 $ &   $ 1.00 \pm 0.19 $ &   $ 8.9 \pm 1.4 $ &   $ 0.39 \pm 0.09 $ \\
\midrule\midrule
$t\bar{t}$+light-jets & $ 91 \pm 46 $ & $ 38 \pm 17 $ &   $ 4.8 \pm 2.4 $ &   $ 5.4 \pm 3.3 $ &   $ 0.99 \pm 0.49 $ \\
\ttcin\ & $ 75 \pm 45 $ & $ 64 \pm 38 $ &   $ 9.5 \pm 5.6 $ &   $ 11.8 \pm 7.5 $ &   $ 2.1 \pm 1.3 $ \\
\ttbin\ & $ 86 \pm 41 $ & $ 215 \pm 83 $ &   $ 32.4 \pm 9.5 $ &   $ 42 \pm 22 $ &   $ 7.1 \pm 2.2 $ \\
$t\bar{t}V$ & $ 9.7 \pm 1.8 $ & $ 11.4 \pm 2.4 $ &   $ 1.73 \pm 0.39 $ &   $ 2.46 \pm 0.53 $ &   $ 0.41 \pm 0.10 $ \\
$t\bar{t}H$ & $ 4.90 \pm 0.78 $ & $ 15.0 \pm 2.8 $ &   $ 3.79 \pm 0.65 $ &   $ 2.84 \pm 0.62 $ &   $ 1.19 \pm 0.20 $ \\
$W$+jets & $ 9.4 \pm 4.4 $ & $ 8.2 \pm 4.2 $ &   $ 0.69 \pm 0.50 $ &   $ 1.32 \pm 0.71 $ &   $ 0.54 \pm 0.48 $ \\
$Z$+jets & $ 1.31 \pm 0.64 $ & $ 0.95 \pm 0.48 $ &   $ 0.10 \pm 0.07 $ &   $ 0.13 \pm 0.08 $ &   $ 0.06 \pm 0.05 $ \\
Single top & $ 13.1 \pm 5.5 $ & $ 16.6 \pm 7.0 $ &   $ 1.69 \pm 0.76 $ &   $ 1.97 \pm 0.95 $ &   $ 0.26 \pm 0.21 $ \\
Diboson & $ 1.8 \pm 1.1 $ & $ 0.99 \pm 0.55 $ &   $ 0.11 \pm 0.09 $ &   $ 0.22 \pm 0.14 $ &   $ 0.01 \pm 0.04 $ \\
$\fourtop$ (SM) & $ 2.82 \pm 0.86 $ & $ 4.9 \pm 1.6 $ &   $ 1.12 \pm 0.36 $ &   $ 2.55 \pm 0.82 $ &   $ 0.23 \pm 0.07 $ \\
\midrule
Total background & $ 299 \pm 83 $ & $ 380 \pm 110 $ &   $ 56 \pm 13 $ &   $ 71 \pm 25 $ &   $ 12.9 \pm 3.2 $ \\
\midrule
Data & 353 & 428  & 60  & 78  & 18  \\
\bottomrule\bottomrule     \\
\end{tabular}
\vspace{0.1cm}
\end{center}
\vspace{-0.5cm}
\caption{Predicted and observed yields in the 1-lepton channel in five of the most sensitive search regions
(depending on the signal scenario) considered.
The multijet background is estimated to be negligible in these regions and thus not shown.
The background prediction is shown before the fit to data. Also shown are the signal predictions for different benchmark scenarios considered.
The quoted uncertainties are the sum in quadrature of statistical and systematic uncertainties in the yields,
excluding the normalisation uncertainty of the $\ttbin$ background, which is determined via a likelihood fit to data.}
\label{tab:Prefit_Yields_1L_unblind}
\end{table}
\begin{table}[h!]
\begin{center}
\begin{tabular}{l*{5}{c}}
\toprule\toprule
0-lepton channel & $\geq$2tH, & 1t, 1H, & $\geq$2t, 0--1H,  & 1t, 0H, & $\geq$2tH,  \\
& $\geq$7j, 2b, HM & $\geq$7j, 3b, HM & $\geq$7j, 3b, HM & $\geq$7j, $\geq$4b, HM & $\geq$7j, $\geq$4b\\
\midrule\midrule
\multicolumn{5}{l}{$T\bar{T}$  ($m_{T}=1~\tev$)} \\
{\,} ${\mathcal{B}}(T\to Zt)=1$  & $ 22.3 \pm 2.3 $ &   $ 2.60 \pm 0.57 $ &   $ 6.02 \pm 0.61 $ &   $ 4.72 \pm 0.66 $ &   $ 6.94 \pm 0.98 $ \\
{\,} $T$ doublet & $ 16.0 \pm 1.1 $ &   $ 4.22 \pm 0.34 $ &   $ 5.92 \pm 0.49 $ &  $ 5.32 \pm 0.61 $  &   $ 18.7 \pm 2.0 $ \\
{\,} $T$ singlet  & $ 8.52 \pm 0.61 $ &   $ 1.81 \pm 0.16 $ &   $ 2.63 \pm 0.22 $ &  $ 2.32 \pm 0.29 $  &   $ 6.91 \pm 0.80 $ \\
\midrule\midrule
$t\bar{t}$+light-jets & $ 17.8 \pm 9.8 $ &   $ 0.72 \pm 0.40 $ &   $ 0.80 \pm 0.53 $ &  $ 1.30 \pm 0.72 $  &   $ 1.71 \pm 0.98 $ \\
\ttcin\ & $ 9.7 \pm 6.4 $ &   $ 0.92 \pm 0.65 $ &   $ 0.95 \pm 0.71 $ &  $ 2.4 \pm 1.6 $  &   $ 3.2 \pm 2.0 $ \\
\ttbin\ & $ 6.3 \pm 4.2 $ &   $ 1.17 \pm 0.59 $ &   $ 1.78 \pm 0.74 $ &  $ 9.4 \pm 3.2 $  &   $ 11.4 \pm 4.1 $ \\
$t\bar{t}V$ & $ 5.5 \pm 1.0 $ &   $ 0.49 \pm 0.12 $ &   $ 0.88 \pm 0.19 $ &  $ 1.19 \pm 0.27 $  &   $ 1.01 \pm 0.24 $ \\
$t\bar{t}H$ & $ 0.61 \pm 0.12 $ &   $ 0.17 \pm 0.05 $ &   $ 0.13 \pm 0.04 $ &  $ 0.85 \pm 0.17 $  &   $ 1.08 \pm 0.25 $ \\
$W$+jets & $ 9.6 \pm 4.1 $ &   $ 0.52 \pm 0.27 $ &   $ 0.80 \pm 0.37 $ &  $ 0.81 \pm 0.40 $  &   $ 0.56 \pm 0.28 $ \\
$Z$+jets & $ 8.6 \pm 4.5 $ &   $ 0.59 \pm 0.28 $ &   $ 0.8 \pm 2.1 $ &  $ 0.80 \pm 0.40 $  &   $ 0.63 \pm 0.42 $ \\
Single top & $ 8.3 \pm 4.4 $ &   $ 0.69 \pm 0.43 $ &   $ 0.97 \pm 0.59 $ &  $ 1.8 \pm 1.0 $  &   $ 1.10 \pm 0.61 $ \\
Diboson & $ 2.9 \pm 1.9 $ &   $ 0.11 \pm 0.20 $ &   $ 0.55 \pm 0.66 $ &   $ 0.24 \pm 0.25 $  &   $ 0.14 \pm 0.15 $ \\
$\fourtop$ (SM) & $ 0.22 \pm 0.07 $ &   $ 0.06 \pm 0.02 $ &   $ 0.12 \pm 0.04 $ &  $ 0.31 \pm 0.10 $  &   $ 0.77 \pm 0.25 $ \\
Multijet & $ 3.9 \pm 3.9 $ &   $ 0.13 \pm 0.17 $ &   $ 0.20 \pm 0.24 $ &  $ 0.64 \pm 0.68 $  &   $ 2.8 \pm 2.8 $ \\
\midrule
Total background & $ 73 \pm 19 $ &   $ 5.6 \pm 1.4 $ &   $ 8.0 \pm 3.7 $ &  $ 19.7 \pm 5.0 $  &   $ 24.4 \pm 6.3 $ \\
\midrule
Data & 87  & 8  & 7  &  18  & 29  \\
\bottomrule\bottomrule     \\
\end{tabular}
\vspace{0.1cm}
\end{center}
\vspace{-0.5cm}
\caption{Predicted and observed yields in the 0-lepton channel in five of the most sensitive search regions
(depending on the signal scenario) considered.
The background prediction is shown before the fit to data. Also shown are the signal predictions for different benchmark scenarios considered.
The quoted uncertainties are the sum in quadrature of statistical and systematic uncertainties in the yields,
excluding the normalisation uncertainty of the $\ttbin$ background, which is determined via a likelihood fit to data.}
\label{tab:Prefit_Yields_0L_unblind}
\end{table}
% End of text imported from the .//sections/strategy.tex input file
 
\clearpage
% The next lines are included from the .//sections/systematic_uncertainties.tex input file
\section{Systematic uncertainties}
\label{sec:systematics}
 
Several sources of systematic uncertainty are considered that affect the normalisation of signal
and background and/or the shape of their $\meff$ distributions.
Each source of systematic uncertainty is considered to be uncorrelated with the other sources.
Correlations for a given systematic uncertainty are maintained across processes and channels, unless explicitly stated otherwise.

The leading sources of systematic uncertainty vary depending on the analysis region considered.
For example, the total systematic uncertainty of the background normalisation in the highest-sensitivity
search region in the 1-lepton channel ($\geq$0t, $\geq$2H, $\geq$6j, $\geq$4b) is 25\%,
with the largest contributions originating from uncertainties in $\ttbar$+HF modelling and
flavour tagging efficiencies ($b$, $c$, and light). The above uncertainty does
not include the uncertainty in the $\ttbin$ normalisation, which is allowed to vary freely in the fit to data.
However, as discussed previously, the joint fit to data across the 34 search regions considered in total in
the 1-lepton and 0-lepton channels allows the overall background uncertainty to be reduced significantly,
e.g., in the case of the search region specified above, down to 10\% (including the uncertainty in the $\ttbin$ normalisation).
Such a reduction results from the significant constraints that the data places on some
systematic uncertainties, as well as the correlations among systematic uncertainties built
into the likelihood model.
 
The following sections describe the systematic uncertainties considered in this analysis.
 
\subsection{Luminosity}
\label{sec:syst_lumi}
 
The uncertainty in the integrated luminosity is 2.1\%, affecting the overall normalisation of
all processes estimated from the simulation. It is derived, following a methodology similar to that detailed in Ref.~\cite{Aaboud:2016hhf},
from a calibration of the luminosity scale using $x$--$y$ beam-separation scans performed in August 2015 and May 2016.
 
\subsection{Reconstructed objects}
\label{sec:syst_objects}

Uncertainties associated with leptons arise from the trigger, reconstruction, identification, and isolation
efficiencies, as well as the lepton momentum scale and resolution. These are measured in data using $Z\to \ell^+\ell^-$ and
$J/\psi\to \ell^+\ell^-$ events~\cite{ATLAS-CONF-2016-024,Aad:2016jkr}.
The combined effect of all these uncertainties results in an overall normalisation
uncertainty in signal and background of approximately 1\%.
 
Uncertainties associated with jets arise from the jet energy scale
and resolution, and the efficiency to pass the JVT requirement.
The largest contribution results from the jet energy scale,
whose uncertainty dependence on jet $\pt$ and $\eta$, jet flavour, and pile-up treatment is split into 21 uncorrelated components
that are treated independently in the analysis~\cite{Aaboud:2017jcu}.
 
The leading uncertainties associated with reconstructed objects in this analysis originate from the modelling
of the $b$-, $c$-, and light-jet-tagging efficiencies in the simulation, which is corrected
to match the efficiencies measured in data control samples~\cite{Aad:2015ydr}.
Uncertainties in these corrections include a total of six independent sources
affecting $b$-jets and four independent sources affecting $c$-jets.
Each of these uncertainties has a different dependence on jet $\pt$.
Seventeen sources of uncertainty affecting light jets are considered, which depend on jet $\pt$ and $\eta$.
The sources of systematic uncertainty listed above are taken as uncorrelated between $b$-jets,  $c$-jets, and light-jets.
An additional uncertainty is included due to the extrapolation of these corrections to jets
with $\pt$ beyond the kinematic reach of the data calibration samples used ($\pt>300~\gev$ for $b$- and $c$-jets,
and $\pt>750~\gev$ for light-jets); it is taken to be correlated among the three jet flavours.
This uncertainty is evaluated in the simulation by comparing the tagging efficiencies while varying e.g. the fraction of tracks with
shared hits in the silicon detectors or the fraction of fake tracks resulting from random combinations of hits, both of which typically
increase at high $\pt$ due to growing track multiplicity and density of hits within the jet.
Finally, an uncertainty related to the application of $c$-jet scale factors to $\tau$-jets is considered,
but has a negligible impact in this analysis.
The combined effect of these uncertainties results in an uncertainty in the \ttbar\ background normalisation ranging from 4\% to 12\% depending on the analysis region. The corresponding uncertainty range for signal is 2--12\%, assuming $T\bar{T}$ production in the weak-isospin doublet scenario and $m_{T} = 1$~\tev.
 
\subsection{Background modelling}
\label{sec:syst_bkgmodeling}
 
A number of sources of systematic uncertainty affecting the modelling of $\ttbar$+jets are considered.
An uncertainty of  6\% is assigned to the inclusive $\ttbar$ production
cross section~\cite{Czakon:2011xx}, including contributions from varying the factorisation and renormalisation
scales, and from uncertainties in the PDF, $\alpha_{\textrm{S}}$, and the top quark mass, all added in quadrature.
Since several search regions have a sufficiently large number of events of \ttbin\ background,
its normalisation is completely determined by the data during the fit procedure.
In the case of the  \ttcin\ normalisation, since the fit to the data is unable to precisely determine it and
the analysis has very limited sensitivity to its uncertainty, a normalisation uncertainty of 50\% is assumed.
 
Alternative $\ttbar$ samples were generated using {\powheg} interfaced to {\herwigpp}~2.7.1~\cite{Bahr:2008pv}
and {\amcatnlo} 2.2.1 interfaced to {\herwigpp}~2.7.1 in order to estimate systematic uncertainties related to the
modelling of this background.
The effects of initial- and final-state radiation (ISR/FSR) are explored using two alternative {\powheg}+{\pythia}
samples, one with $h_{\textrm{damp}}$ set to $2 m_t$, the renormalisation and factorisation scales set to half the nominal value
and using the P2012 radHi UE tune, giving more radiation (referred to as ``radHi''), and one with the P2012 radLo UE tune, $h_{\textrm{damp}}=m_t$
and the renormalisation and factorisation scales set to twice the nominal value, giving less radiation (referred to as ``radLow'')~\cite{ATL-PHYS-PUB-2016-004}.
 
Uncertainties affecting the modelling of \ttbin\ production include shape uncertainties (including inter-category migration effects)
associated with the NLO prediction from {\ShOL}, which is used for reweighting the nominal  {\powheg}+{\pythia} 6 \ttbin\ prediction.
These uncertainties include different scale variations,  a different shower-recoil model scheme, and two
alternative PDF sets (see Ref.~\cite{Aaboud:2017rss} for details), and are significantly smaller than those estimated by comparing different event generators.
An uncertainty due to the choice of generator is assessed by comparing the \ttbin\ predictions
obtained after reweighting {\powheg}+{\pythia} 6 to the NLO calculation from {\ShOL} and to an equivalent
NLO calculation from {\amcatnlo}+{\pythia} 8, which differs in the procedure used to match the NLO matrix element
calculation and the parton shower (see Section~1.6.8 of Ref.~\cite{deFlorian:2016spz}).
The uncertainty from the parton shower and hadronisation model is taken from the difference between the {\amcatnlo}
calculation showered with either {\pythia}8 or {\herwigpp}.
Additional uncertainties are assessed for the contributions to the \ttbin\ background originating from multiple parton interactions
or final-state radiation from top quark decay products, which are not part of the NLO prediction.
The latter are assessed via the alternative ``radHi'' and ``radLow'' samples, as discussed below.
The nominal NLO corrections, as well as their variations used to propagate the theoretical uncertainties in the NLO prediction, are adjusted so that
the particle-level cross section of the \ttbin\ background (i.e.~prior to reconstruction-level selection requirements)
is fixed to the nominal prediction, i.e.~effectively only migrations across categories and distortions to the shape of the kinematic distributions are considered.
 
In the following, uncertainties affecting all $\ttbar$+jets processes are discussed.
Uncertainties associated with the modelling of ISR/FSR are obtained from the comparison of the \textsc{Powheg}-\textsc{Box}+{\pythia} 6 ``radHi'' and ``radLow'' samples
(see Section~\ref{sec:bkg_model}) with the nominal {\powheg}+{\pythia} 6 sample. An uncertainty associated with the choice of NLO generator is derived
by comparing two $\ttbar$ samples, one generated with {\powheg}+{\herwigpp} and another generated with {\amcatnlo}+{\herwigpp},
and propagating the resulting fractional difference to the nominal {\powheg}+{\pythia} 6 prediction.
An uncertainty due to the choice of parton shower and hadronisation model
is derived by comparing events produced by {\powheg} interfaced to {\pythia} 6 or {\herwigpp}.
Finally, the uncertainty in the modelling of the top quark's $\pt$, affecting only the $\ttbar$+light-jets and \ttcin\ processes,
is evaluated by taking the full difference between applying and not applying the reweighting to match the NNLO prediction.
The above uncertainties are taken as uncorrelated between the $\ttbar$+light-jets, \ttcin\ and \ttbin\ processes.
In the case of \ttbin, in all instances the various HF categories and the corresponding partonic kinematics for the alternative MC samples
are reweighted to match the NLO prediction of  {\ShOL} so that only effects other than distortions to the inter-normalisation of
the various \ttbin\ topologies and their parton-level kinematics are propagated. In the case of $\ttbar$+light-jets and \ttcin\, the full effect of these uncertainties is propagated.
Similarly to the treatment of the NLO corrections and uncertainties associated with \ttbin\ discussed above, in the case of the additional
uncertainties derived by comparing alternative $\ttbar$ samples, the overall normalisation of the \ttcin\ and \ttbin\ background
at the particle level is fixed to the nominal prediction. In this way, only migrations across categories and distortions to the shape of the kinematic
distributions are considered. In order to maintain the inclusive $\ttbar$ cross section, the $\ttbar$+light-jets background is adjusted accordingly.
 
Uncertainties affecting the modelling of the single-top-quark background include a
$+5\%$/$-4\%$ uncertainty in the total cross section estimated as a weighted average
of the theoretical uncertainties in $t$-, $Wt$- and $s$-channel production~\cite{Kidonakis:2011wy,Kidonakis:2010ux,Kidonakis:2010tc}.
Additional uncertainties associated with the modelling of ISR/FSR are assessed by comparing the nominal
samples with alternative samples where generator parameters were varied (i.e.~``radHi'' and ``radLow'').
For the $t$- and $Wt$-channel processes, an uncertainty due to the choice of parton shower and hadronisation model is derived
by comparing events produced by {\powheg} interfaced to {\pythia} 6 or {\herwigpp}.
These uncertainties are treated as fully correlated among single-top production processes, but uncorrelated with the
corresponding uncertainty in the $\ttbar$+jets background.
The sum in quadrature of the above uncertainties on the single top normalisation at the preselection level
is 20\% in the 1-lepton channel and 20\%(25\%) in LM(HM) regions of the 0-lepton channel, respectively.
An additional systematic uncertainty on $Wt$-channel production concerning the separation
between $t\bar{t}$ and $Wt$ at NLO~\cite{Frixione:2008yi} is assessed by comparing
the nominal sample, which uses the so-called ``diagram subtraction'' scheme, with an alternative sample
using the ``diagram removal'' scheme. This uncertainty,  which is taken to be single-sided, has a strong shape dependence and
affects the $Wt$ normalisation by about $-50\%$ in the 1-lepton channel and LM regions of the 0-lepton channel,
and by about $-75\%$ in HM regions of the 0-lepton channel.
Due to the small size of the simulated samples, and hence limited statistical precision, these uncertainties cannot be reliably estimated in each analysis
region and so their estimates at the preselection level are used instead.
They are treated as uncorrelated across regions with different top-tagged jet and Higgs-tagged jet multiplicities and between the 1-lepton and 0-lepton channels.
 
Uncertainties affecting the normalisation of the $V$+jets background are estimated for the sum
of $W$+jets and $Z$+jets, and separately for $V$+light-jets, $V$+$\geq$1$c$+jets, and $V$+$\geq$1$b$+jets subprocesses.
The total normalisation uncertainty of $V$+jets processes is estimated by comparing the data and total background prediction in
the different analysis regions considered, but requiring exactly 0 $b$-tagged jets. Agreement between data and predicted background
in these modified regions, which are dominated by $V$+light-jets, is found to be within approximately 30\%. This bound is taken to
be the normalisation uncertainty, correlated across all $V$+jets subprocesses.
Since {\sherpa}~2.2 has been found to underestimate $V$+heavy-flavour by about a factor
of 1.3~\cite{Aaboud:2017xsd}, additional 30\% normalisation uncertainties are assumed for $V$+$\geq$1$c$+jets and $V$+$\geq$1$b$+jets
subprocesses, considered uncorrelated between them. These uncertainties are treated as uncorrelated across regions with
different top-/Higgs-tagged jet multiplicities and between the 1-lepton and 0-lepton channels.
 
Uncertainties in the diboson background normalisation include 5\% from the NLO theory cross sections~\cite{Campbell:1999ah},
as well as an additional 24\% normalisation uncertainty added in quadrature for each additional inclusive jet-multiplicity bin, based on a
comparison among different algorithms for merging LO matrix elements and parton showers~\cite{Alwall:2007fs}.
Therefore, normalisation uncertainties of $5\% \oplus \sqrt{3}\times 24\% = 42\%$ and $5\% \oplus \sqrt{4}\times 24\% = 48\%$
are assigned for events with exactly 5 jets and $\geq$6 jets, respectively (this assumes that two jets come from the $W/Z$ decay,
as in $WW/WZ \to \ell \nu jj$). Recent comparisons between data and {\sherpa}~2.1.1 for $WZ(\to \ell\nu\ell\ell) + \geq$4 jets show
agreement within the experimental uncertainty of approximately 40\%~\cite{Aaboud:2016yus}, which further justifies the above uncertainty.
This uncertainty is taken to be uncorrelated across regions with different top-/Higgs-tagged jet multiplicities and between the 1-lepton and 0-lepton channels
 
Uncertainties in the $\ttbar V$ and $\ttbar H$ cross sections are 15\% and $+10\%$/$-13\%$, respectively,
from the uncertainties in their respective NLO theoretical cross sections~\cite{Campbell:2012dh,Garzelli:2012bn,Dittmaier:2011ti}.
Finally, an uncertainty of 30\% is estimated for the NLO prediction of the SM $\fourtop$ cross section~\cite{Alwall:2014hca}.
Since no additional modelling uncertainties are taken into account for these backgrounds, and the 1-lepton and 0-lepton
channels cover different kinematic phase spaces, the above uncertainties in the $\ttbar V$, $\ttbar H$, and SM $\fourtop$
cross sections are taken to be uncorrelated between the two channels.
 
Uncertainties in the data-driven multijet background estimate receive
contributions from the limited sample size in data, particularly at high jet and $b$-tag multiplicities, as
well as from the uncertainty in the misidentified-lepton rate, measured in
different control regions (e.g.~selected with a requirement on either the maximum $\met$ or $\mtw$).
Based on the comparisons between data and total prediction in multijet-rich selections, the normalisation uncertainties assumed for this
background are 50\% (100\%) for electrons with $|\eta_{\textrm{cluster}}|\leq 1$ ($|\eta_{\textrm{cluster}}|>1$), and 50\% for muons,
taken to be uncorrelated across regions with different top-/Higgs-tagged jet multiplicities and between events containing electrons and events containing muons.
In the case of the 0-lepton channel, the normalisation uncertainty assigned to the multijet background is 100\%.
No explicit shape uncertainty is assigned since the large statistical uncertainties associated with the multijet background prediction,
which are uncorrelated between bins in the final discriminant distribution, are assumed to effectively cover possible shape uncertainties.
 
% End of text imported from the .//sections/systematic_uncertainties.tex input file
 
% The next lines are included from the .//sections/statistical_analysis.tex input file
\section{Statistical analysis}
\label{sec:stat_analysis}
 
For each search, the $\meff$ distributions across all regions considered are jointly analysed to test for the presence
of a signal predicted by the benchmark scenarios.
The statistical analysis uses a binned likelihood function ${\cal L}(\mu,\theta)$ constructed as
a product of Poisson probability terms over all bins considered in the search. This function depends
on the signal-strength parameter $\mu$, which
multiplies
the predicted production cross section for signal,
and $\theta$, a set of nuisance parameters that encode the effect of systematic uncertainties
in the signal and background expectations. Therefore, the expected total number of events in a given bin
depends on $\mu$ and
$\theta$. With the exception of the parameter that controls the normalisation of the
\ttbin\ background, all other nuisance parameters are implemented in the likelihood function as Gaussian or
log-normal constraints. The above-mentioned \ttbin\ normalisation factor is a free parameter of the fit.
 
For a given value of $\mu$, the nuisance parameters $\theta$ allow variations of the expectations for signal and background
according to the corresponding systematic uncertainties, and their fitted values result in the deviations from
the nominal expectations that globally provide the best fit to the data.
This procedure allows a reduction of the impact of systematic uncertainties on
the search sensitivity by taking advantage of the highly populated background-dominated regions included in the likelihood fit.
To verify the improved background prediction, fits under the background-only hypothesis are performed,
and differences between the data and the post-fit background prediction are checked
using kinematic variables other than the ones used in the fit.
The $\meff$ distributions in validation regions not used in the fit are also checked.
Statistical uncertainties in each bin of the predicted $\meff$ distributions due to the limited size of the simulated samples
are taken into account by dedicated parameters in the fit.
 
The test statistic $q_\mu$ is defined as the profile likelihood ratio:
$q_\mu = -2\ln({\cal L}(\mu,\hat{\hat{\theta}}_\mu)/{\cal L}(\hat{\mu},\hat{\theta}))$,
where $\hat{\mu}$ and $\hat{\theta}$ are the values of the parameters that
maximise the likelihood function (subject to the constraint $0\leq \hat{\mu} \leq \mu$), and $\hat{\hat{\theta}}_\mu$ are the values of the
nuisance parameters that maximise the likelihood function for a given value of $\mu$.
The test statistic $q_\mu$ is evaluated with the {\textsc RooFit} package~\cite{Verkerke:2003ir,RooFitManual}.
A related statistic is used to determine the probability that the observed data are compatible with the background-only hypothesis (i.e.~the discovery test)
by setting $\mu=0$ in the profile likelihood ratio and leaving $\hat{\mu}$ unconstrained: $q_0 = -2\ln({\cal L}(0,\hat{\hat{\theta}}_0)/{\cal L}(\hat{\mu},\hat{\theta}))$.
The $p$-value (referred to as $p_0$) representing the probability of the data being compatible with the
background-only hypothesis is estimated by integrating
the distribution of $q_0$ from background-only pseudo-experiments, approximated using the asymptotic formulae given in Refs.~\cite{Cowan:2010js,ErratumCowan:2010js}, above the observed value of $q_0$. Some model dependence exists
in the estimation of the $p_0$, as a given signal scenario needs to be assumed in the calculation of the denominator of $q_\mu$, even
if the overall signal normalisation is left floating and fitted to data. The observed $p_0$ is checked for each explored signal scenario.
Upper limits on the signal production cross section for each of the
signal scenarios considered are derived by using $q_\mu$ in the CL$_{\textrm{s}}$ method~\cite{Junk:1999kv,Read:2002hq}.
For a given signal scenario, values of the production cross section (parameterised by $\mu$) yielding
CL$_{\textrm{s}} < 0.05$,
where CL$_{\textrm{s}}$ is computed using the asymptotic approximation~\cite{Cowan:2010js,ErratumCowan:2010js}, are excluded at $\geq 95\%$ CL.

% End of text imported from the .//sections/statistical_analysis.tex input file
 
% The next lines are included from the .//sections/results.tex input file
\section{Results}
\label{sec:results}
 
This section presents the results obtained from searches in the 1-lepton and 0-lepton channels, as
well as their combination, following the statistical analysis discussed in Section~\ref{sec:stat_analysis}.
 
\subsection{Likelihood fits to data}
\label{sec:result_fits}
 
A binned likelihood fit under the background-only hypothesis is performed on the $\meff$ distributions in all search regions considered.
In this section, the results of the simultaneous likelihood fit to the search regions in the 1-lepton and 0-lepton channels are discussed. This combined fit is used to obtain results on $T\bar{T}$ production. In this combination, all common systematic uncertainties are considered fully correlated between
the 1-lepton and 0-lepton channels, with the exception of those affecting non-$\ttbar$ backgrounds.
To obtain the results in the individual channels, separate fits are performed.
In general, good agreement is found among the fitted nuisance parameters in the individual and combined fits.
 
\begin{figure*}[t!]
\begin{center}
\includegraphics[width=0.8\textwidth]{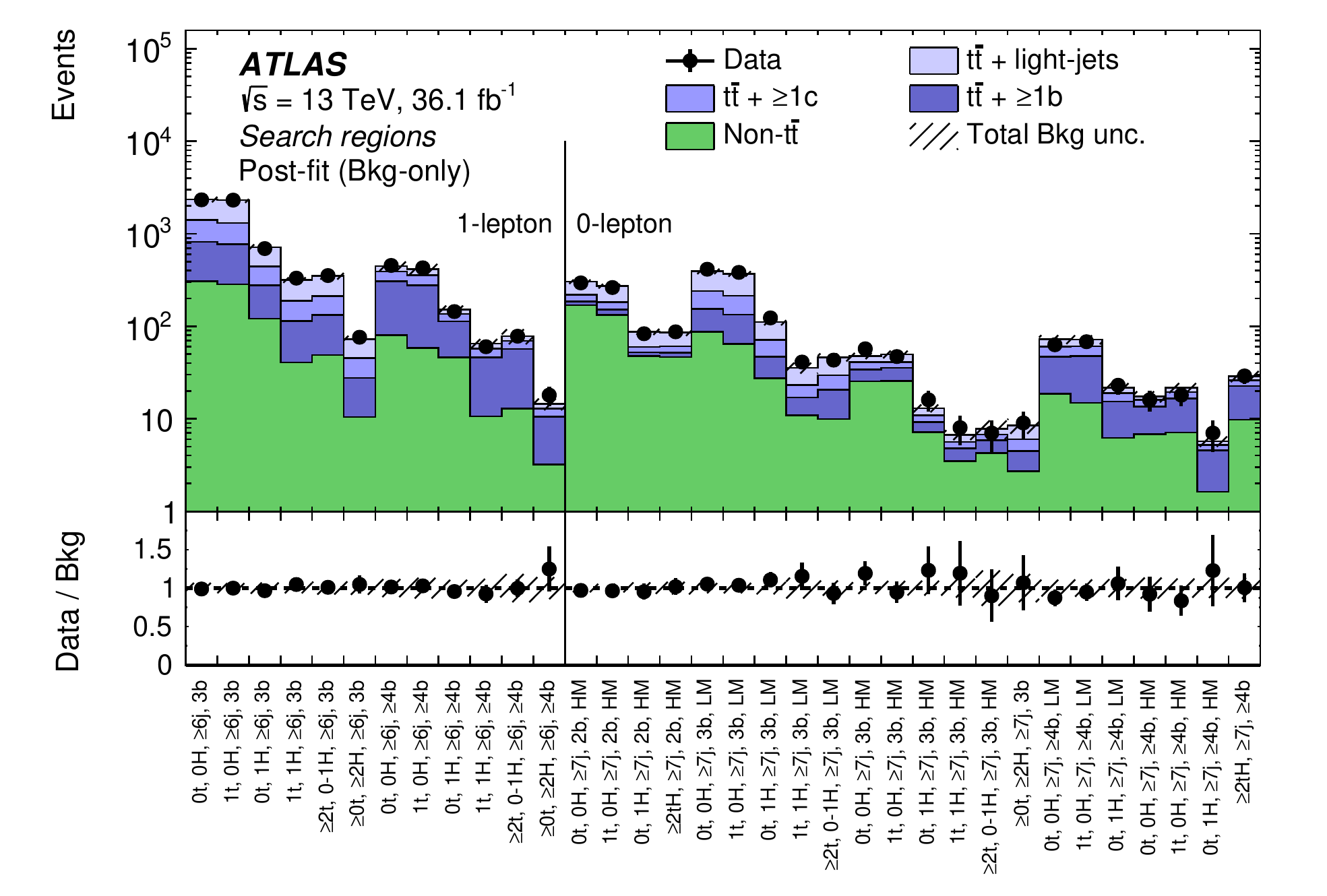}
\caption{\small {Comparison between the data and the background prediction for the yields in the search regions considered in
the 1-lepton and 0-lepton channels, after the combined fit to data (``Post-fit'') under the background-only hypothesis.
The small contributions from $\ttbar V$, $\ttbar H$, single-top, $W/Z$+jets, diboson, and multijet backgrounds are combined into a single background source
referred to as ``Non-$\ttbar$''.
The bottom panel displays the ratio of data to the SM background (``Bkg'') prediction.
The hashed area represents the total uncertainty of the background.}}
\label{fig:Summary_postFit_COMB}
\end{center}
\end{figure*}
 
A comparison of the distribution of observed and expected yields in the search regions in the 1-lepton and 0-lepton channels after the combined fit
is shown in Figure~\ref{fig:Summary_postFit_COMB} (see Figure~\ref{fig:Summary} for the results before the combined fit).
The post-fit yields in five of the most sensitive search regions in the 1-lepton and 0-lepton channels can be found in
Tables~\ref{tab:Postfit_Yields_1L_unblind_COMB} and~\ref{tab:Postfit_Yields_0L_unblind_COMB}, respectively.
For the same search regions, the corresponding $\meff$ distributions, both before and after the fit to data, are shown in
Figures~\ref{fig:prepostfit_unblinded_COMB_1}--\ref{fig:prepostfit_unblinded_COMB_5}.
The binning used for the $\meff$ distributions in the different search regions represents a compromise between preserving enough
discrimination between the background and the different signal hypotheses considered, and keeping the statistical uncertainty on
the background prediction per bin well below 30\%. While some of the systematic uncertainties from individual sources described in Section \ref{sec:systematics} vary across the $\meff$ spectrum, the total pre-fit uncertainty
is largely independent of $\meff$.
The large number of events in the signal-depleted regions, together with their different background compositions, and the
assumptions of the fit model, constrain the combined effect of the sources of systematic uncertainty.
As a result, an improved background prediction is obtained with significantly reduced uncertainty, not only in the
signal-depleted channels, but also in the signal-rich channels such as ($\geq$0t, $\geq$2H, $\geq$6j, $\geq$4b) in the 1-lepton channel.
In the combined fit, the channels with three $b$-tagged jets are effectively used to constrain the leading uncertainties affecting the $\ttbar$+light-jets background prediction,
while the channels with $\geq$4 $b$-tagged jets are sensitive to the uncertainties affecting the $\ttbar$+HF background prediction.
In particular, one of the main corrections
determined in
the fit is a scale factor that multiplies the \ttbin\ normalisation by $0.90 \pm 0.23$ relative
to the nominal
prediction.\footnote{Even though the \ttbin\ normalisation factor is assumed to be the same in all regions, the overall change in \ttbin\ normalisation can be
different across channels due to the different impact of other nuisance parameters affecting the \ttbin\ background, such as those
related to \ttbin\ modelling.}
In addition, the nuisance parameter controlling the \ttcin\ normalisation
is adjusted to scale this background by a factor of $1.3 \pm 0.4$ relative to its nominal prediction.
The fit results in better agreement between data and prediction in the channels with $\geq$3 $b$-tagged jets,
where the $\ttbar$+HF background dominates.
Detailed studies were performed to verify the stability of the fit against variations in the treatment of the systematic uncertainties affecting
the $\ttbar$+HF background (e.g.\ by decorrelating normalisation and shape uncertainties between different \ttbin\ categories,
or by scaling the  \ttbin\ and  \ttcin\ backgrounds by a common factor), finding in all instances a robust post-fit
background prediction.
Furthermore, the impact on the background-only fit of injecting a $T\bar{T}$ signal (with $m_T=1~\tev$) in the doublet configuration was confirmed to be negligible.
Although
there is no single nuisance parameter directly responsible for the normalisation of $\ttbar$+light-jets background, the yields for this
contribution within each region are affected by systematic uncertainties in the $\ttbar$ modelling and the jet flavour tagging,
and thus are changed after the fit.
 
A comparison of the distribution of observed and expected yields in all validation regions considered, before and after the combined fit in the search regions,
is shown in Figure~\ref{fig:Summary_VR_preFit_postFit_COMB}. Agreement between data and prediction in normalisation and shape of the $\meff$ distribution
for these regions, which are not used in the fit, is generally improved after the fit, giving confidence in the overall procedure.
To increase the background yields and strengthen the validation of the fit strategy, comparisons between data and background prediction, before and after the fit, are performed for more-inclusive event selections. As an example, the distributions of two kinematic variables used to define the search strategy can be found in Figures~\ref{fig:prepostfit_RCMHiggs_jets_n} and~\ref{fig:prepostfit_mtbmin}. They display respectively the Higgs-tagged jet multiplicity in the 1-lepton channel, after requiring at least 6 jets and 3 $b$-jets, and the distribution of the $\mtbmin$ variable in the 0-lepton channel for events containing at least 7 jets and 2 $b$-jets, together with at least one top/Higgs-tagged jet.
Although these variables are not directly used in the fit, a good description of the data by the post-fit background
prediction is observed, which further validates the fitting procedure.
The result of the background-only fit to data is used for the background prediction in the computation of the limits presented in the following subsections.
 
\begin{table}
\begin{center}
\begin{tabular}{l*{5}{c}}
\toprule\toprule
1-lepton channel & $\geq$2t, 0--1H, &  1t, 0H,  & 1t, 1H,  & $\geq$2t, 0--1H, & $\geq$0t, $\geq$2H,  \\
& $\geq$6j, 3b & $\geq$6j, $\geq$4b & $\geq$6j, $\geq$4b &  $\geq$6j, $\geq$4b & $\geq$6j, $\geq$4b \\
\midrule\midrule
$t\bar{t}$+light-jets & $ 137 \pm 24 $ & $ 59 \pm 11 $ &   $ 7.6 \pm 1.6 $ &   $ 9.0 \pm 2.0 $ &   $ 1.50 \pm 0.34 $ \\
\ttcin\ & $ 79 \pm 34 $ & $ 81 \pm 26 $ &   $ 11.4 \pm 3.8 $ &   $ 12.4 \pm 5.1 $ &   $ 2.36 \pm 0.84 $ \\
\ttbin\ & $ 84 \pm 20 $ & $ 217 \pm 27 $ &   $ 35.3 \pm 5.6 $ &   $ 44.1 \pm 9.1 $ &   $ 7.4 \pm 1.2 $ \\
$t\bar{t}V$ & $ 10.7 \pm 1.6 $ & $ 13.2 \pm 2.1 $ &   $ 2.12 \pm 0.34 $ &   $ 2.82 \pm 0.46 $ &   $ 0.50 \pm 0.08 $ \\
$t\bar{t}H$ & $ 5.26 \pm 0.61 $ & $ 17.4 \pm 2.3 $ &   $ 4.28 \pm 0.56 $ &   $ 3.25 \pm 0.46 $ &   $ 1.33 \pm 0.17 $ \\
$W$+jets & $ 11.4 \pm 4.0 $ & $ 9.5 \pm 3.4 $ &   $ 0.71 \pm 0.36 $ &   $ 1.68 \pm 0.59 $ &   $ 0.78 \pm 0.31 $ \\
$Z$+jets & $ 1.56 \pm 0.55 $ & $ 1.11 \pm 0.41 $ &   $ 0.08 \pm 0.06 $ &   $ 0.16 \pm 0.06 $ &   $ 0.07 \pm 0.04 $ \\
Single top & $ 11.3 \pm 5.6 $ & $ 10.8 \pm 6.2 $ &   $ 2.01 \pm 0.62 $ &   $ 1.85 \pm 0.90 $ &   $ 0.24 \pm 0.15 $ \\
Diboson & $ 2.20 \pm 0.91 $ & $ 1.10 \pm 0.50 $ &   $ 0.20 \pm 0.08 $ &   $ 0.30 \pm 0.12 $ &   $ 0.03 \pm 0.07 $ \\
$\fourtop$ (SM) & $ 2.83 \pm 0.84 $ & $ 5.3 \pm 1.5 $ &   $ 1.20 \pm 0.35 $ &   $ 2.74 \pm 0.79 $ &   $ 0.24 \pm 0.07 $ \\
\midrule
Total background & $ 349 \pm 20 $ & $ 416 \pm 18 $ &   $ 64.9 \pm 4.7 $ &   $ 78.2 \pm 8.0 $ &   $ 14.4 \pm 1.2 $ \\
\midrule
Data & 353 & 428  & 60  & 78  & 18  \\
\bottomrule\bottomrule     \\
\end{tabular}
\vspace{0.1cm}
\end{center}
\vspace{-0.5cm}
\caption{Predicted and observed yields in the 1-lepton channel in five of the most sensitive search regions
(depending on the signal scenario) considered.
The multijet background is considered negligible in these regions and thus not shown.
The background prediction is shown after the combined fit to data in the 0-lepton and 1-lepton channels under the background-only hypothesis.
The quoted uncertainties are the sum in quadrature of statistical and systematic uncertainties in the yields,
computed taking into account correlations among nuisance parameters and among processes.}
\label{tab:Postfit_Yields_1L_unblind_COMB}
\end{table}
\begin{table}
\begin{center}
\begin{tabular}{l*{5}{c}}
\toprule\toprule
0-lepton channel & $\geq$2tH, & 1t, 1H, & $\geq$2t, 0--1H,  & 1t, 0H, & $\geq$2tH,  \\
& $\geq$7j, 2b, HM & $\geq$7j, 3b, HM & $\geq$7j, 3b, HM & $\geq$7j, $\geq$4b, HM & $\geq$7j, $\geq$4b\\
\midrule\midrule
$t\bar{t}$+light-jets & $ 24.7 \pm 5.0 $ &   $ 1.08 \pm 0.20 $ &   $ 1.04 \pm 0.25 $ &  $ 2.20 \pm 0.43 $  &   $ 2.91 \pm 0.57 $ \\
\ttcin\ &  $ 9.2 \pm 4.9 $ &   $ 0.85 \pm 0.44 $ &   $ 0.89 \pm 0.48 $ &  $ 2.9 \pm 1.1 $  &   $ 3.4 \pm 1.4 $ \\
\ttbin\ &  $ 5.3 \pm 1.9 $ &   $ 1.31 \pm 0.39 $ &   $ 1.58 \pm 0.55 $ &  $ 9.4 \pm 1.3 $  &   $ 12.8 \pm 2.4 $ \\
$t\bar{t}V$ & $ 5.96 \pm 0.88 $ &   $ 0.59 \pm 0.09 $ &   $ 1.00 \pm 0.15 $ &  $ 1.46 \pm 0.23 $  &   $ 1.25 \pm 0.19 $  \\
$t\bar{t}H$ &  $ 0.61 \pm 0.08 $ &   $ 0.19 \pm 0.03 $ &   $ 0.13 \pm 0.02 $ &  $ 1.02 \pm 0.13 $  &   $ 1.16 \pm 0.17 $ \\
$W$+jets & $ 12.0 \pm 3.2 $ &   $ 0.63 \pm 0.22 $ &   $ 0.92 \pm 0.34 $ &  $ 0.71 \pm 0.27 $  &   $ 0.86 \pm 0.22 $ \\
$Z$+jets & $ 10.6 \pm 3.1 $ &   $ 0.69 \pm 0.26 $ &   $ 0.4 \pm 1.3 $ &  $ 0.65 \pm 0.29 $  &   $ 0.94 \pm 0.29 $ \\
Single top & $ 8.9 \pm 3.2 $ &   $ 0.77 \pm 0.36 $ &   $ 0.95 \pm 0.48 $ &  $ 1.84 \pm 0.82 $  &   $ 1.17 \pm 0.47$ \\
Diboson & $ 3.9 \pm 1.6 $ &   $ 0.41 \pm 0.39 $ &   $ 0.53 \pm 0.44 $ &  $ 0.37 \pm 0.15 $  &   $ 0.23 \pm 0.10 $   \\
$\fourtop$ (SM) & $ 0.20 \pm 0.07 $ &   $ 0.05 \pm 0.02 $ &   $ 0.12 \pm 0.04 $ &  $ 0.36 \pm 0.10 $  &   $ 0.87 \pm 0.24 $ \\
Multijet & $ 4.1 \pm 3.7 $ &   $ 0.14 \pm 0.13 $ &   $ 0.18 \pm 0.19 $ &  $ 0.67 \pm 0.62 $  &   $ 3.3 \pm 2.6 $ \\
\midrule
Total background & $ 85.5 \pm 6.8 $ &   $ 6.70 \pm 0.75 $ &   $ 7.8 \pm 1.7 $ &  $ 21.6 \pm 1.4 $  &   $ 28.8 \pm 3.1 $  \\
\midrule
Data & 87  & 8  & 7  &  18  & 29  \\
\bottomrule\bottomrule     \\
\end{tabular}
\vspace{0.1cm}
\end{center}
\vspace{-0.5cm}
\caption{Predicted and observed yields in the 0-lepton channel in five of the most sensitive search regions
(depending on the signal scenario) considered.
The background prediction is shown after the combined fit to data in the 0-lepton and 1-lepton channels under the background-only hypothesis.
The quoted uncertainties are the sum in quadrature of statistical and systematic uncertainties in the yields,
computed taking into account correlations among nuisance parameters and among processes.}
\label{tab:Postfit_Yields_0L_unblind_COMB}
\end{table}

\begin{figure*}[t!]
\begin{center}
\subfloat[]{\includegraphics[width=0.40\textwidth]{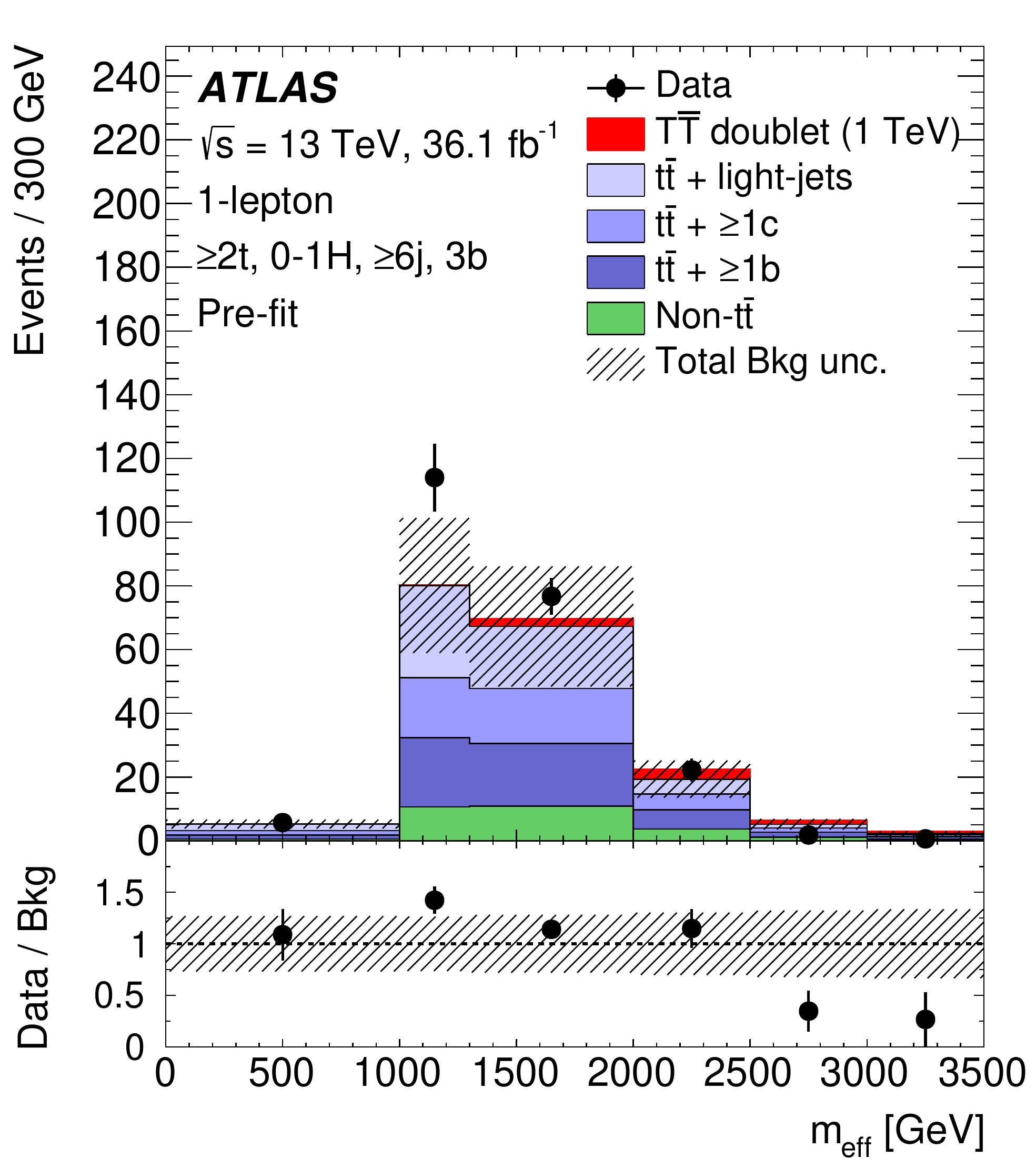}}
\subfloat[]{\includegraphics[width=0.40\textwidth]{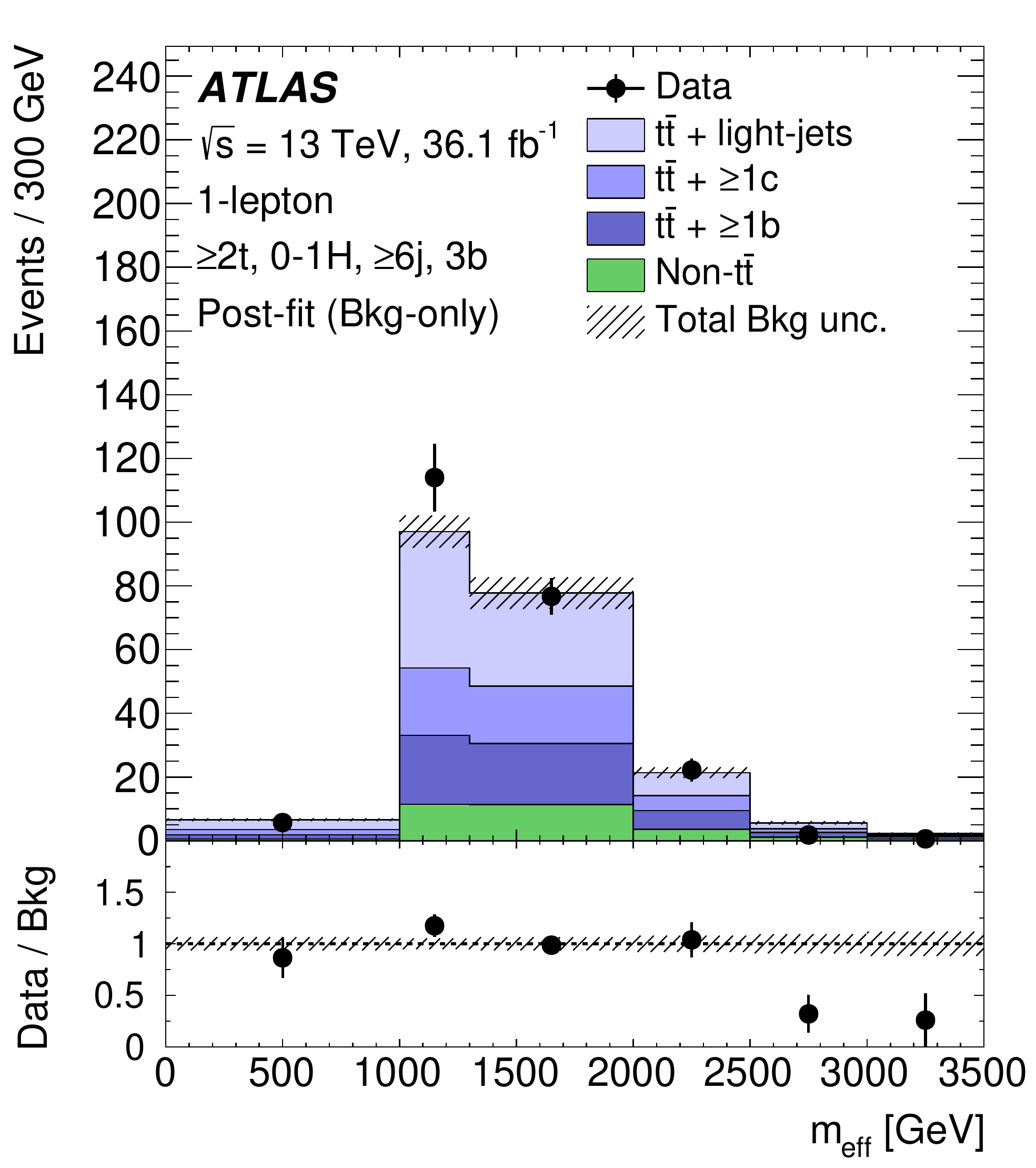}} \\
\subfloat[]{\includegraphics[width=0.40\textwidth]{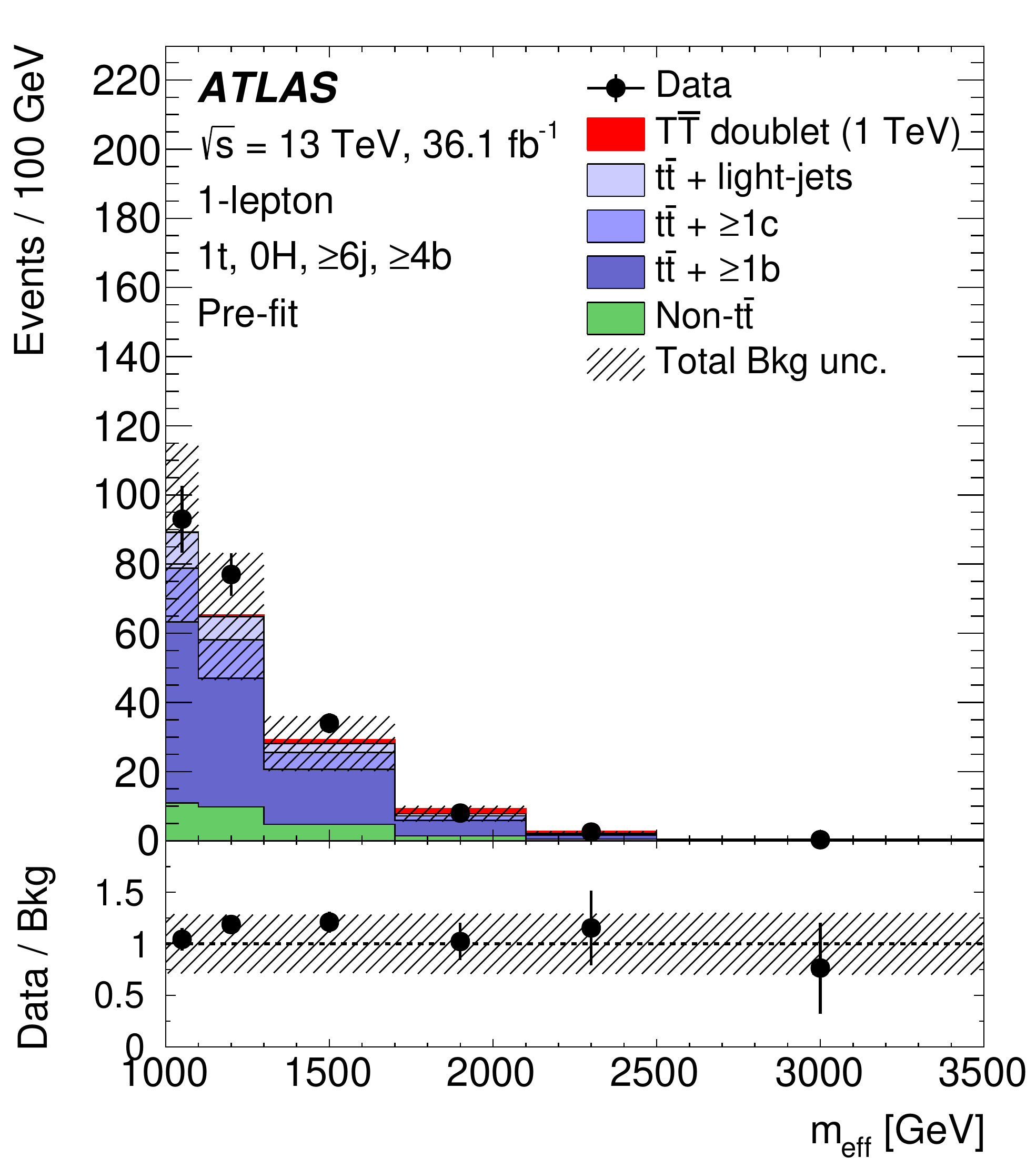}}
\subfloat[]{\includegraphics[width=0.40\textwidth]{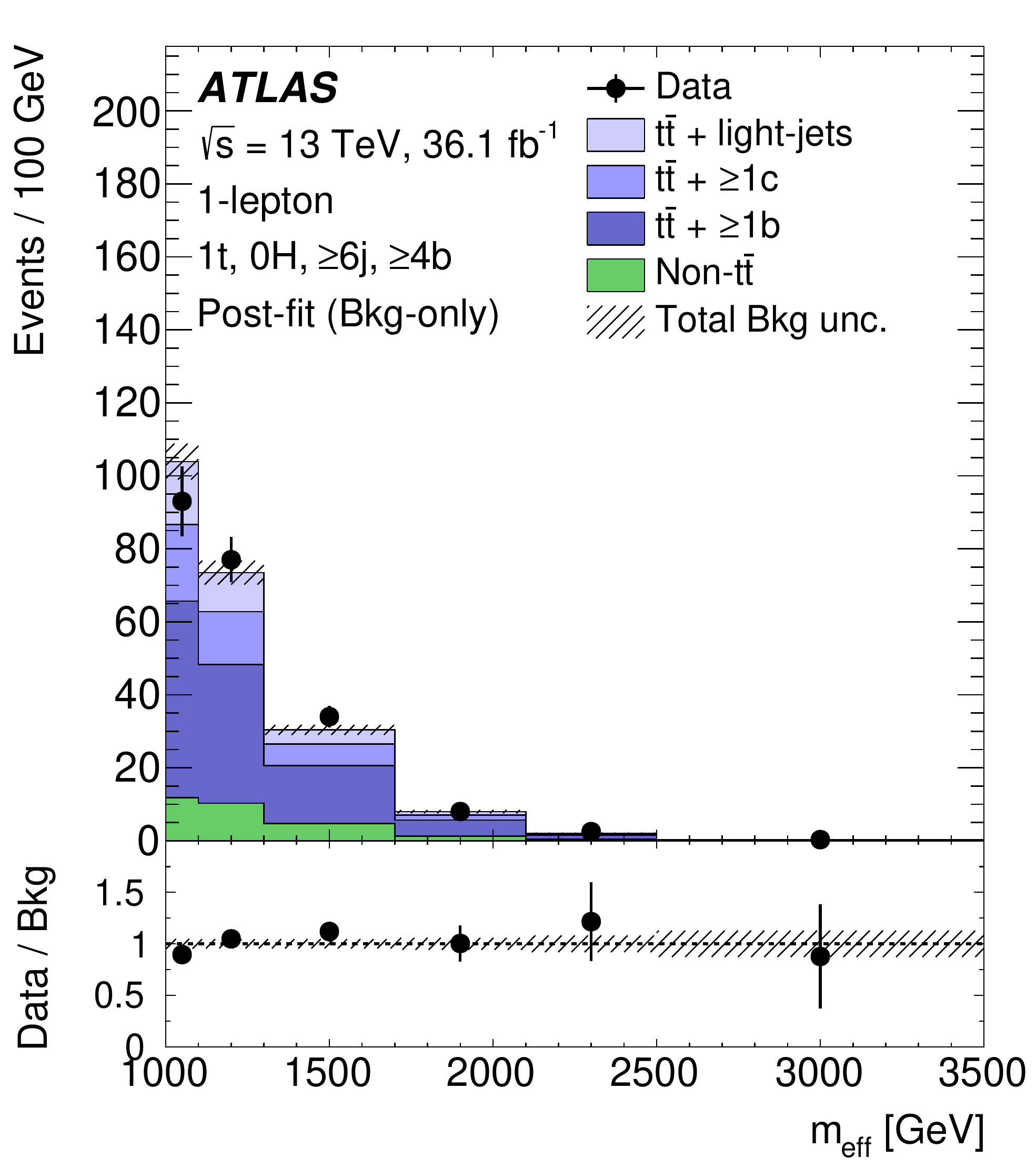}} \\
\caption{\small{Comparison between the data and prediction for the $\meff$ distribution in some of the most sensitive search regions in the 1-lepton channel,
before and after performing the combined fit to data in the 0-lepton and 1-lepton channels (``Pre-fit'' and ``Post-fit'', respectively) under the background-only hypothesis.
Shown are the ($\geq$2t, 0--1H, $\geq$6j, 3b) region (a) pre-fit and (b) post-fit,  and
the (1t, 0H, $\geq$6j, $\geq$4b) region (c) pre-fit and (d) post-fit.
In the pre-fit figures the expected $T\bar{T}$ signal (solid red) corresponding to $m_{T}=1~\tev$ in the $T$ doublet scenario is also shown,
added on top of the background prediction.
The small contributions from $\ttbar V$, $\ttbar H$, single-top, $W/Z$+jets, diboson, and multijet backgrounds are combined into a single background source
referred to as ``Non-$\ttbar$''. The last bin in all figures contains the overflow.
The bottom panels display the ratios of data to the total background prediction (``Bkg'').
The hashed area represents the total uncertainty of the background.
In the case of the pre-fit background uncertainty, the normalisation uncertainty of the $\ttbin$ background is not included.}}
\label{fig:prepostfit_unblinded_COMB_1}
\end{center}
\end{figure*}
\begin{figure*}[t!]
\begin{center}
\subfloat[]{\includegraphics[width=0.40\textwidth]{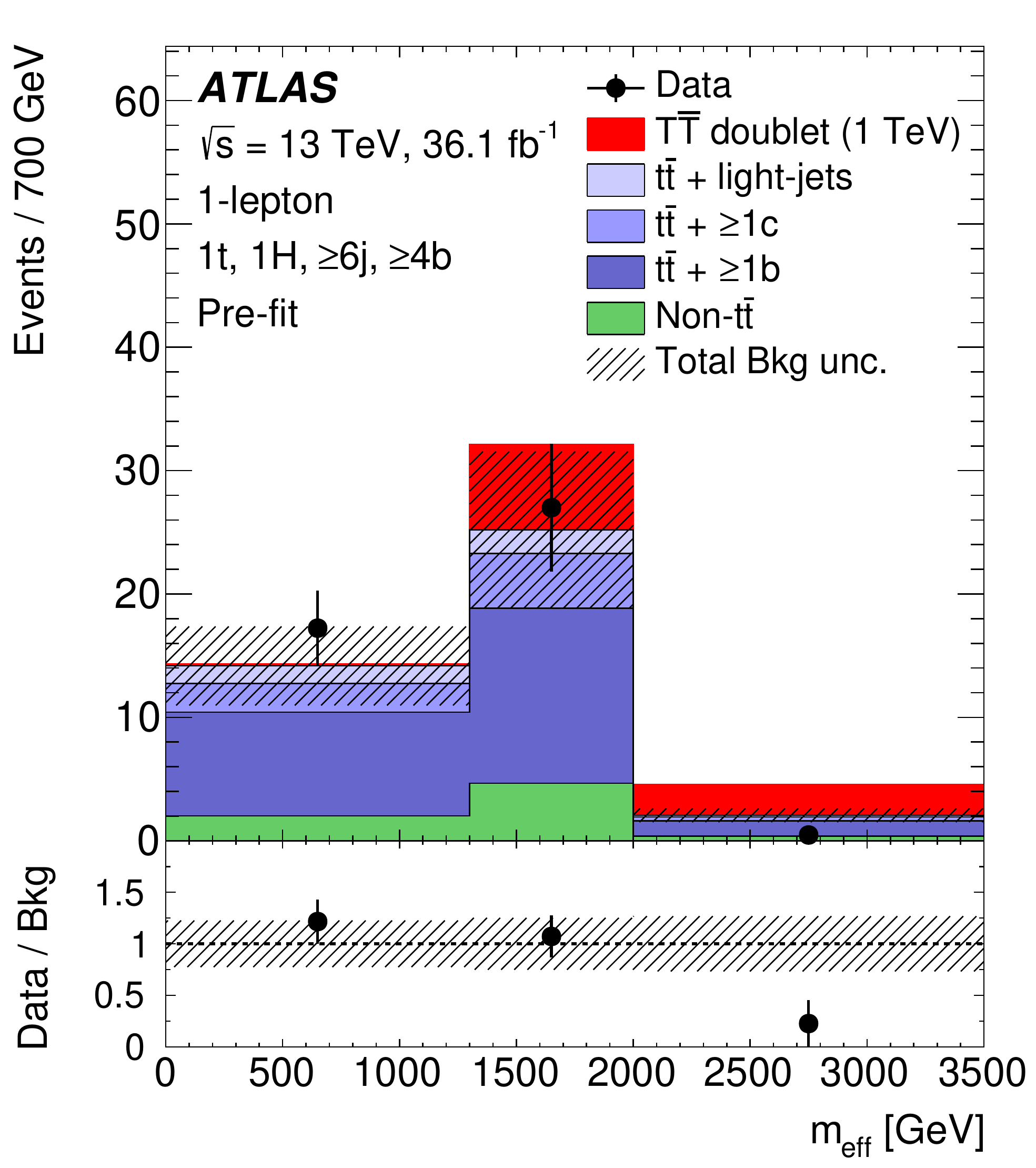}}
\subfloat[]{\includegraphics[width=0.40\textwidth]{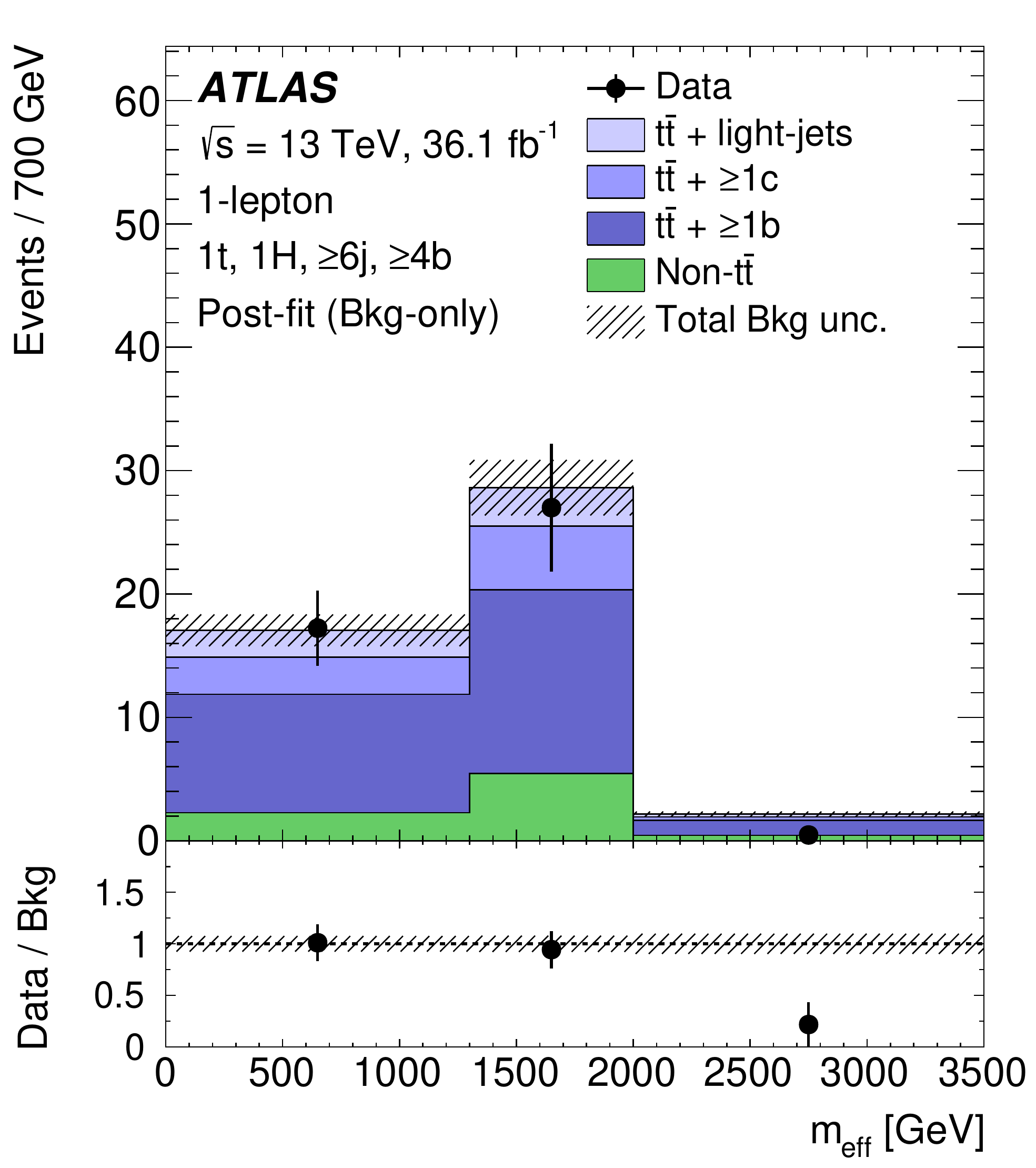}} \\
\subfloat[]{\includegraphics[width=0.40\textwidth]{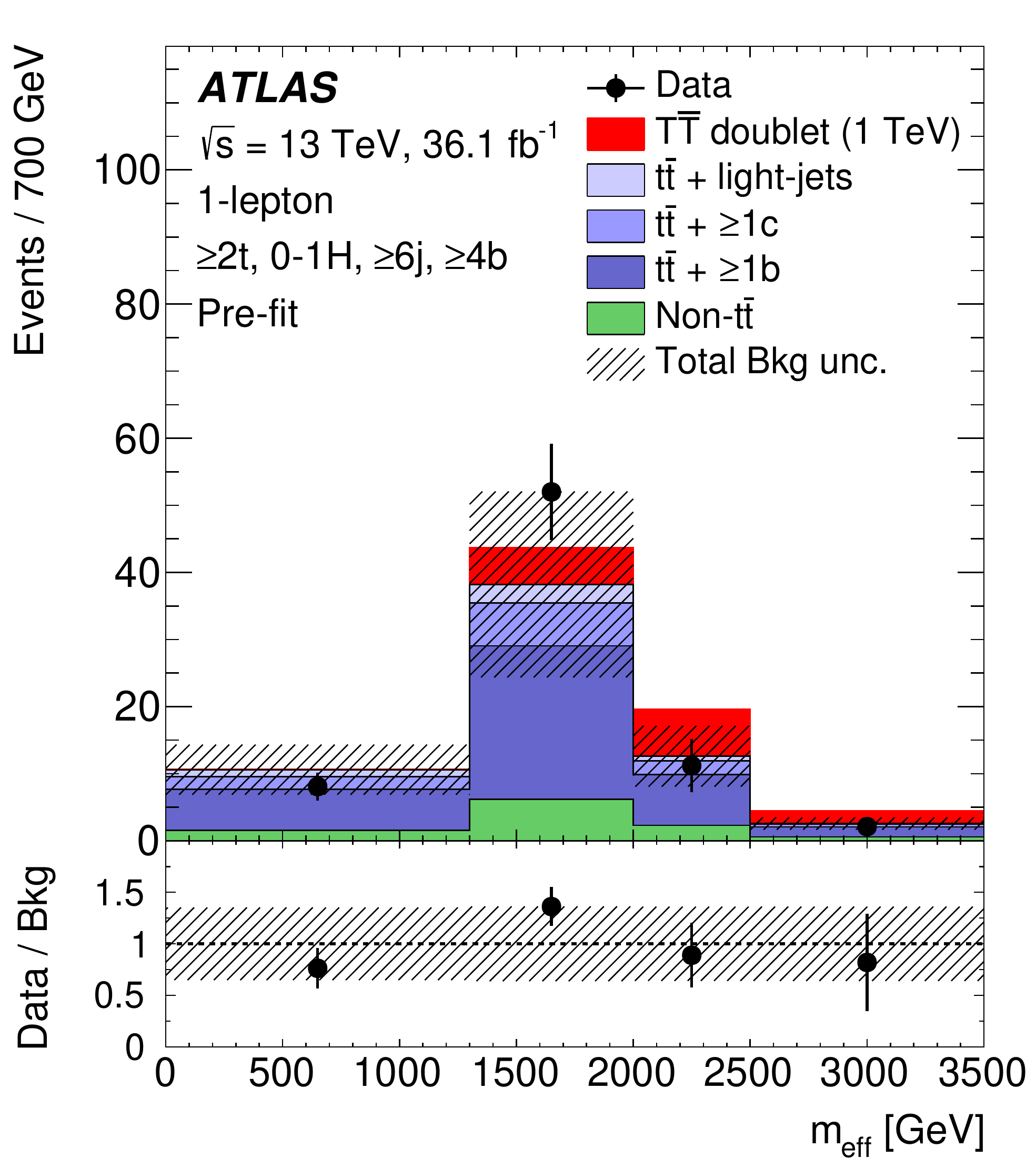}}
\subfloat[]{\includegraphics[width=0.40\textwidth]{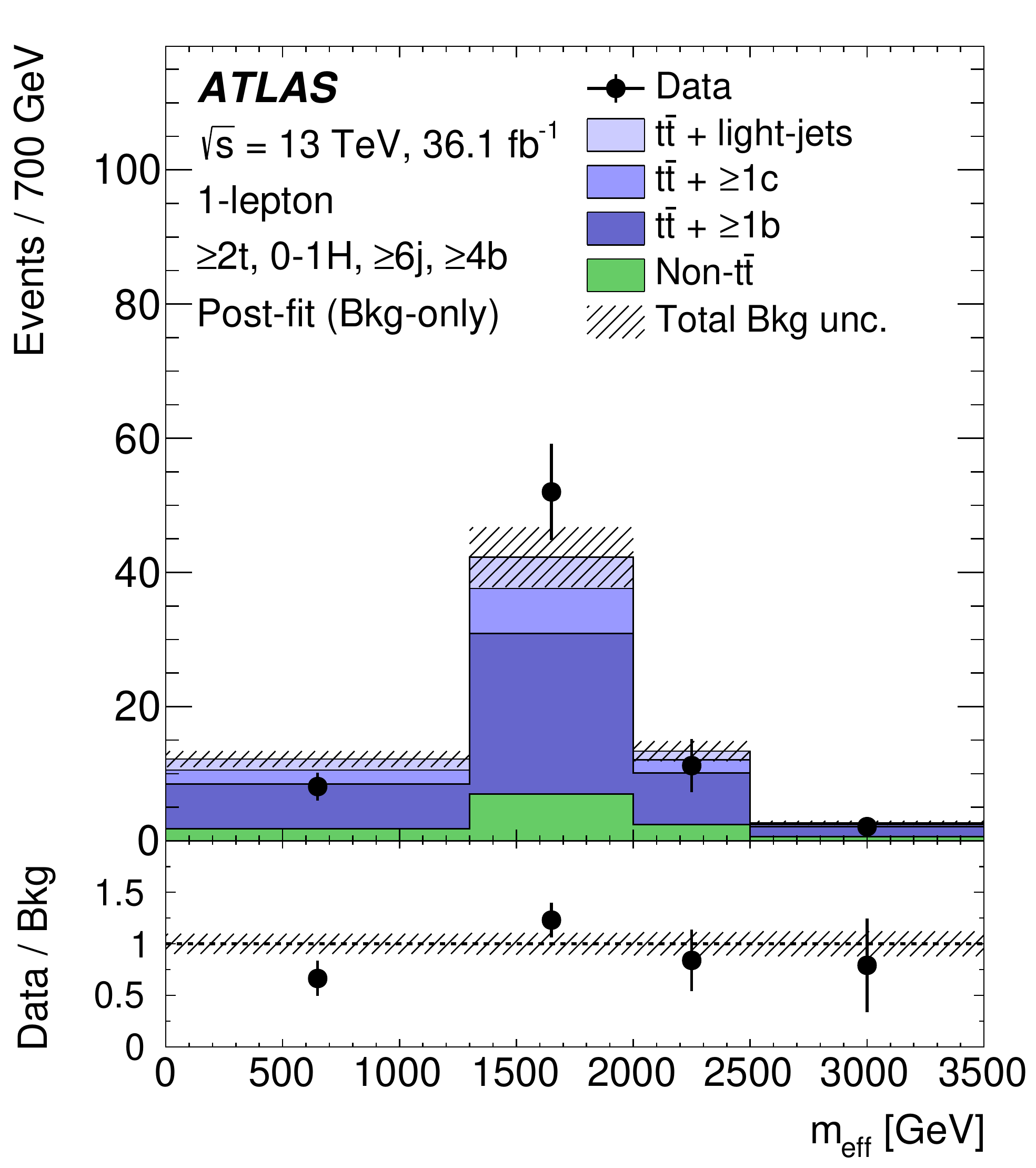}} \\
\caption{\small{Comparison between the data and prediction for the $\meff$ distribution in some of the most sensitive search regions in the 1-lepton channel,
before and after performing the combined fit to data in the 0-lepton and 1-lepton channels (``Pre-fit'' and ``Post-fit'', respectively) under the background-only hypothesis.
Shown are the (1t, 1H, $\geq$6j, $\geq$4b) region (a) pre-fit and (b) post-fit,  and
the ($\geq$2t, 0--1H, $\geq$6j, $\geq$4b) region (c) pre-fit and (d) post-fit.
In the pre-fit figures the expected $T\bar{T}$ signal (solid red) corresponding to $m_{T}=1~\tev$ in the $T$ doublet scenario is also shown,
added on top of the background prediction.
The small contributions from $\ttbar V$, $\ttbar H$, single-top, $W/Z$+jets, diboson, and multijet backgrounds are combined into a single background source
referred to as ``Non-$\ttbar$''. The last bin in all figures contains the overflow.
The bottom panels display the ratios of data to the total background prediction (``Bkg'').
The blue triangles indicate points that are outside the vertical range of the figure.
The hashed area represents the total uncertainty of the background.
In the case of the pre-fit background uncertainty, the normalisation uncertainty of the $\ttbin$ background is not included.}}
\label{fig:prepostfit_unblinded_COMB_2}
\end{center}
\end{figure*}
\begin{figure*}[t!]
\begin{center}
\subfloat[]{\includegraphics[width=0.40\textwidth]{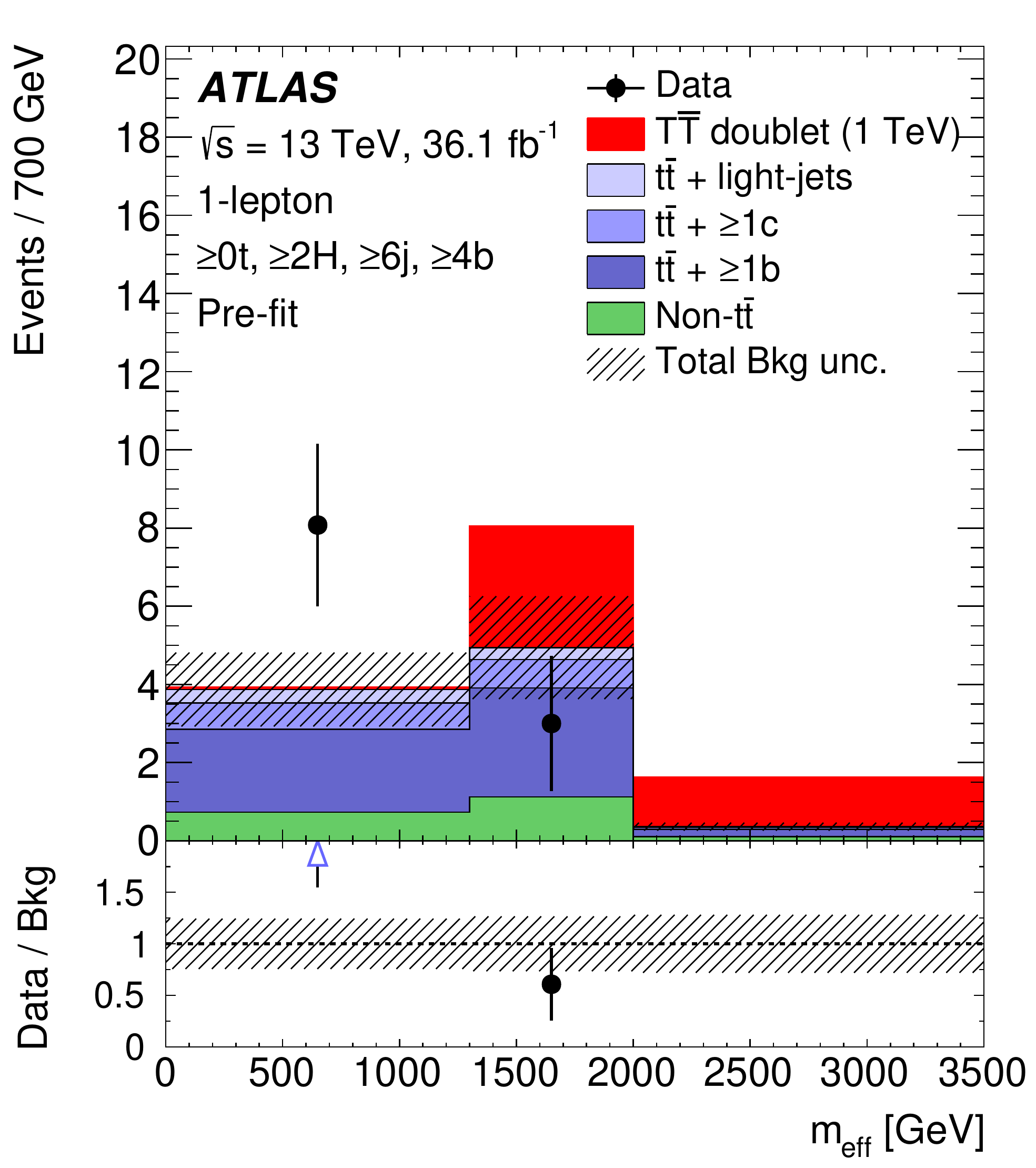}}
\subfloat[]{\includegraphics[width=0.40\textwidth]{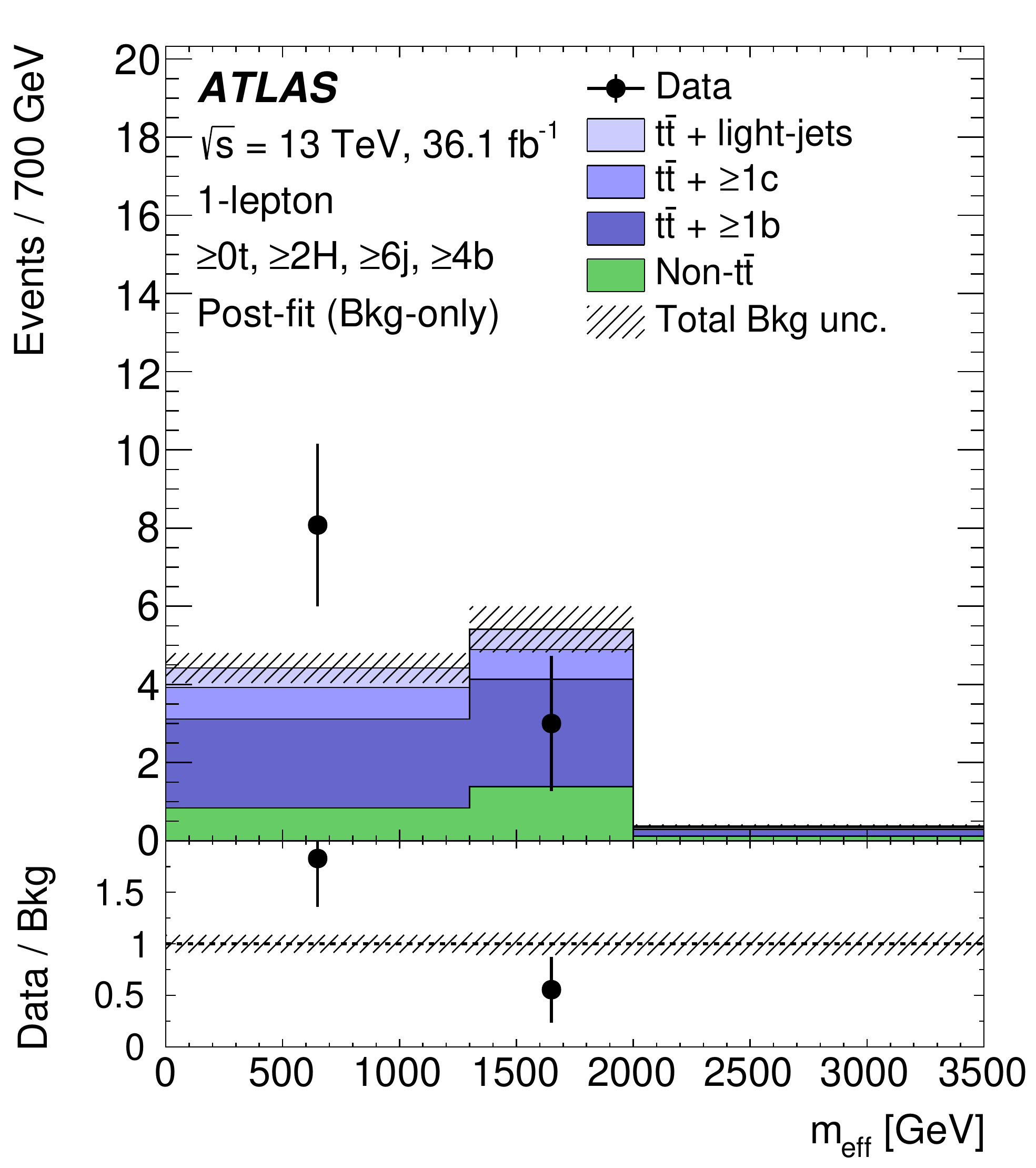}} \\
\subfloat[]{\includegraphics[width=0.40\textwidth]{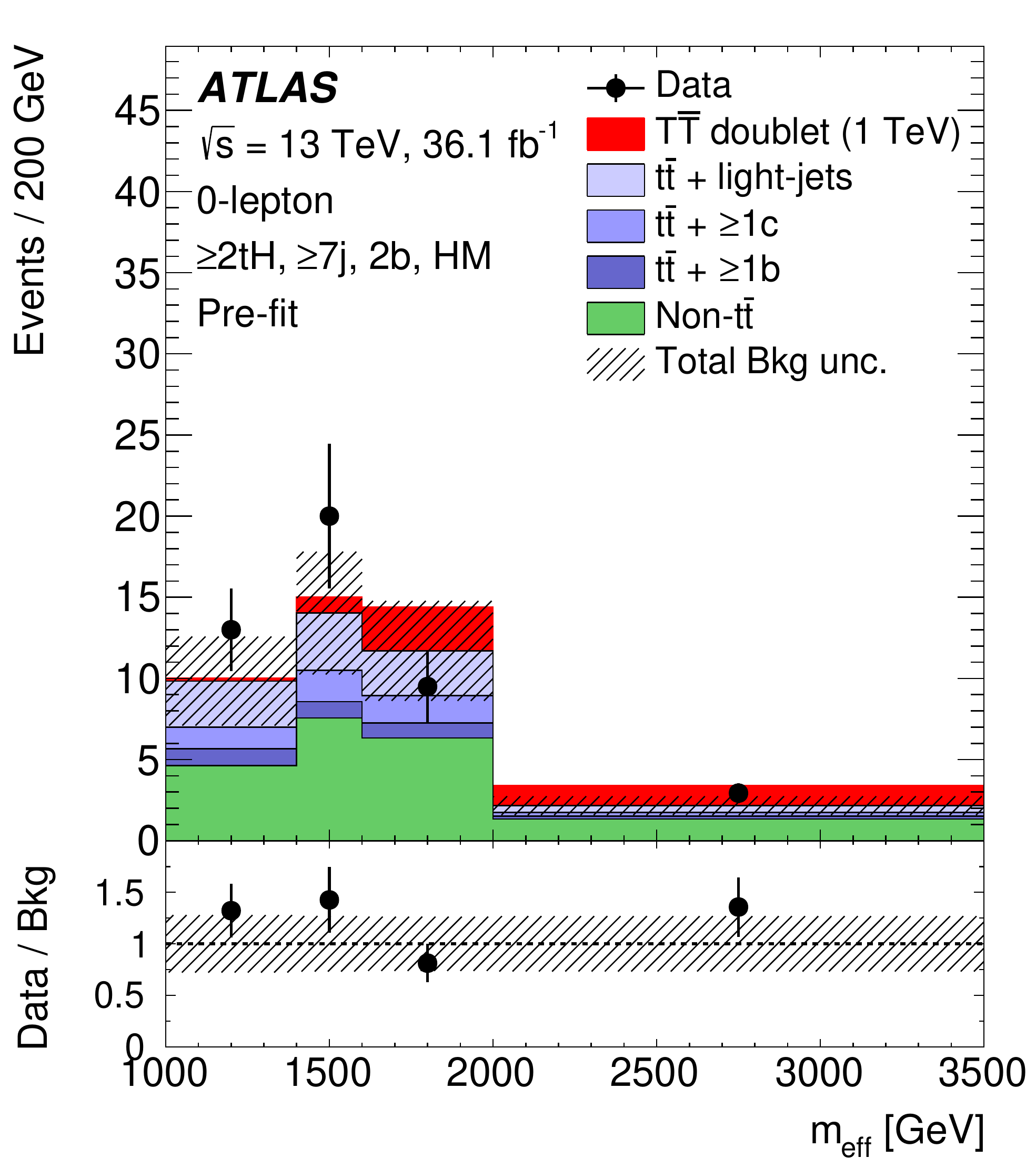}}
\subfloat[]{\includegraphics[width=0.40\textwidth]{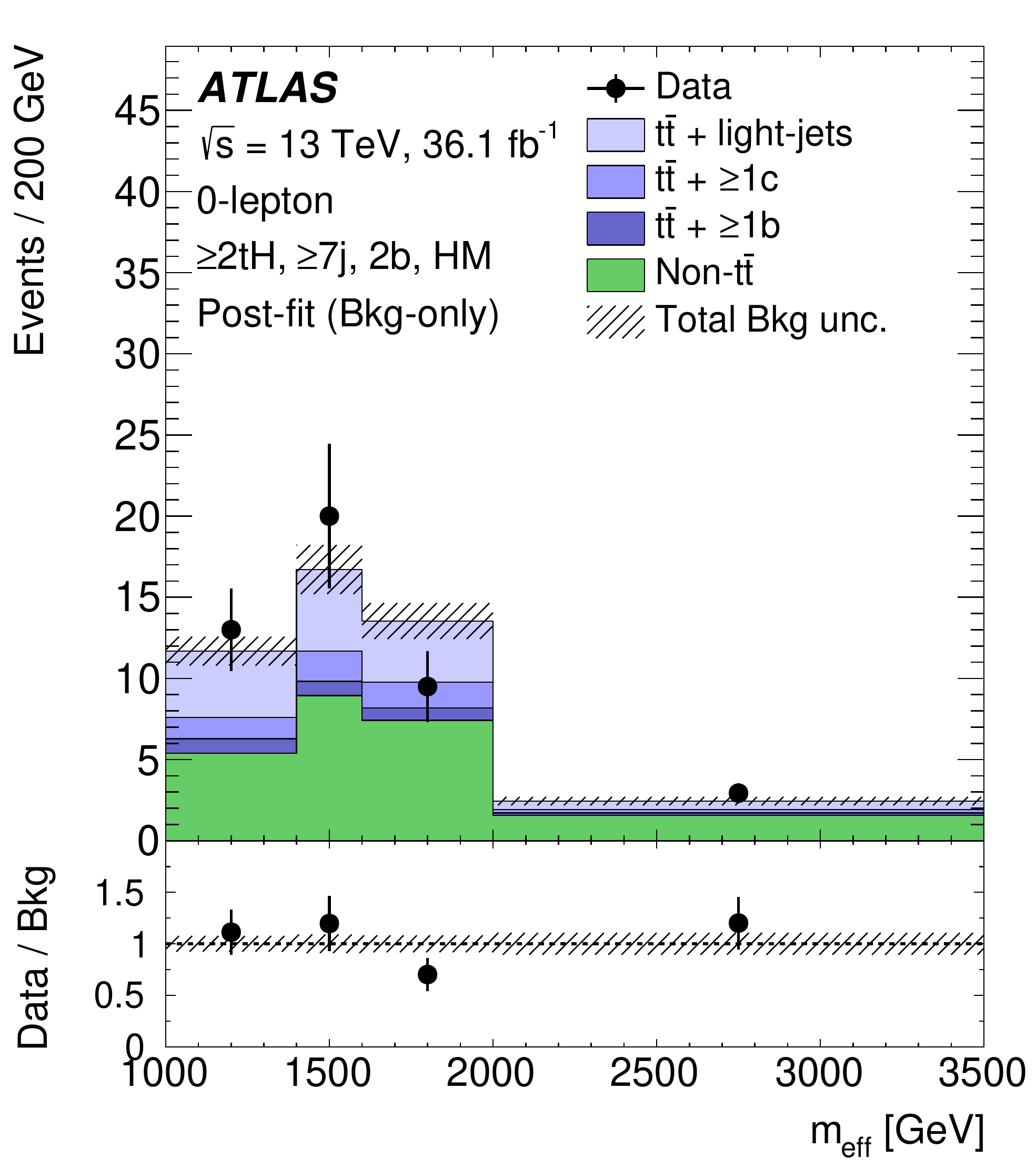}} \\
\caption{\small{Comparison between the data and prediction for the $\meff$ distribution in some of the most sensitive search regions,
before and after performing the combined fit to data in the 0-lepton and 1-lepton channels (``Pre-fit'' and ``Post-fit'', respectively) under the background-only hypothesis.
Shown are the ($\geq$2H, $\geq$6j, $\geq$4b) region in the 1-lepton channel (a) pre-fit and (b) post-fit,  and
the ($\geq$2tH, $\geq$7j, 2b, HM)  region in the 0-lepton channel (c) pre-fit and (d) post-fit.
In the pre-fit figures the expected $T\bar{T}$ signal (solid red) corresponding to $m_{T}=1~\tev$ in the $T$ doublet scenario is also shown,
added on top of the background prediction.
The small contributions from $\ttbar V$, $\ttbar H$, single top, $W/Z$+jets, diboson, and multijet backgrounds are combined into a single background source
referred to as ``Non-$\ttbar$''. The last bin in all figures contains the overflow.
The bottom panels display the ratios of data to the total background prediction (``Bkg'').
The blue triangles indicate points that are outside the vertical range of the figure.
The hashed area represents the total uncertainty of the background.
In the case of the pre-fit background uncertainty, the normalisation uncertainty of the $\ttbin$ background is not included.}}
\label{fig:prepostfit_unblinded_COMB_3}
\end{center}
\end{figure*}
 
\begin{figure*}[h!]
\begin{center}
\subfloat[]{\includegraphics[width=0.40\textwidth]{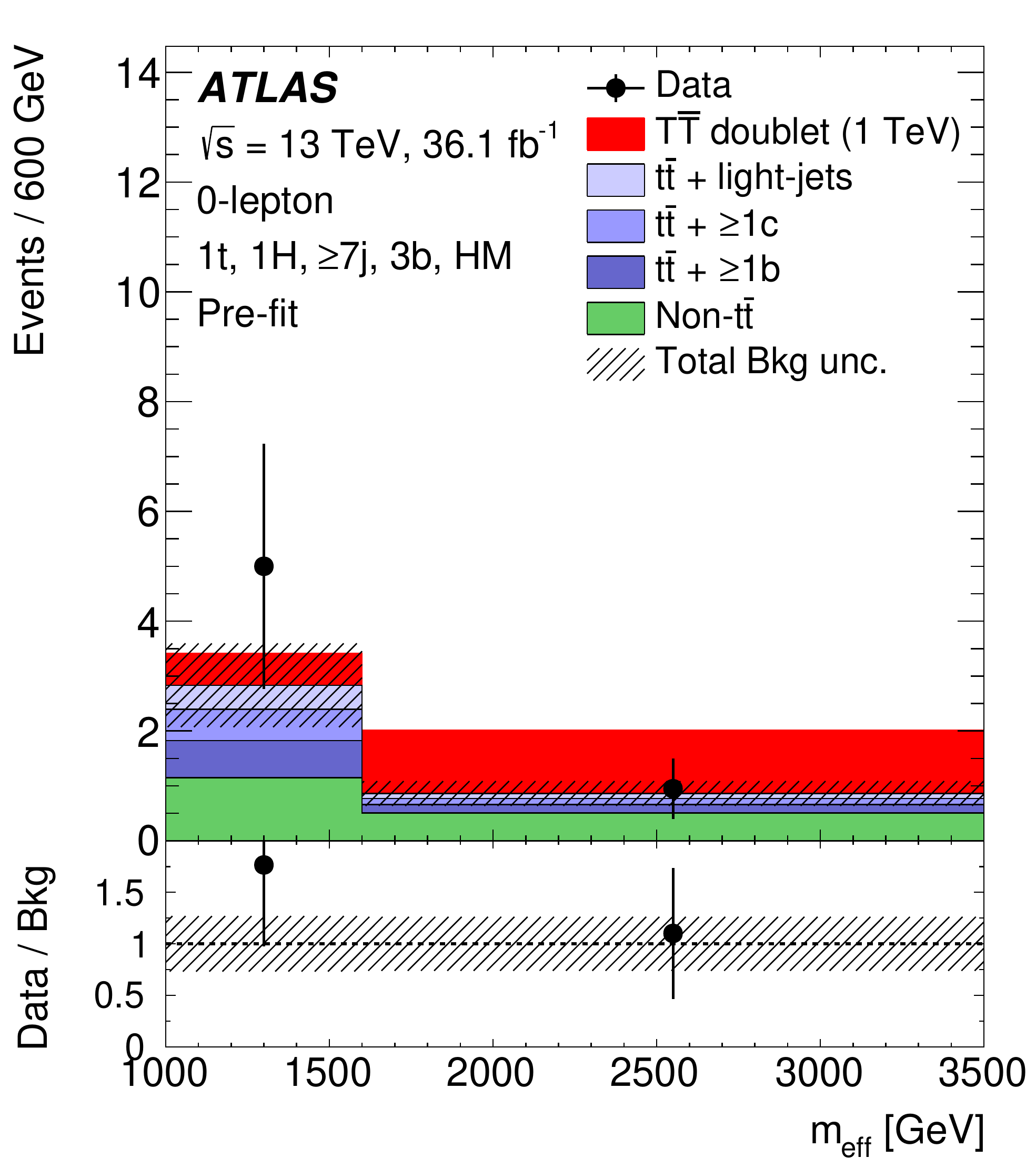}}
\subfloat[]{\includegraphics[width=0.40\textwidth]{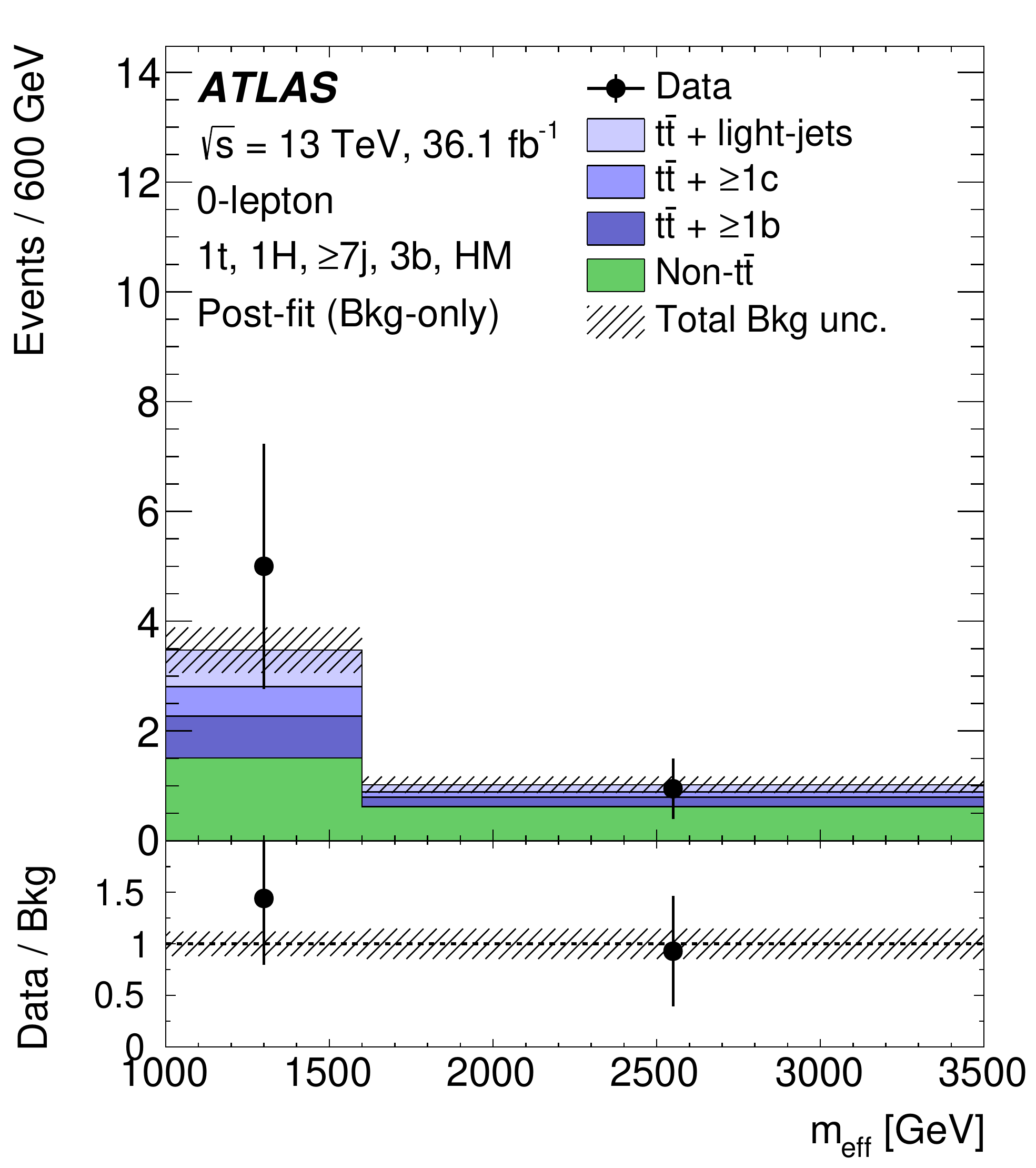}} \\
\subfloat[]{\includegraphics[width=0.40\textwidth]{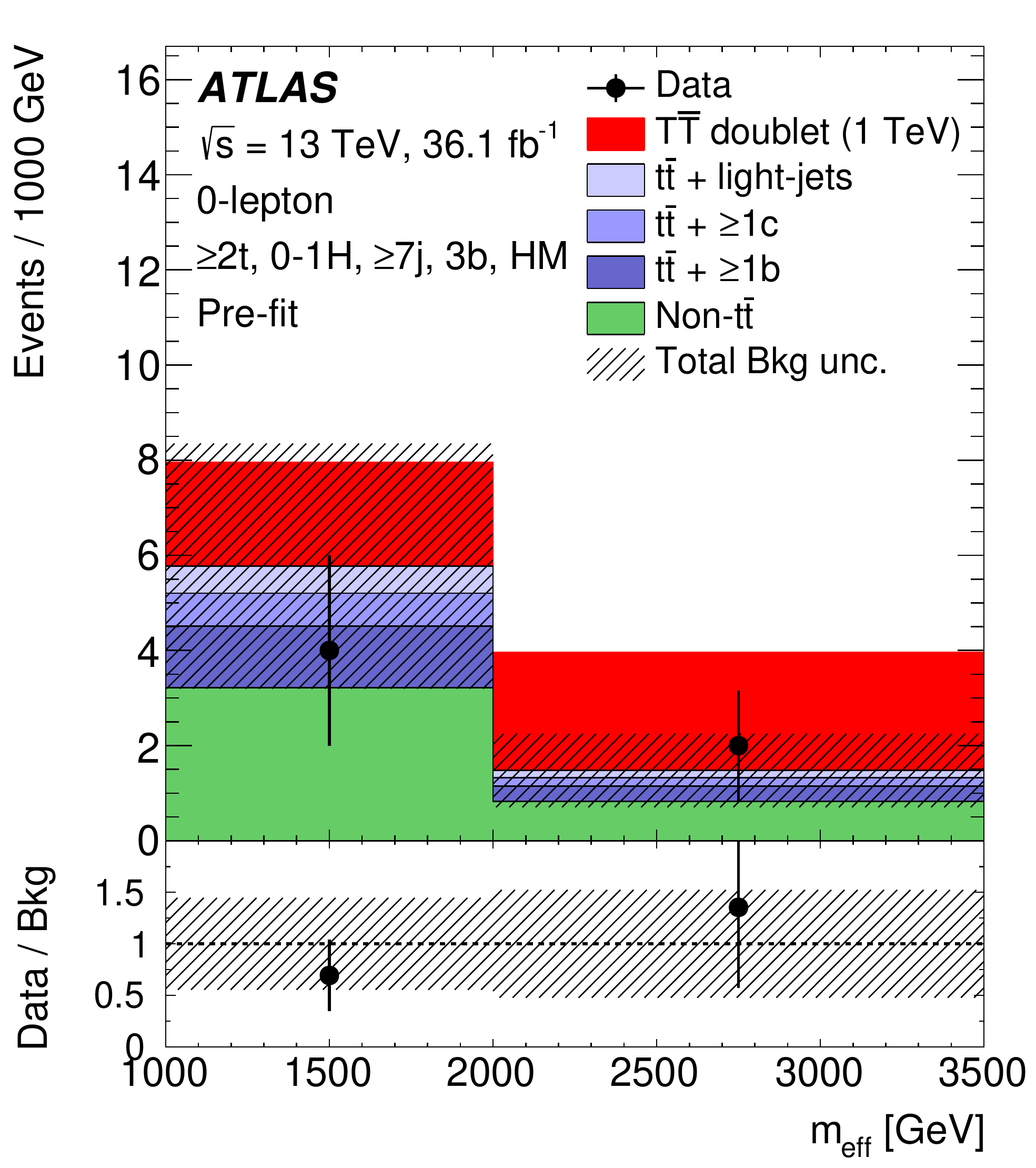}}
\subfloat[]{\includegraphics[width=0.40\textwidth]{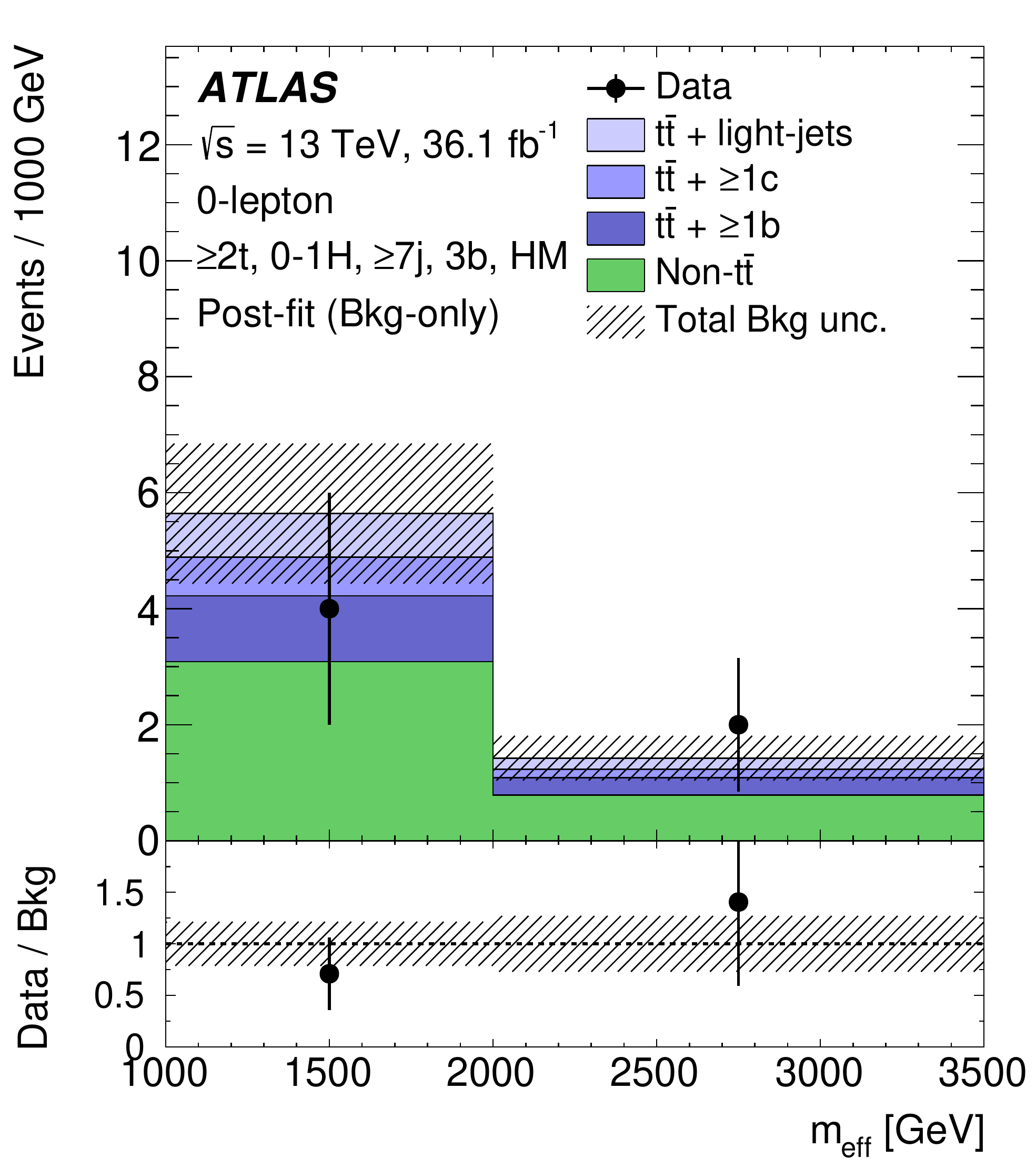}} \\
\caption{\small{Comparison between the data and prediction for the $\meff$ distribution in some of the most sensitive search regions in the 0-lepton channel,
before and after performing the combined fit to data in the 0-lepton and 1-lepton channels (``Pre-fit'' and ``Post-fit'', respectively) under the background-only hypothesis.
Shown are the (1t, 1H, $\geq$7j, 3b, HM) region (a) pre-fit and (b) post-fit,  and
the ($\geq$2t, 0--1H, $\geq$7j, 3b, HM) region (c) pre-fit and (d) post-fit.
In the pre-fit figures the expected $T\bar{T}$ signal (solid red) corresponding to $m_{T}=1~\tev$ in the $T$ doublet scenario is also shown,
added on top of the background prediction.
The small contributions from $\ttbar V$, $\ttbar H$, single-top, $W/Z$+jets, diboson, and multijet backgrounds are combined into a single background source
referred to as ``Non-$\ttbar$''. The last bin in all figures contains the overflow.
The bottom panels display the ratios of data to the total background prediction (``Bkg'').
The hashed area represents the total uncertainty of the background.
In the case of the pre-fit background uncertainty, the normalisation uncertainty of the $\ttbin$ background is not included.}}
\label{fig:prepostfit_unblinded_COMB_4}
\end{center}
\end{figure*}
 
\begin{figure*}[h!]
\begin{center}
\subfloat[]{\includegraphics[width=0.40\textwidth]{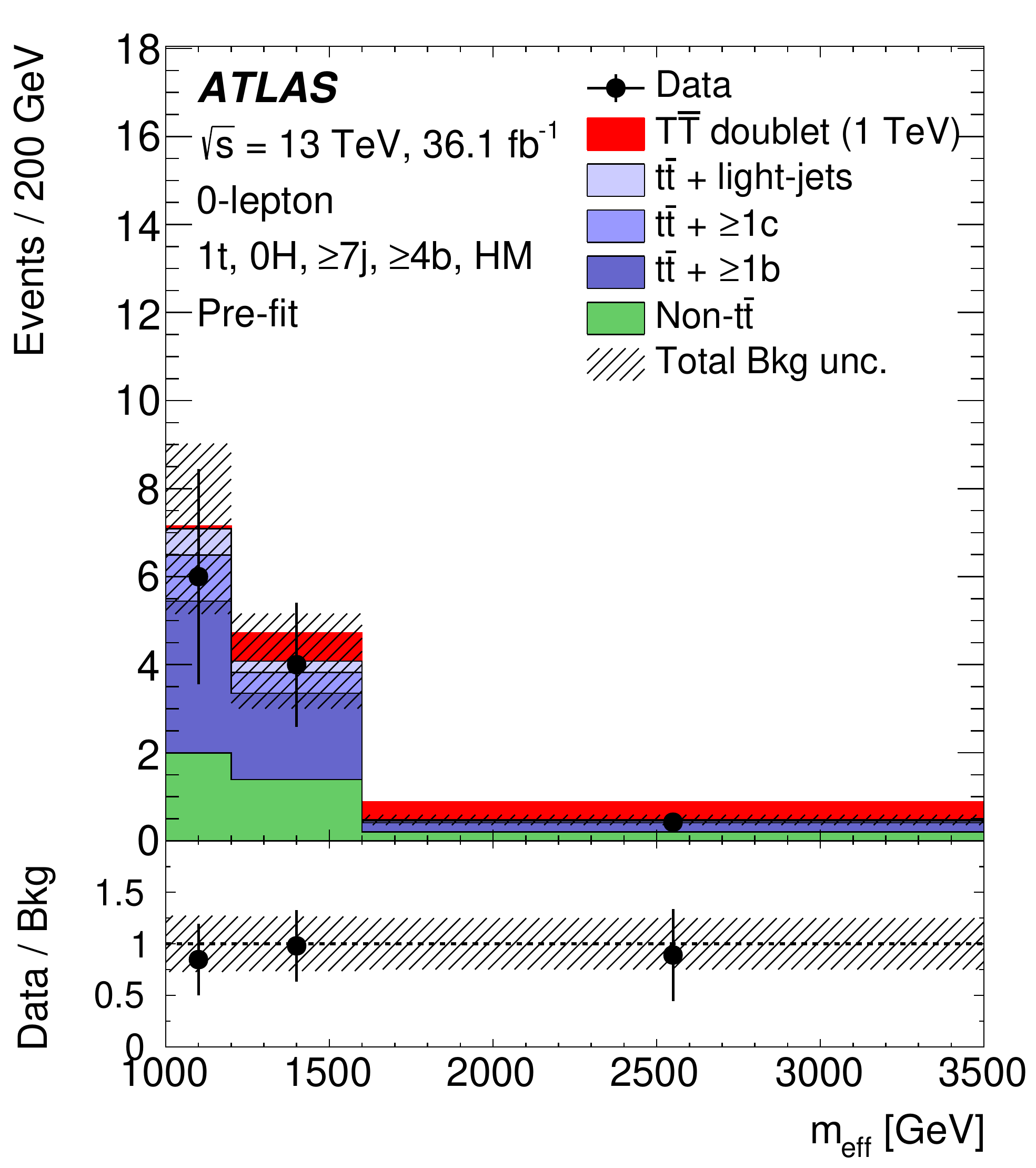}}
\subfloat[]{\includegraphics[width=0.40\textwidth]{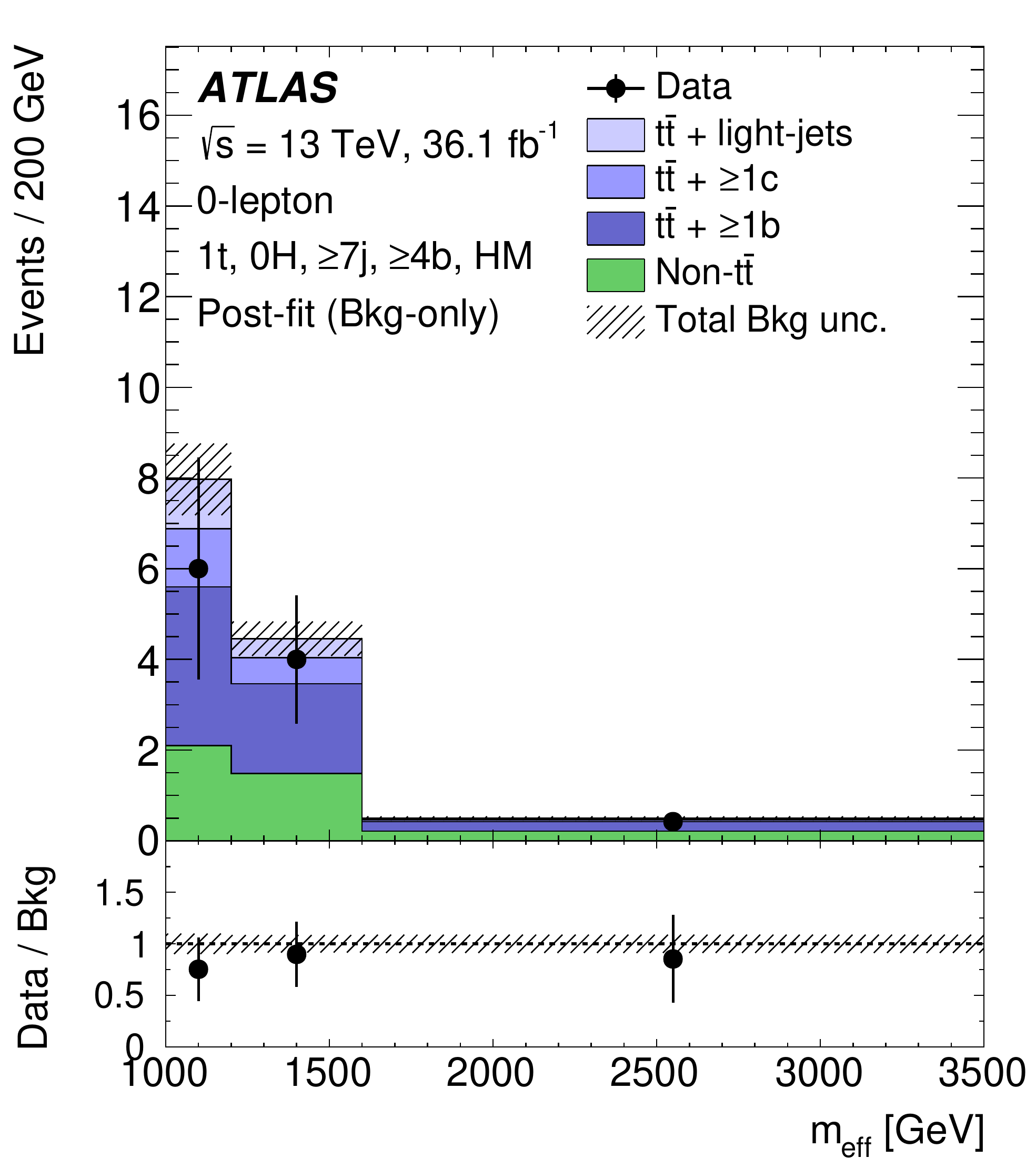}} \\
\subfloat[]{\includegraphics[width=0.40\textwidth]{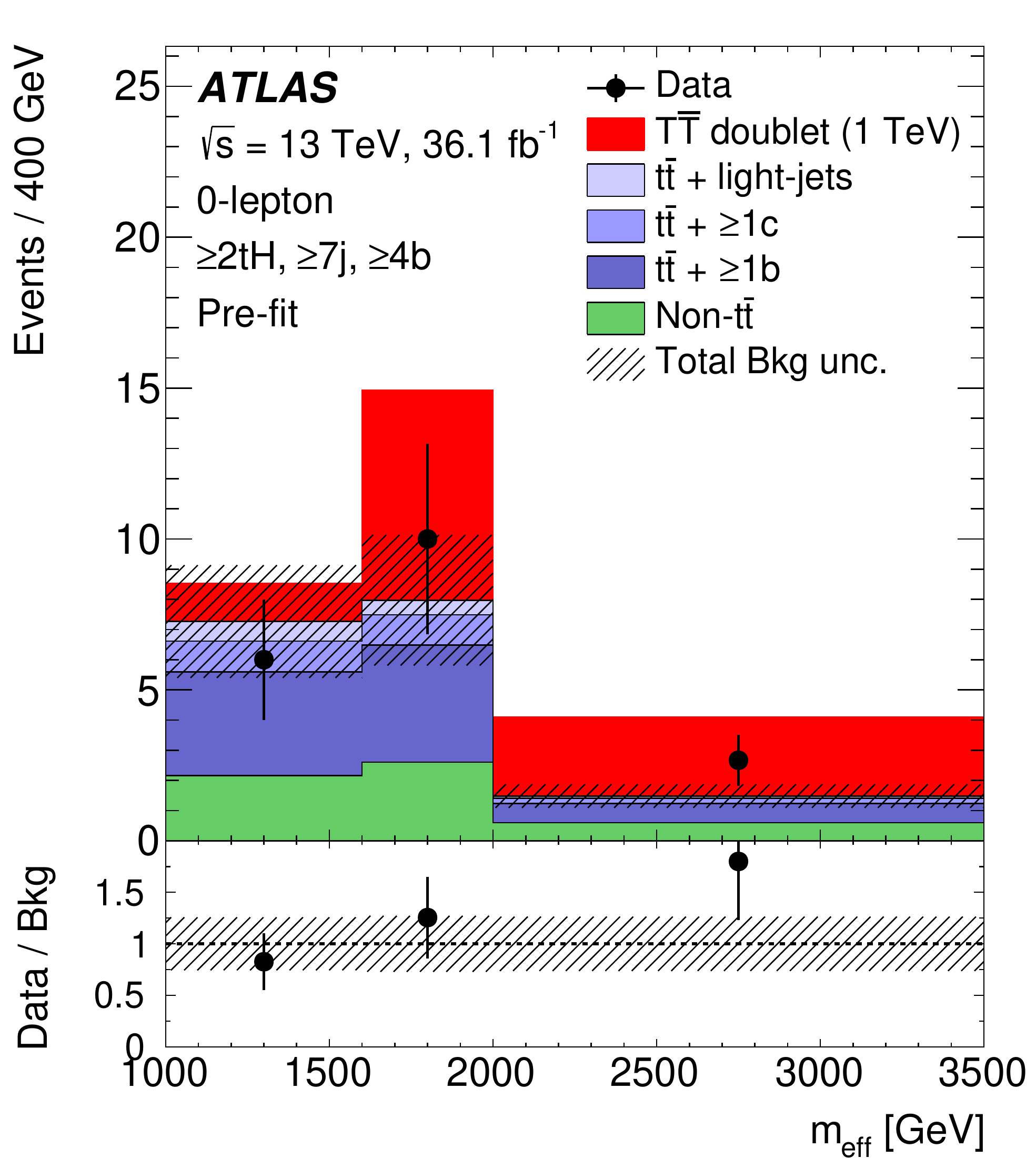}}
\subfloat[]{\includegraphics[width=0.40\textwidth]{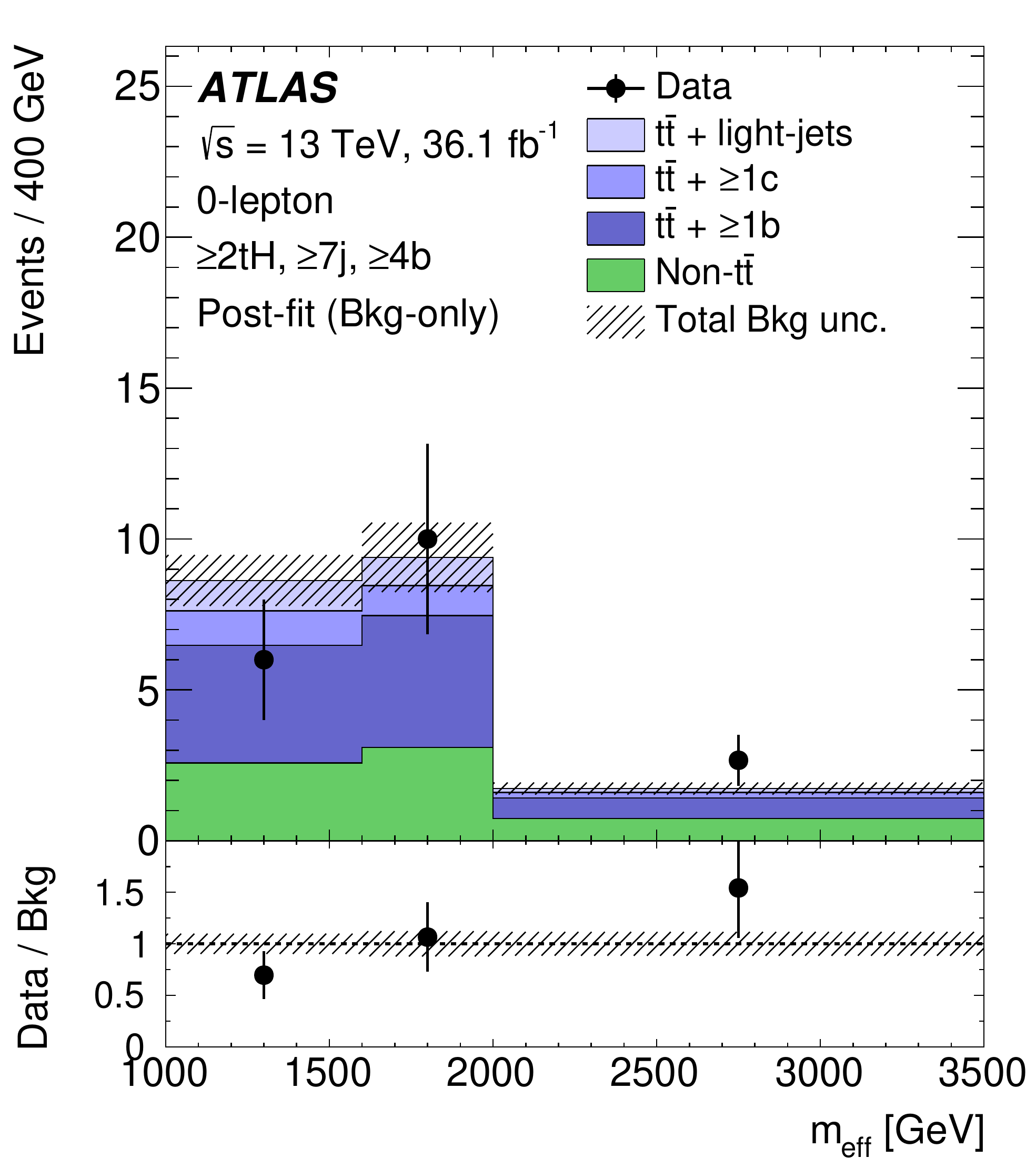}} \\
\caption{\small{Comparison between the data and prediction for the $\meff$ distribution in some of the most sensitive search regions in the 0-lepton channel,
before and after performing the combined fit to data in the 0-lepton and 1-lepton channels (``Pre-fit'' and ``Post-fit'', respectively) under the background-only hypothesis.
Shown are the (1t, 0H, $\geq$7j, $\geq$4b, HM) region (a) pre-fit and (b) post-fit,  and
the ($\geq$2tH, $\geq$7j, $\geq$4b) region (c) pre-fit and (d) post-fit.
In the pre-fit figures the expected $T\bar{T}$ signal (solid red) corresponding to $m_{T}=1~\tev$ in the $T$ doublet scenario is also shown,
added on top of the background prediction.
The small contributions from $\ttbar V$, $\ttbar H$, single-top, $W/Z$+jets, diboson, and multijet backgrounds are combined into a single background source
referred to as ``Non-$\ttbar$''. The last bin in all figures contains the overflow.
The bottom panels display the ratios of data to the total background prediction (``Bkg'').
The hashed area represents the total uncertainty of the background.
In the case of the pre-fit background uncertainty, the normalisation uncertainty of the $\ttbin$ background is not included.}}
\label{fig:prepostfit_unblinded_COMB_5}
\end{center}
\end{figure*}
 
\begin{figure*}[h!]
\begin{center}
\subfloat[]{\includegraphics[width=0.8\textwidth]{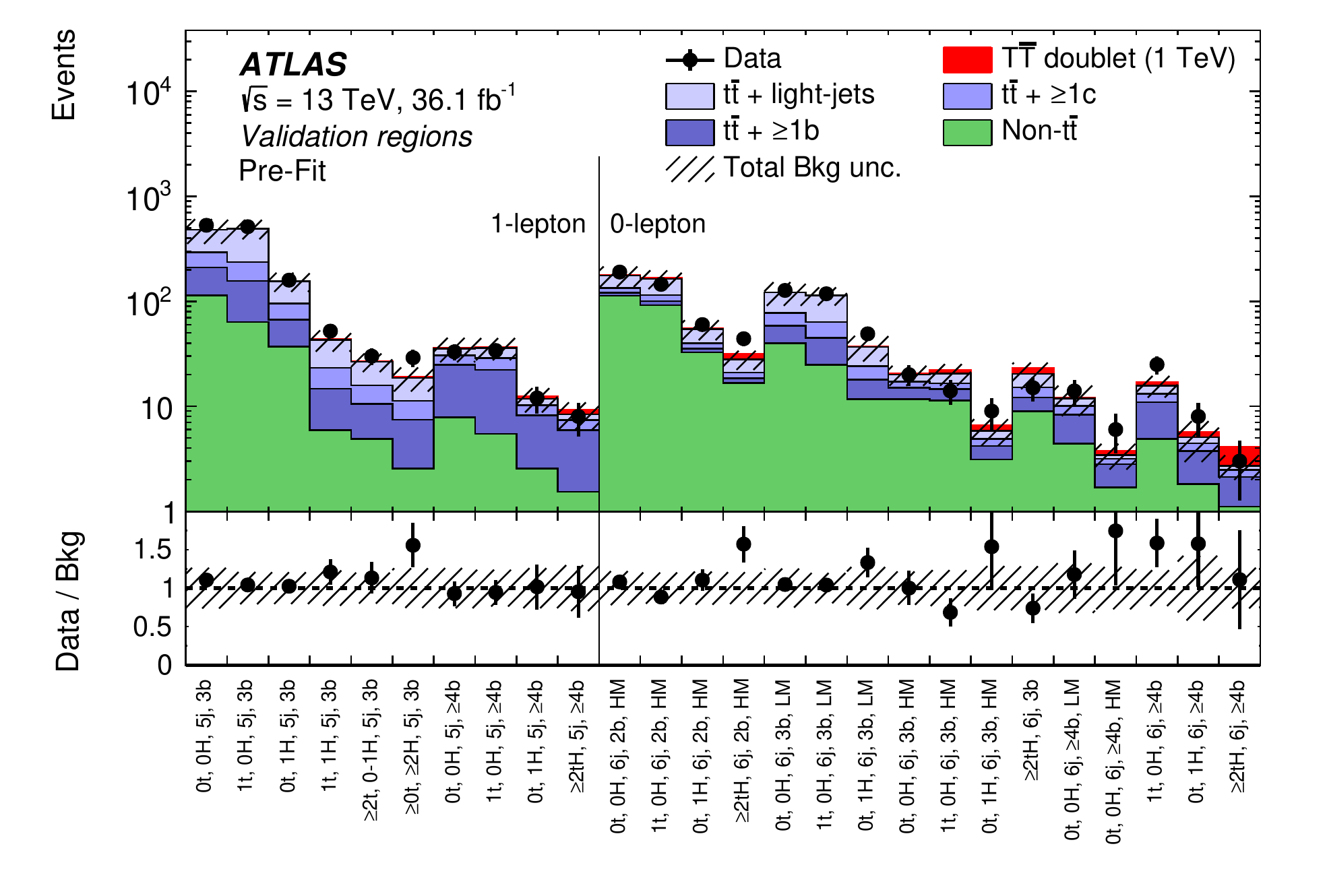}} \\
\subfloat[]{\includegraphics[width=0.8\textwidth]{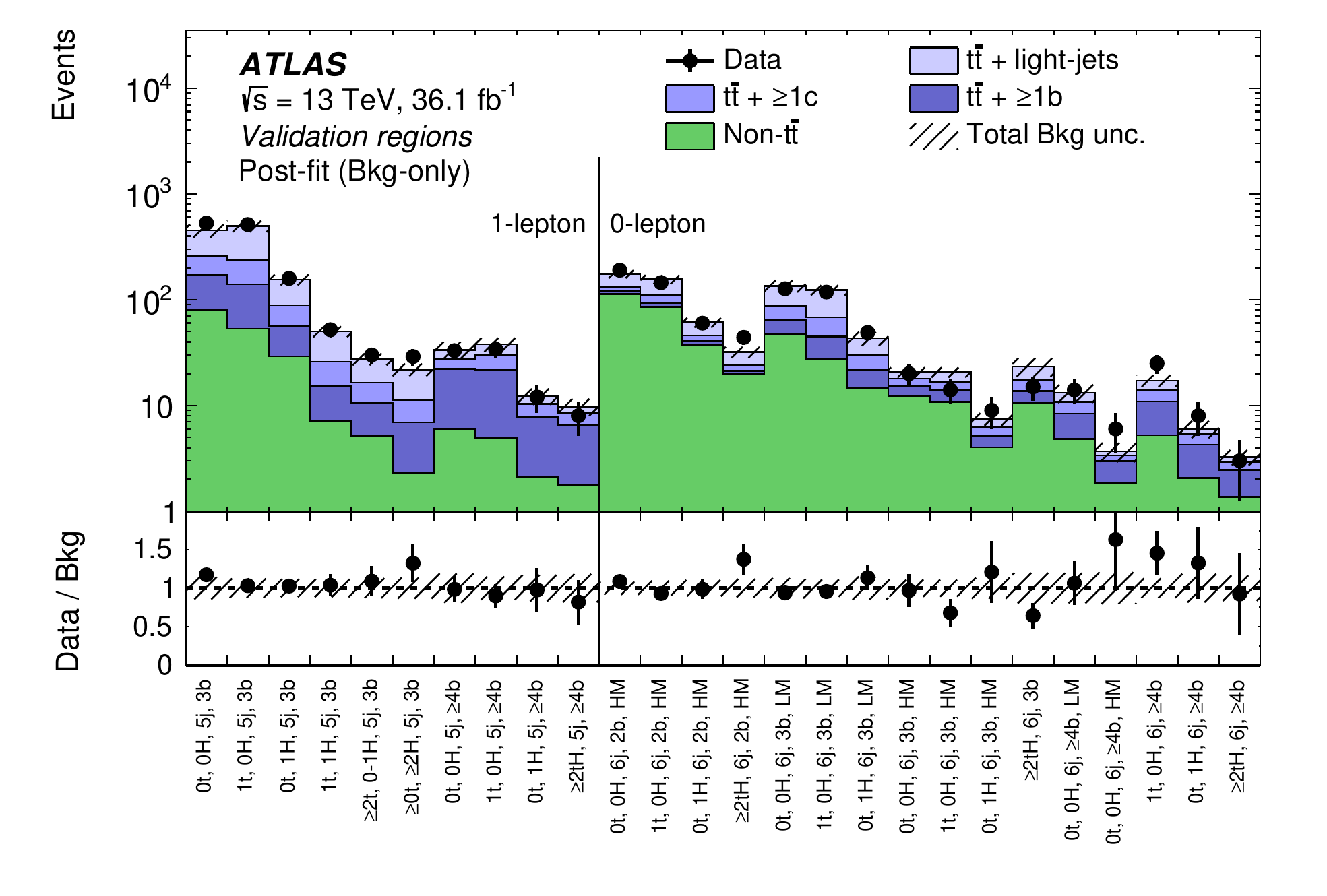}}
\caption{\small {Comparison between the data and background prediction for the yields in each of the validation regions considered in
the 1-lepton and 0-lepton channels (a) before the fit (``Pre-fit'') and (b) after the fit (``Post-fit'').
The fit is performed on the data in 1-lepton and 0-lepton channels under the background-only hypothesis considering only the search regions.
In the pre-fit figure the expected $T\bar{T}$ signal (solid red) corresponding to $m_{T}=1~\tev$ in the $T$ doublet scenario is also shown,
added on top of the background prediction.
The small contributions from $\ttbar V$, $\ttbar H$, single-top, $W/Z$+jets, diboson, and multijet backgrounds are combined into a single background source
referred to as ``Non-$\ttbar$''.
The bottom panels display the ratios of data to the total background prediction (``Bkg'').
The hashed area represents the total uncertainty of the background.
In the case of the pre-fit background uncertainty, the normalisation uncertainty of the $\ttbin$ background is not included.}}
\label{fig:Summary_VR_preFit_postFit_COMB}
\end{center}
\end{figure*}
 
\begin{figure*}[h!]
\begin{center}
\subfloat[]{\includegraphics[width=0.40\textwidth]{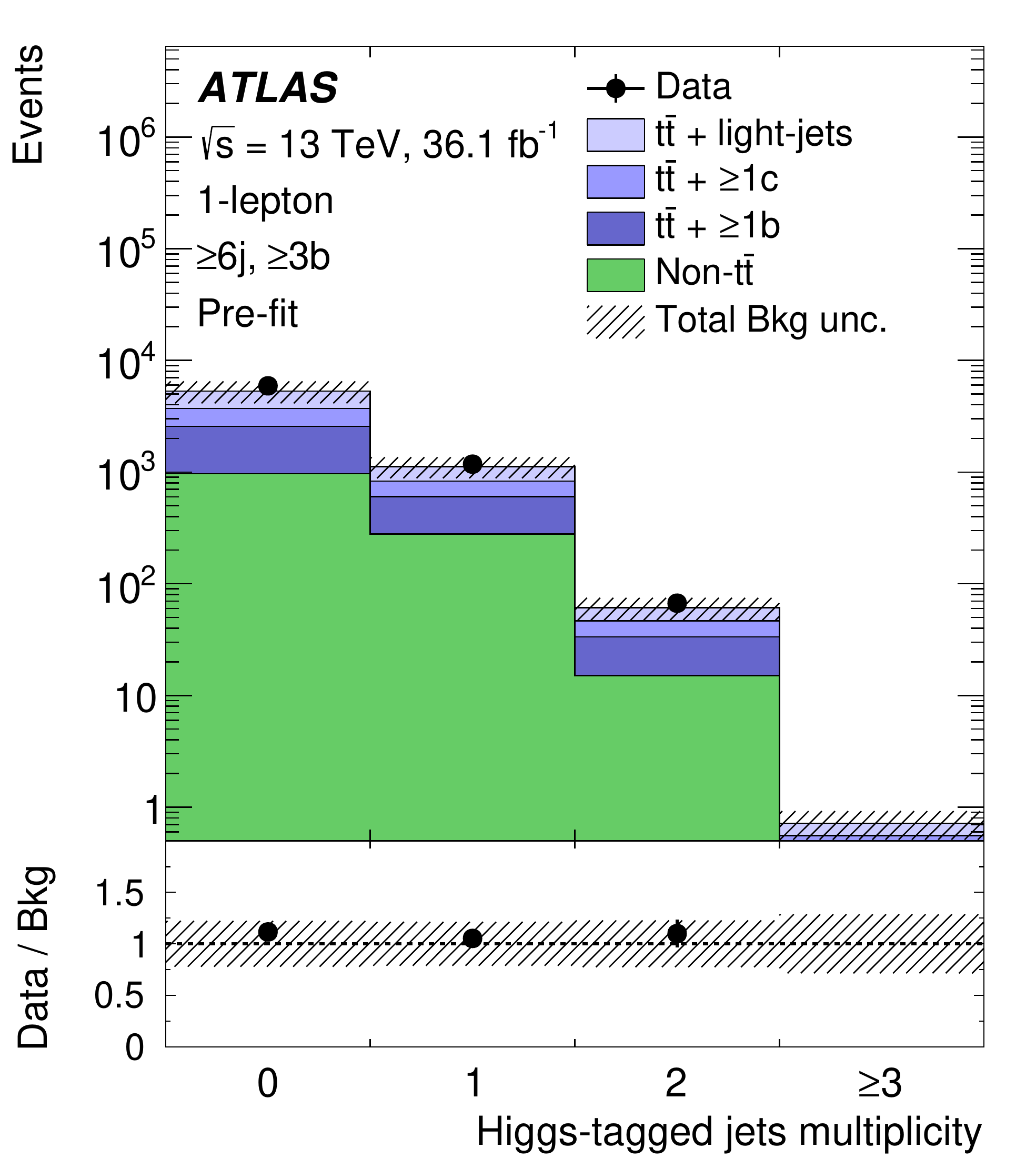}}
\subfloat[]{\includegraphics[width=0.40\textwidth]{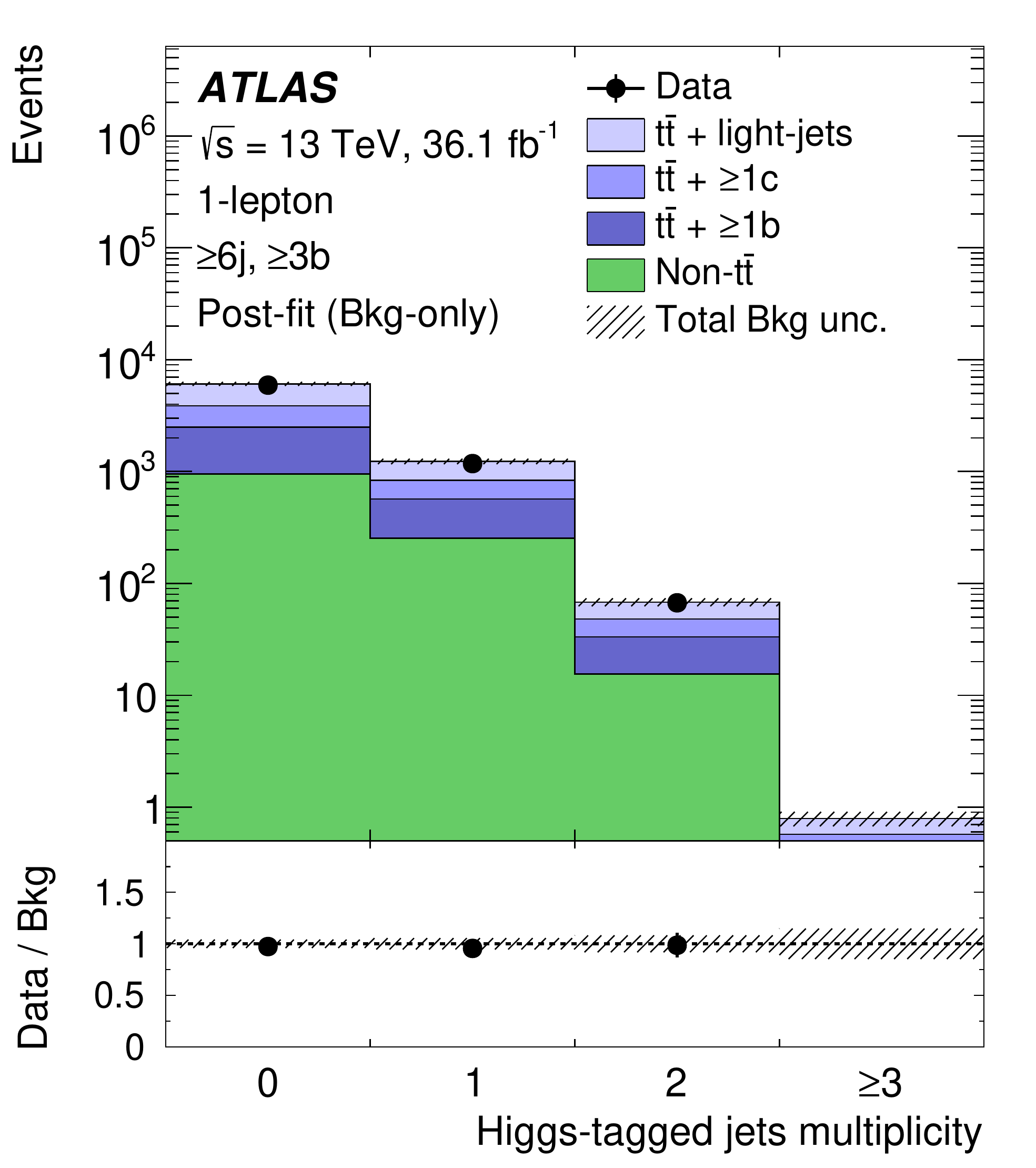}} \\
\caption{\small{Comparison between the data and prediction for the Higgs-tagged jet multiplicity
in the 1-lepton channel after preselection plus the requirement of $\geq$6 jets and $\geq$3 $b$-tagged jets,
(a) before and (b) after performing the combined fit of the $\meff$ spectrum to data in the 0-lepton and 1-lepton channels search regions (``Pre-fit'' and ``Post-fit'', respectively) under the background-only hypothesis.
The small contributions from $\ttbar V$, $\ttbar H$, single-top, $W/Z$+jets, diboson, and multijet backgrounds are combined into a single background source
referred to as ``Non-$\ttbar$''.
The last bin in all figures contains the overflow.
The bottom panels display the ratios of data to the total background prediction (``Bkg'').
The hashed area represents the total uncertainty of the background.
In the case of the pre-fit background uncertainty, the normalisation uncertainty of the $\ttbin$ background is not included.}}
\label{fig:prepostfit_RCMHiggs_jets_n}
\end{center}
\end{figure*}
\begin{figure*}[h!]
\begin{center}
\subfloat[]{\includegraphics[width=0.40\textwidth]{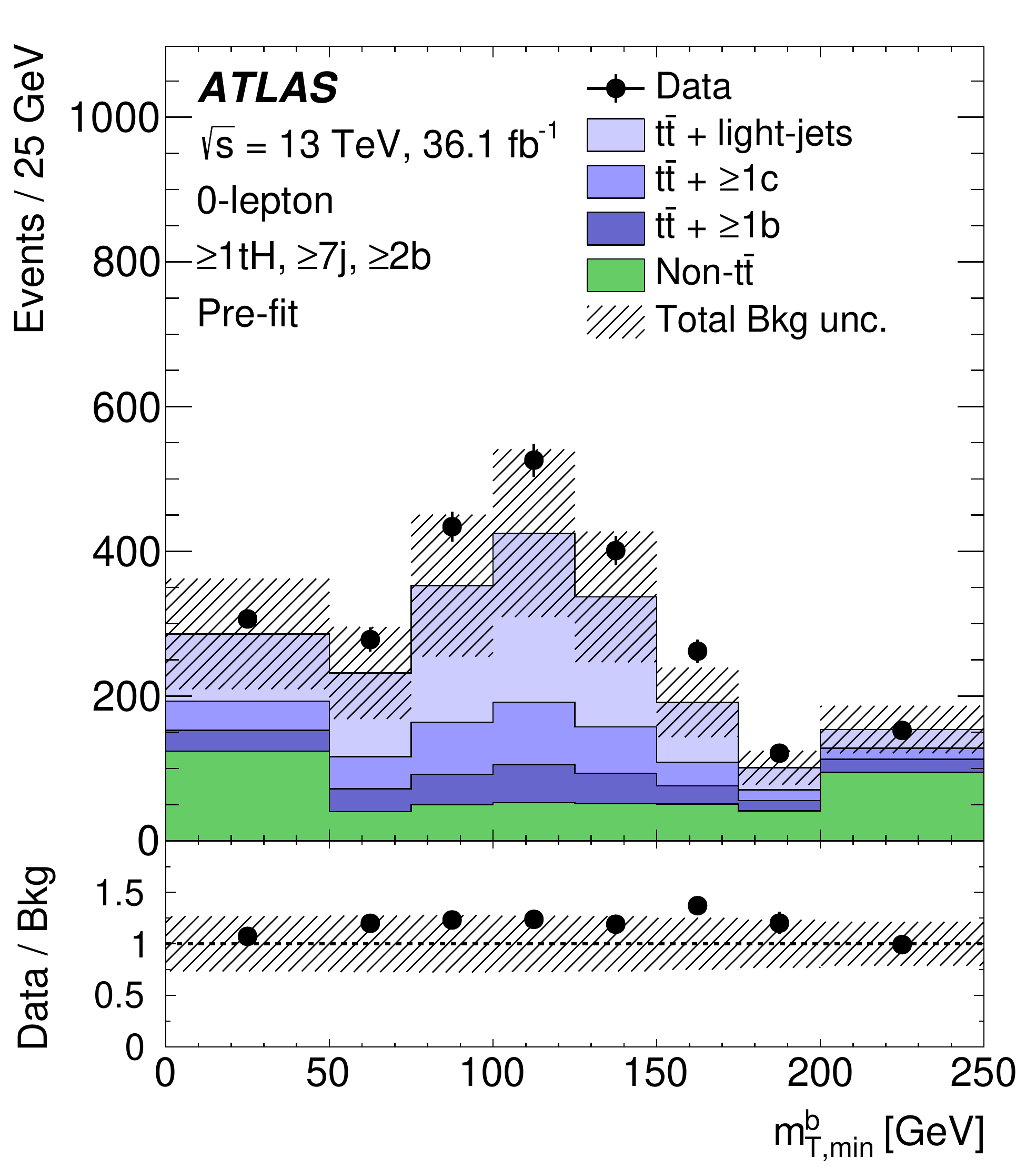}}
\subfloat[]{\includegraphics[width=0.40\textwidth]{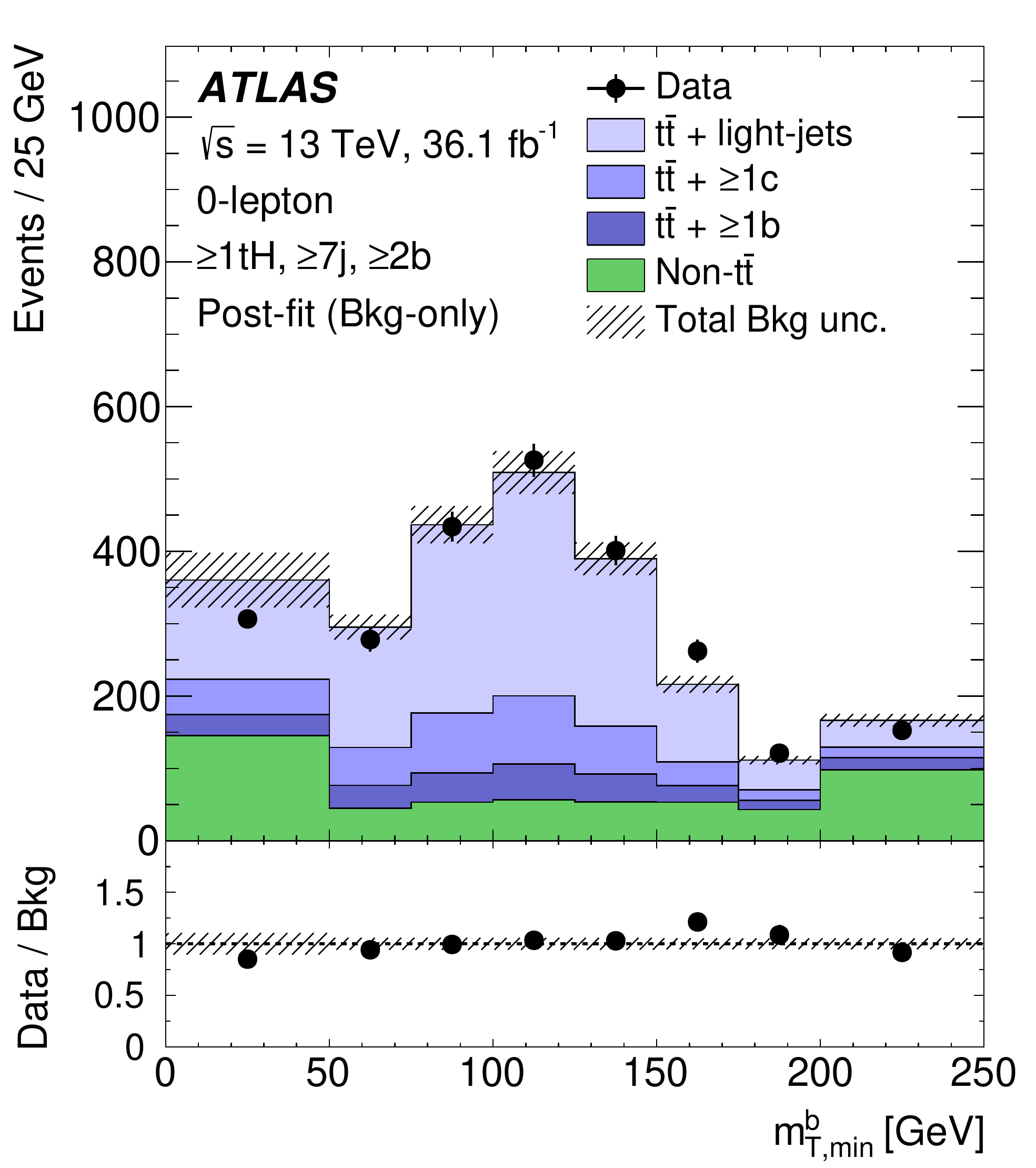}} \\
\caption{\small{Comparison between the data and prediction for the distribution of the minimum transverse mass of $\met$ and
any of the three leading $b$-tagged jets in the event ($\mtbmin$) in the ($\geq$1tH, $\geq$7j, $\geq$2b) region of the 0-lepton channel
(a) before and (b) after performing the combined fit of the $\meff$ spectrum to data in the 0-lepton and 1-lepton channels search regions (``Pre-fit'' and ``Post-fit'', respectively) under the background-only hypothesis.
The small contributions from $\ttbar V$, $\ttbar H$, single-top, $W/Z$+jets, diboson, and multijet backgrounds are combined into a single background source
referred to as ``Non-$\ttbar$''.
The last bin in all figures contains the overflow.
The bottom panels display the ratios of data to the total background prediction (``Bkg'').
The hashed area represents the total uncertainty of the background.
In the case of the pre-fit background uncertainty, the normalisation uncertainty of the $\ttbin$ background is not included.}}
\label{fig:prepostfit_mtbmin}
\end{center}
\end{figure*}
 
\clearpage
\subsection{Limits on vector-like quark pair production}
 
No significant excess above the SM expectation is found in any of the search regions.
Upper limits at 95\% CL on the $T\bar{T}$ production cross section are set in several benchmark scenarios as a function of the
$T$ quark mass $m_{T}$ and are compared to the theoretical prediction from {\textsc Top++}.
The resulting lower limits on $m_{T}$ correspond to the central value of the theoretical cross section.
The scenarios considered involve different assumptions about the decay branching ratios.
The search in the 1-lepton (0-lepton) channel is particularly sensitive to the benchmark scenario
of ${\mathcal{B}}(T \to Ht)=1$ (${\mathcal{B}}(T \to Zt)=1$).
In contrast, both the 1-lepton and the 0-lepton searches have comparable sensitivity to the weak-isospin doublet and singlet scenarios,
and thus their combination represents an improvement of 60--70~$\gev$ on the expected $T$ quark mass exclusion
over the most sensitive individual search.
The limits corresponding to the weak-isospin doublet and singlet scenarios obtained for the combination of the 1-lepton and 0-lepton searches
are shown in Figure~\ref{fig:limits1D_TT_COMB2}.
A summary of the observed and expected lower limits on the $T$ quark mass in the different benchmark scenarios for
the individual 1-lepton and 0-lepton searches, as well as their combination, is given in Table~\ref{tab:masslimits}.
As can be seen, the observed mass limits for the 1-lepton search are above the expected limits in all benchmark scenarios.
Detailed studies on the statistical model found no sources of systematic bias and showed that the results are consistent with downward statistical fluctuations in data in some of the highest $\meff$ bins in three search regions:
(1t, 1H, $\geq$6j, $\geq$4b), ($\geq$2t, 0--1H, $\geq$6j, 3b), and ($\geq$0t, $\geq$2H, $\geq$6j, $\geq$4b). Several other regions with similar event kinematics and background composition to these three search regions show good agreement between data and expectations. In particular, additional regions with larger event yields were constructed to test this agreement by merging signal regions in certain categories, but retaining similar multiplicities of b-tagged jets or boosted objects as the original signal regions.
 
Table~\ref{tab:masslimits} also includes a comparison to the limits obtained by the ATLAS Run-1 $T\bar{T}\to Ht$+X search in
the 1-lepton channel~\cite{Aad:2015kqa}: the current results extend the expected $T$ quark mass exclusion by $\sim$390--490~$\gev$, depending on the assumed benchmark scenario.

\begin{figure*}[tbp]
\centering
\subfloat[]{\includegraphics[width=0.48\textwidth]{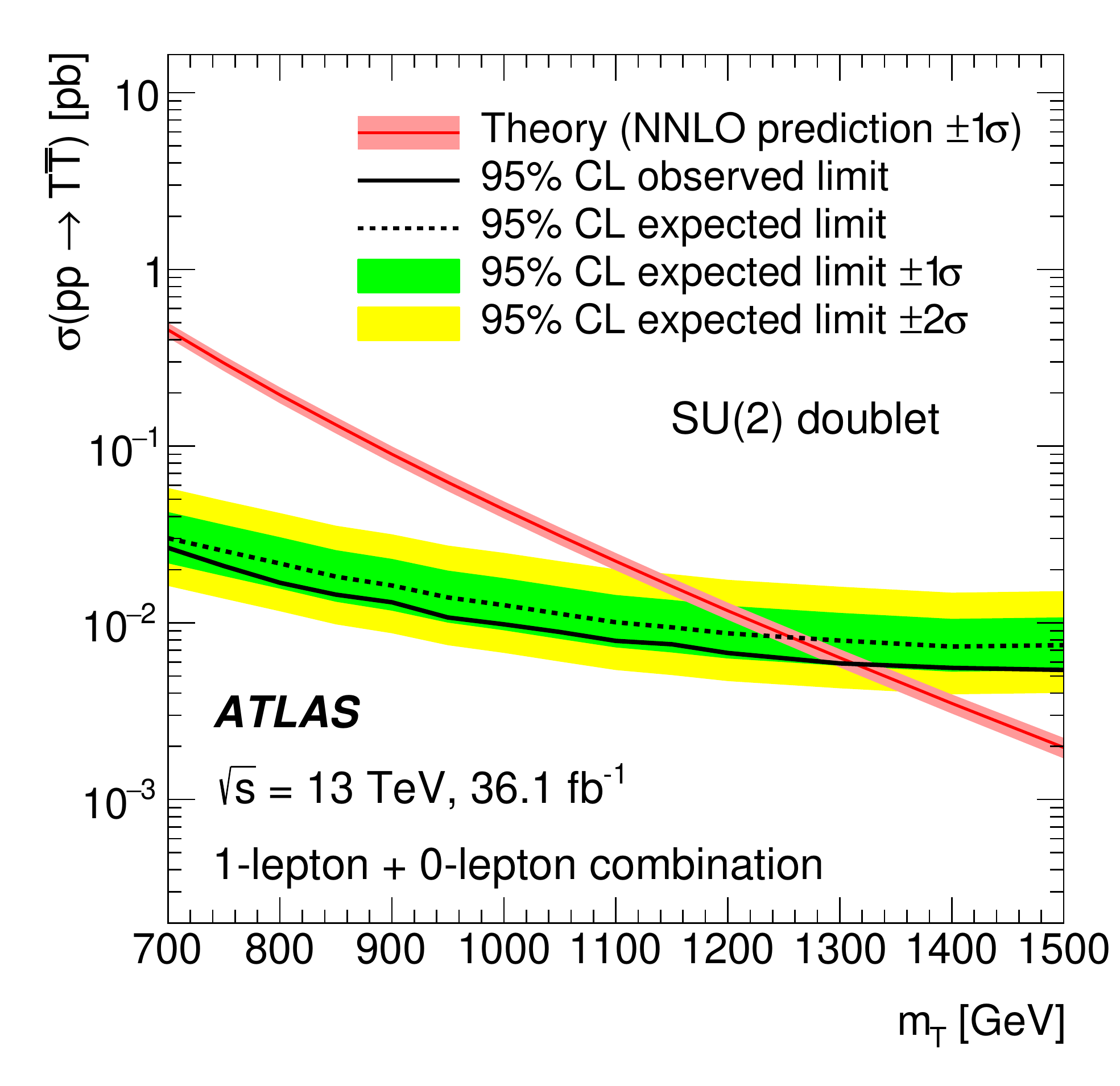}}
\subfloat[]{\includegraphics[width=0.48\textwidth]{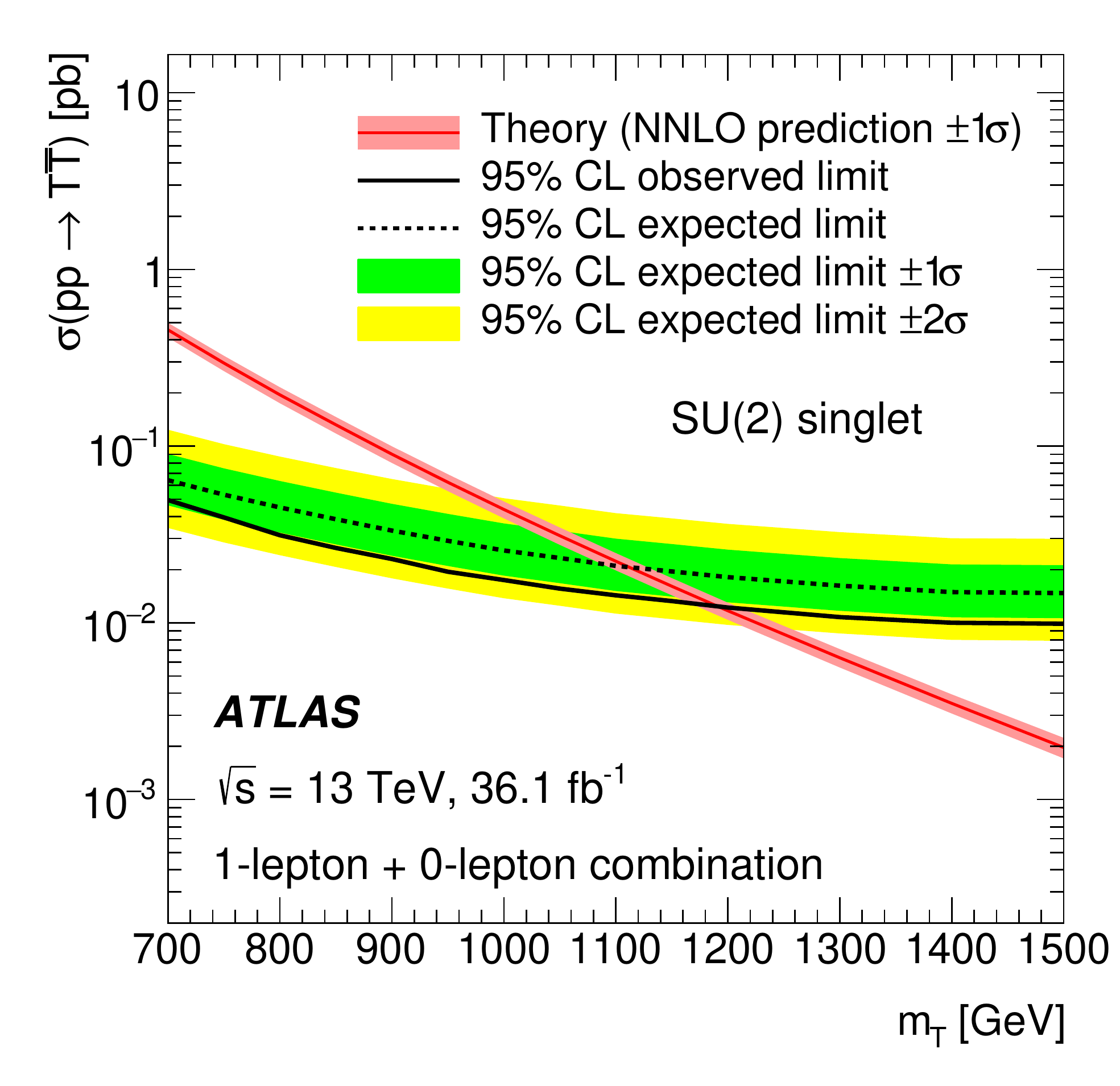}}
\caption{Observed (solid line) and expected (dashed line) 95\% CL upper limits on the $T\bar{T}$ cross section as a function of the $T$ quark mass
for the combination of the 1-lepton and 0-lepton searches (a) for a $T$ quark doublet, and (b) for a $T$ quark singlet. The background estimate used
in the computation of the limits is the result obtained from the background-only fit to data.
The surrounding shaded bands correspond to $\pm 1$ and $\pm 2$ standard deviations around the expected limit.
The thin red line and band show the theoretical prediction and its $\pm 1$ standard deviation uncertainty.}
\label{fig:limits1D_TT_COMB2}
\end{figure*}
 
\begin{table}[h!]
\begin{center}
\begin{tabular}{lccccc}
\toprule\toprule
\multicolumn{5}{c}{95\% CL lower limits on $T$ quark mass [\tev]} \\
\midrule
Search & ${\mathcal{B}}(T \to Ht)=1$ &  ${\mathcal{B}}(T \to Zt)=1$ & Doublet & Singlet \\
\midrule
1-lepton channel &  $1.47\,(1.30)$ & $1.12\,(0.91)$      & $1.36\,(1.16)$ & $1.23\,(1.02)$ \\
0-lepton channel &  $1.11\,(1.20)$   & $1.12\,(1.17)$ & $1.12\,(1.19)$ & $0.99\,(1.05)$ \\
\midrule
\textbf{Combination} & \textbf{$1.43\,(1.34)$}  & \textbf{$1.17\,(1.18)$} & \textbf{$1.31\,(1.26)$} & \textbf{$1.19\,(1.11)$} \\
\bottomrule
\\
\toprule
\toprule
\multicolumn{5}{l}{Previous Run-1 ATLAS $T\bar{T}\to Ht$+X search~\cite{Aad:2015kqa}}  \\
\midrule
1-lepton channel & $0.95\,(0.88)$ & $0.75\,(0.69)$ & $0.86\,(0.82)$ & $0.76\,(0.72)$  \\
\bottomrule\bottomrule
\end{tabular}
\caption{\small{Summary of observed (expected) 95\% CL lower limits on $T$ quark mass (in $\tev$) for the 1-lepton and 0-lepton channels, as well as their combination,
with different assumptions about the decay branching ratios. The background estimate used
in the computation of the limits is the result obtained from the background-only fit to data. Also shown are the corresponding limits obtained by the Run-1 ATLAS
$T\bar{T}\to Ht$+X search in the 1-lepton channel~\cite{Aad:2015kqa}.}}
\label{tab:masslimits}
\end{center}
\end{table}
 
The same analyses are used to derive exclusion limits on vector-like $T$ quark production, for different
values of $m_{T}$ and as a function of ${\mathcal{B}}(T\to W b)$ and ${\mathcal{B}}(T\to Ht)$, assuming that
${\mathcal{B}}(T\to Wb)+{\mathcal{B}}(T\to Zt)+{\mathcal{B}}(T\to Ht)=1$.
To probe this branching ratio plane, the signal samples are reweighted by the ratio
of the desired branching ratio to the original branching ratio in \textsc{Protos}, and the complete analysis is repeated.
Owing to the complementarity of the 1-lepton and 0-lepton searches in probing the branching ratio plane,
their combination represents a significant improvement over the individual results, as illustrated in Figure~\ref{fig:limits2D_COMB}.
In this case, the observed lower limits on the $T$ quark mass range between $0.99~\tev$ and $1.43~\tev$
depending on the values of the branching ratios into the three decay modes.
In particular, a vector-like $T$ quark with mass below $0.99~\tev$ is excluded for any values of the branching ratios into the three decay modes.
The corresponding range of expected lower limits is between $0.91~\tev$ and $1.34~\tev$.
Figure~\ref{fig:limits2D_temp_1L0L}  presents the corresponding observed and expected $T$ quark mass limits
in the plane of ${\mathcal{B}}(T \to Ht)$ versus ${\mathcal{B}}(T \to Wb)$, obtained by linear interpolation
of the calculated CL$_{\textrm{s}}$ versus $m_{T}$.

\begin{figure*}[tbp]
\centering
\includegraphics[width=0.9\textwidth]{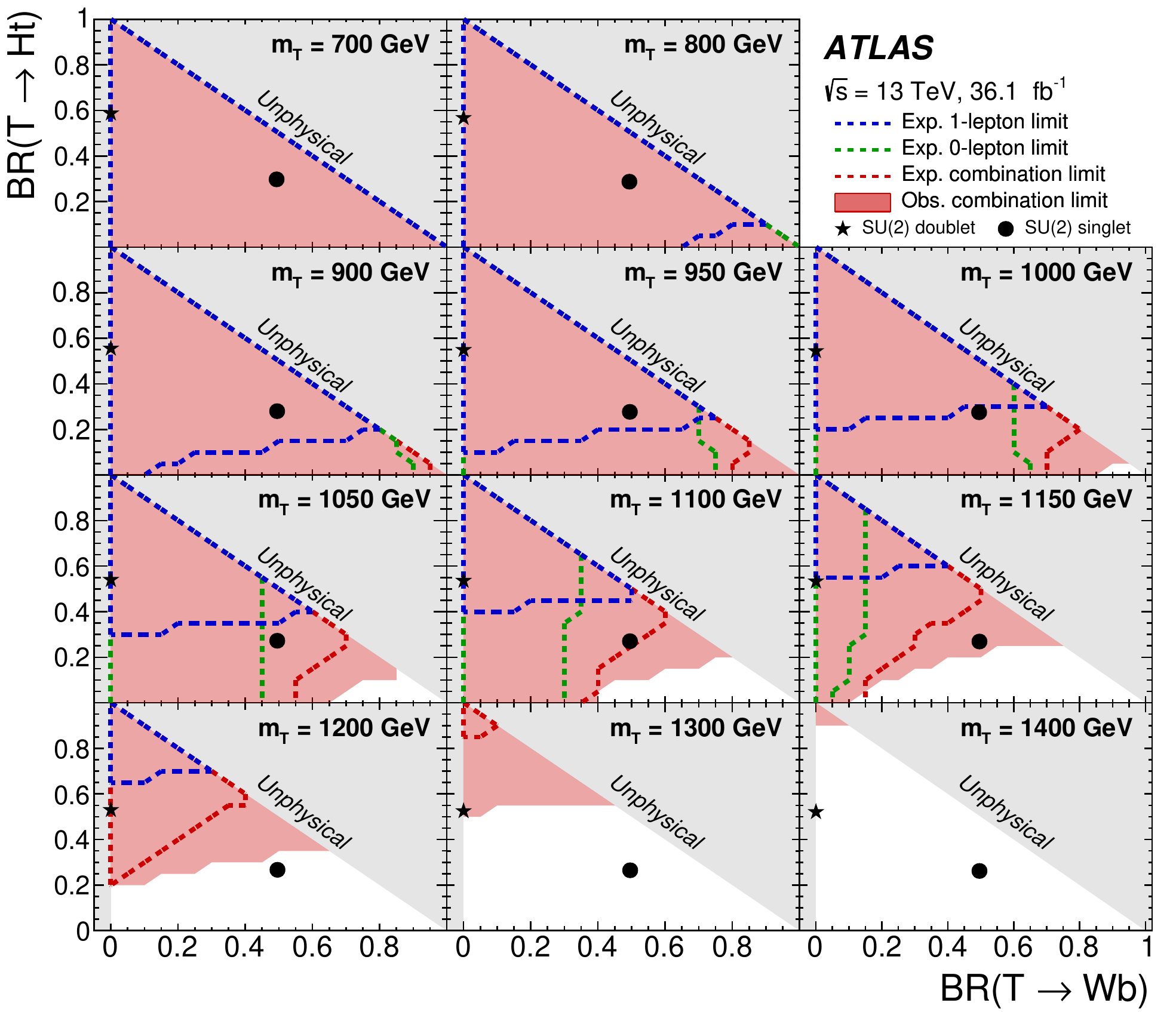}
\caption{
Observed (red filled area) and expected (red dashed line) 95\% CL exclusion in the plane
of ${\mathcal{B}}(T \to Wb)$ versus ${\mathcal{B}}(T \to Ht)$, for different values of
the vector-like $T$ quark mass for the combination of the 1-lepton and 0-lepton searches.
In the figure, the branching ratio is denoted ``BR''.
The background estimate used in the computation of the limits is the result obtained from the background-only fit to data.
Also shown are the expected exclusions by the individual searches, which can be compared to that obtained through their combination.
The grey (light shaded) area corresponds to the unphysical region where the sum of branching ratios exceeds unity, or is smaller than zero.
The default branching ratio values from the \textsc{Protos} event generator for the weak-isospin singlet and doublet cases
are shown as plain circle and star symbols, respectively.
\label{fig:limits2D_COMB}}
\end{figure*}
 
\begin{figure*}[tbp]
\centering
\subfloat[]{\includegraphics[width=0.48\textwidth]{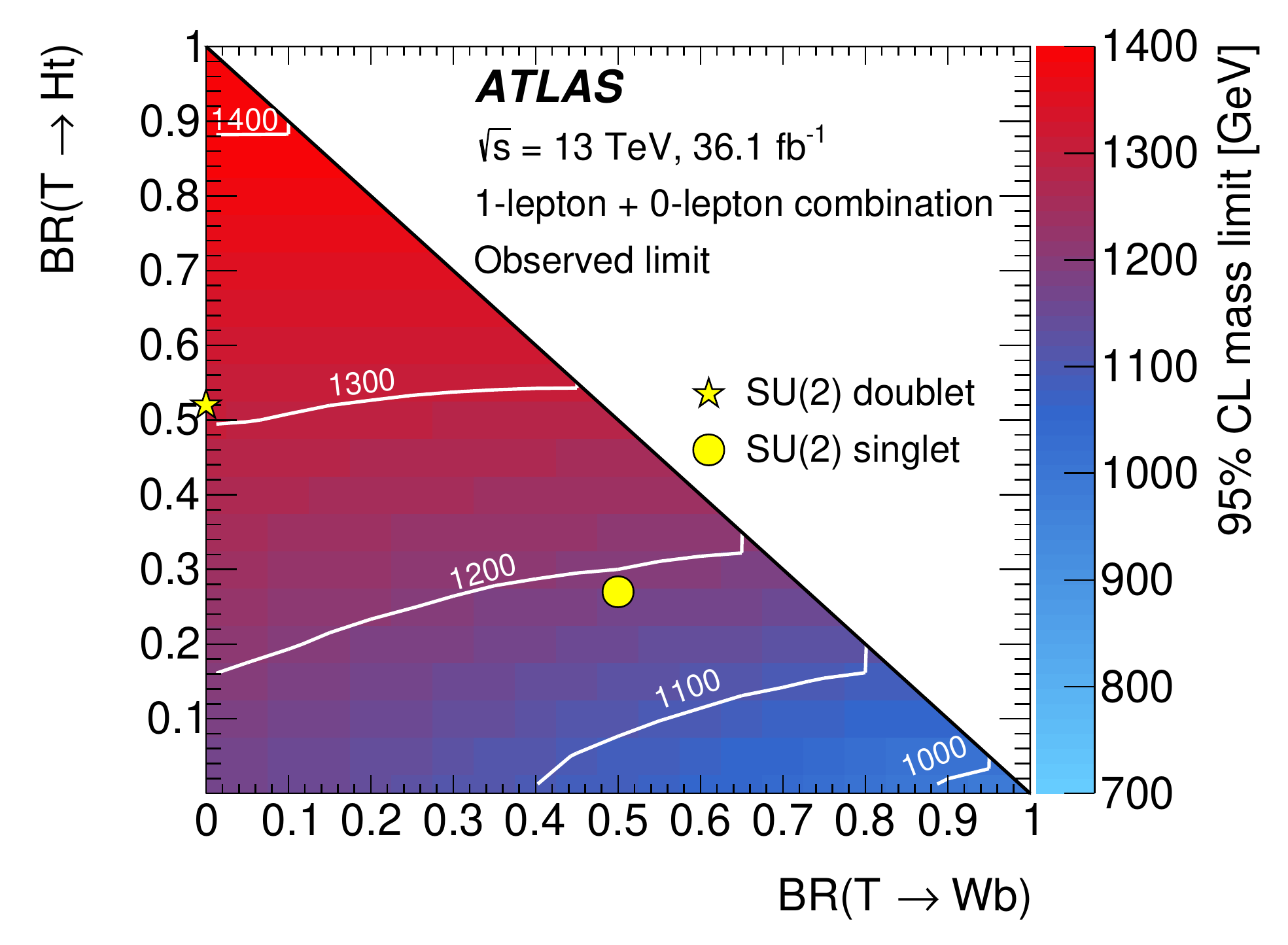}}
\subfloat[]{\includegraphics[width=0.48\textwidth]{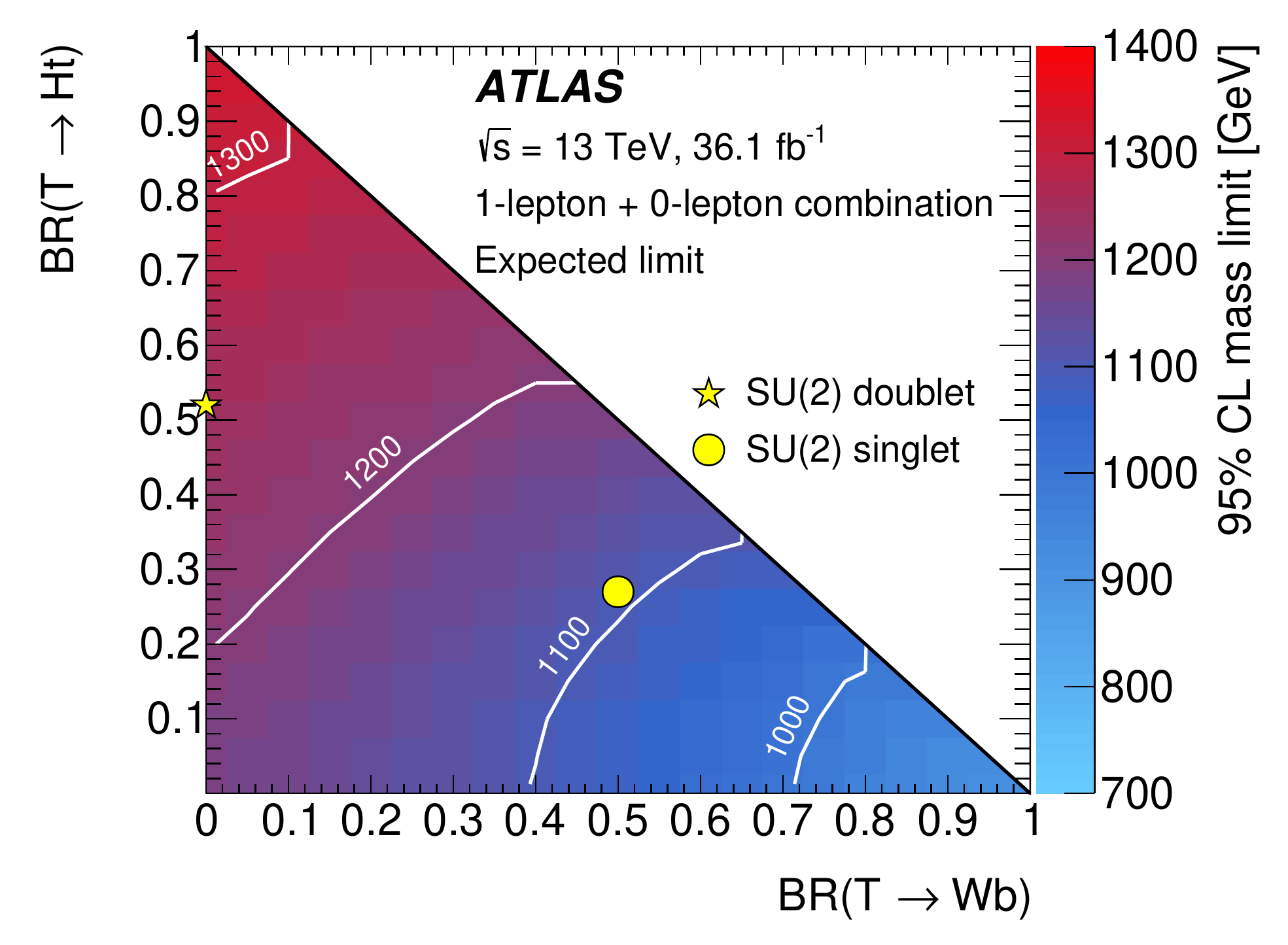}}
\caption{(a) Observed and (b) expected limit (95\% CL) on the mass of the $T$ quark in the plane of
${\mathcal{B}}(T \to Ht)$ versus ${\mathcal{B}}(T \to Wb)$ for the combination of the 1-lepton and
0-lepton searches.
In the figure, the branching ratio is denoted ``BR''.
The background estimate used in the computation of the limits is the result obtained from the
background-only fit to data. Contour lines are provided to guide the eye. The yellow markers indicate
the branching ratios for the SU(2) singlet and doublet scenarios with masses above $\approx 800$~\gev,
where they are approximately independent of the $T$ quark mass.
\label{fig:limits2D_temp_1L0L}}
\end{figure*}
 
\subsection{Limits on four-top-quark production}

The 1-lepton search is used to set limits on BSM four-top-quark production by considering different signal benchmark
scenarios (see Section~\ref{sec:signal_model} for details).
In the case of $\fourtop$ production via an EFT model with a four-top-quark contact interaction,
the observed (expected) 95\% CL upper limit on the production cross section is 16~fb ($31^{+12}_{-9}$~fb).
The upper limit on the production cross section can be translated into an observed (expected)
limit on the free parameter of the model $|C_{4t}|/\Lambda^2<1.6~\tev^{-2}\;(2.3\pm 0.4~\tev^{-2})$.
In the context of the 2UED/RPP model, the observed and expected upper limits on the production cross section
times branching ratio are shown in Figure~\ref{fig:limits_ued_10} as a function of $m_{\KK}$
for the symmetric case ($\xi=R_4/R_5=1$), assuming production by tier (1,1) alone.
The comparison to the LO theoretical cross section translates into an observed (expected) 95\%  CL  limit
on $m_{KK}$ of $1.8~\tev$ ($1.7~\tev$).

\begin{figure*}[h!]
\centering
\includegraphics[width=0.48\textwidth]{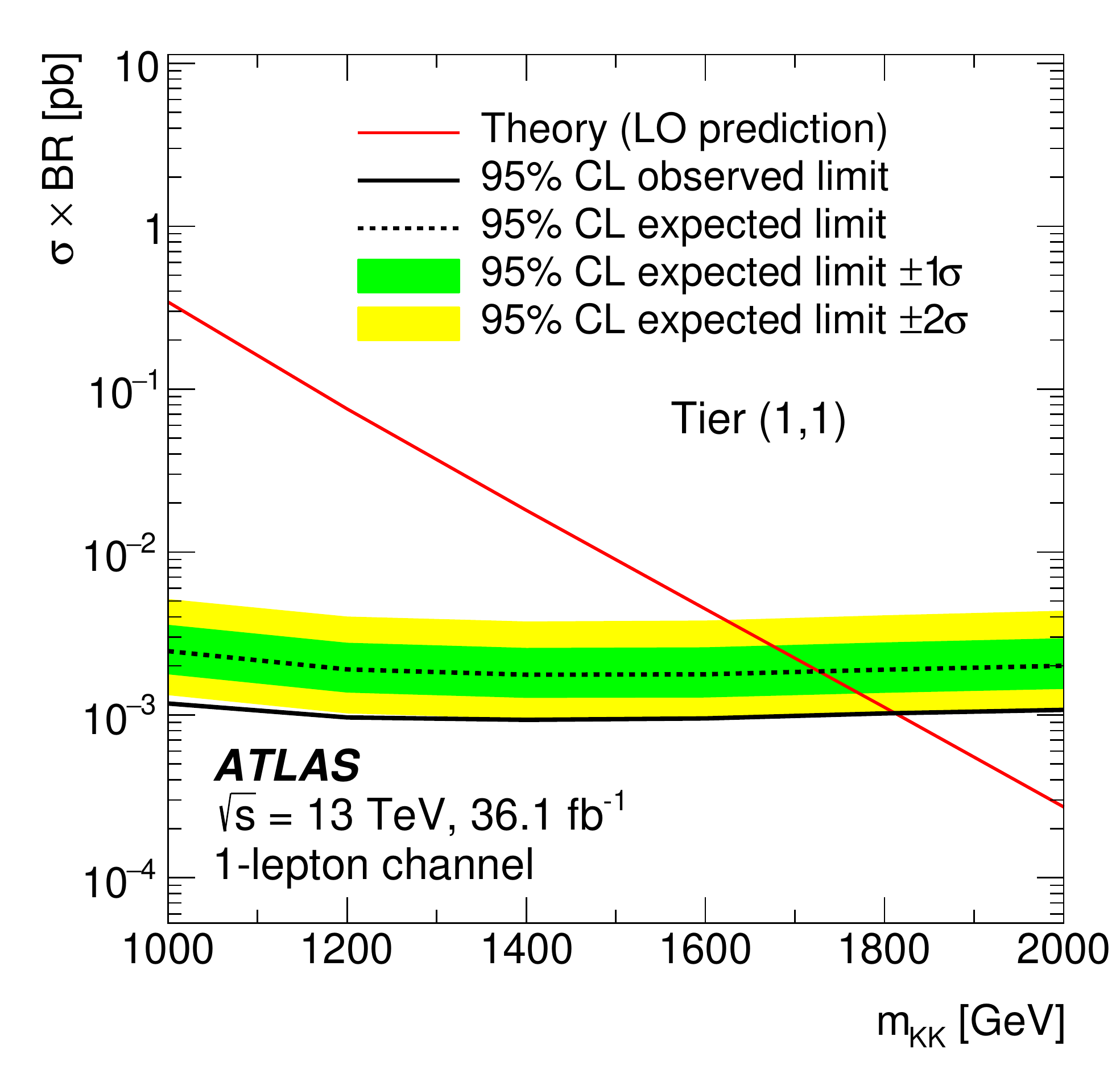}
\caption{
Observed (solid line) and expected (dashed line) 95\% CL upper limits on the production cross section times branching ratio
of four-top-quark events as a function of the Kaluza--Klein mass ($m_{\KK}$) from tier (1,1) in the symmetric case ($\xi=R_4/R_5=1$).
The background estimate used in the computation of the limits is the result obtained from the
background-only fit to data.
The surrounding shaded bands correspond to $\pm 1$ and $\pm 2$ standard deviations around the expected limit.
The thin red line shows the theoretical prediction, computed at LO in QCD, for the production cross section of four-top-quark events
by tier (1,1) assuming ${\mathcal{B}}(A^{(1,1)}\to t\bar{t})=1$, where the heavy photon $A^{(1,1)}$ is the lightest particle of this tier.
\label{fig:limits_ued_10}}
\end{figure*}
 
% End of text imported from the .//sections/results.tex input file
 
\FloatBarrier
 
% The next lines are included from the .//sections/conclusion.tex input file
\section{Conclusion}
\label{sec:conclusion}
 
A search for pair production of up-type vector-like quarks ($T$) with significant branching ratio into a top quark and either a
Standard Model Higgs boson or a $Z$ boson is presented. The same analysis is also used to search for
four-top-quark production, in several new physics scenarios.
The search is based on $pp$ collisions at $\sqrt{s}=13~\tev$ recorded in 2015 and 2016 with the ATLAS detector at the
CERN Large Hadron Collider and corresponds to an integrated luminosity of 36.1 fb$^{-1}$.
Data are analysed in the lepton+jets final state, characterised by an isolated
electron or muon with high transverse momentum, large missing transverse momentum and multiple jets,
as well as the jets+$E_{\text{T}}^{\text{miss}}$ final state, characterised by multiple jets and large missing transverse momentum.
The search exploits the high multiplicity of $b$-jets, the high scalar sum of transverse momenta of all final-state objects, and
the presence of boosted, hadronically decaying top quarks and Higgs bosons reconstructed as large-radius jets, characteristic of signal events.
 
No significant excess of events above the Standard Model expectation is observed, and 95\% CL lower limits are placed on the
mass of the vector-like $T$ quark under several branching ratio hypotheses assuming contributions only from $T \to Wb$, $Zt$, $Ht$.
The 95\% CL observed lower limits on the $T$ quark mass lie
between $0.99~\tev$ and $1.43~\tev$ depending on the
values of the branching ratios into the three decay modes.
Assuming ${\mathcal{B}}(T \to Ht)=1$ and ${\mathcal{B}}(T \to Zt)=1$,  observed (expected) 95\% CL limits of $m_{T}>1.43~\tev\,(1.34~\tev)~\tev$
and $m_{T}>1.17\,(1.18)~\tev$, respectively, are obtained.
The observed (expected) 95\% CL limits for a weak-isospin doublet and singlet are $m_{T}>1.31\,(1.26)~\tev$ and $m_{T}>1.19\,(1.11)~\tev$, respectively.
Additionally, upper limits on the four-top-quark production cross section are set in several new physics scenarios.
In the case of $\fourtop$ production from a contact interaction in an EFT model, the observed (expected) 95\% CL upper limit on the production
cross section is 16~fb ($31^{+12}_{-9}$~fb).
In the context of a 2UED/RPP model, 95\% CL observed (expected) lower limits on $m_{KK}$ of $1.8~\tev$ ($1.7~\tev$) are derived.
% End of text imported from the .//sections/conclusion.tex input file
 
\section*{Acknowledgements}
 
% The next lines are included from the .//acknowledgements/Acknowledgements.tex input file
 
We thank CERN for the very successful operation of the LHC, as well as the
support staff from our institutions without whom ATLAS could not be
operated efficiently.
 
We acknowledge the support of ANPCyT, Argentina; YerPhI, Armenia; ARC, Australia; BMWFW and FWF, Austria; ANAS, Azerbaijan; SSTC, Belarus; CNPq and FAPESP, Brazil; NSERC, NRC and CFI, Canada; CERN; CONICYT, Chile; CAS, MOST and NSFC, China; COLCIENCIAS, Colombia; MSMT CR, MPO CR and VSC CR, Czech Republic; DNRF and DNSRC, Denmark; IN2P3-CNRS, CEA-DRF/IRFU, France; SRNSFG, Georgia; BMBF, HGF, and MPG, Germany; GSRT, Greece; RGC, Hong Kong SAR, China; ISF, I-CORE and Benoziyo Center, Israel; INFN, Italy; MEXT and JSPS, Japan; CNRST, Morocco; NWO, Netherlands; RCN, Norway; MNiSW and NCN, Poland; FCT, Portugal; MNE/IFA, Romania; MES of Russia and NRC KI, Russian Federation; JINR; MESTD, Serbia; MSSR, Slovakia; ARRS and MIZ\v{S}, Slovenia; DST/NRF, South Africa; MINECO, Spain; SRC and Wallenberg Foundation, Sweden; SERI, SNSF and Cantons of Bern and Geneva, Switzerland; MOST, Taiwan; TAEK, Turkey; STFC, United Kingdom; DOE and NSF, United States of America. In addition, individual groups and members have received support from BCKDF, the Canada Council, CANARIE, CRC, Compute Canada, FQRNT, and the Ontario Innovation Trust, Canada; EPLANET, ERC, ERDF, FP7, Horizon 2020 and Marie Sk{\l}odowska-Curie Actions, European Union; Investissements d'Avenir Labex and Idex, ANR, R{\'e}gion Auvergne and Fondation Partager le Savoir, France; DFG and AvH Foundation, Germany; Herakleitos, Thales and Aristeia programmes co-financed by EU-ESF and the Greek NSRF; BSF, GIF and Minerva, Israel; BRF, Norway; CERCA Programme Generalitat de Catalunya, Generalitat Valenciana, Spain; the Royal Society and Leverhulme Trust, United Kingdom.
 
The crucial computing support from all WLCG partners is acknowledged gratefully, in particular from CERN, the ATLAS Tier-1 facilities at TRIUMF (Canada), NDGF (Denmark, Norway, Sweden), CC-IN2P3 (France), KIT/GridKA (Germany), INFN-CNAF (Italy), NL-T1 (Netherlands), PIC (Spain), ASGC (Taiwan), RAL (UK) and BNL (USA), the Tier-2 facilities worldwide and large non-WLCG resource providers. Major contributors of computing resources are listed in Ref.~\cite{ATL-GEN-PUB-2016-002}.
 
% End of text imported from the .//acknowledgements/Acknowledgements.tex input file
\printbibliography

\clearpage
\addcontentsline{toc}{part}{Auxiliary material}

\clearpage % ATLAS Collaboration author list
% Reference date of EXOT-2016-13 is 2017-12-18
% Author list last updated on date 17-JUL-18
% Data extracted on 17-Jul-2018 for paper reference EXOT-2016-13
% at 1:16pm
 
\begin{flushleft}
{\Large The ATLAS Collaboration}

\bigskip

M.~Aaboud$^\textrm{\scriptsize 34d}$,    
G.~Aad$^\textrm{\scriptsize 99}$,    
B.~Abbott$^\textrm{\scriptsize 124}$,    
O.~Abdinov$^\textrm{\scriptsize 13,*}$,    
B.~Abeloos$^\textrm{\scriptsize 128}$,    
S.H.~Abidi$^\textrm{\scriptsize 165}$,    
O.S.~AbouZeid$^\textrm{\scriptsize 143}$,    
N.L.~Abraham$^\textrm{\scriptsize 153}$,    
H.~Abramowicz$^\textrm{\scriptsize 159}$,    
H.~Abreu$^\textrm{\scriptsize 158}$
% ...
\bigskip
\\

% ...
$^{13}$Institute of Physics, Azerbaijan Academy of Sciences, Baku; Azerbaijan.\\
% ...
$^{34}$$^{(a)}$Facult\'e des Sciences Ain Chock, R\'eseau Universitaire de Physique des Hautes Energies - Universit\'e Hassan II, Casablanca;$^{(b)}$Centre National de l'Energie des Sciences Techniques Nucleaires (CNESTEN), Rabat;$^{(c)}$Facult\'e des Sciences Semlalia, Universit\'e Cadi Ayyad, LPHEA-Marrakech;$^{(d)}$Facult\'e des Sciences, Universit\'e Mohamed Premier and LPTPM, Oujda;$^{(e)}$Facult\'e des sciences, Universit\'e Mohammed V, Rabat; Morocco.\\
% ...
$^{99}$CPPM, Aix-Marseille Universit\'e, CNRS/IN2P3, Marseille; France.\\
% ...
$^{124}$Homer L. Dodge Department of Physics and Astronomy, University of Oklahoma, Norman OK; United States of America.\\
% ...
$^{128}$LAL, Universit\'e Paris-Sud, CNRS/IN2P3, Universit\'e Paris-Saclay, Orsay; France.\\
% ...
$^{143}$Santa Cruz Institute for Particle Physics, University of California Santa Cruz, Santa Cruz CA; United States of America.\\
% ...
$^{153}$Department of Physics and Astronomy, University of Sussex, Brighton; United Kingdom.\\
% ...
$^{158}$Department of Physics, Technion, Israel Institute of Technology, Haifa; Israel.\\
$^{159}$Raymond and Beverly Sackler School of Physics and Astronomy, Tel Aviv University, Tel Aviv; Israel.\\
% ...
$^{165}$Department of Physics, University of Toronto, Toronto ON; Canada.\\
% ...
$^{a}$ Also at  Department of Physics, University of Malaya, Kuala Lumpur; Malaysia.\\
$^{b}$ Also at Borough of Manhattan Community College, City University of New York, NY; United States of America.\\
$^{c}$ Also at Centre for High Performance Computing, CSIR Campus, Rosebank, Cape Town; South Africa.\\
$^{d}$ Also at CERN, Geneva; Switzerland.\\
$^{e}$ Also at CPPM, Aix-Marseille Universit\'e, CNRS/IN2P3, Marseille; France.\\
% ...
$^{*}$ Deceased

\end{flushleft}

% Created with Glance <Atlas.Glance@cern.ch>

\end{document}